\documentclass[twoside,12pt]{article}
\usepackage{epsfig}

\usepackage[a4paper, margin=2cm]{geometry}
\usepackage{geometry}                		
\usepackage{graphicx}				
\usepackage{amssymb}
\usepackage{amsmath}
\usepackage{bm}
\usepackage{mathtools}
\usepackage{flexisym}
\usepackage{color}
\usepackage{float}
\usepackage[figuresright]{rotating} 		
\usepackage{comment}				

\begin{document}

\begin{center}
\LARGE\bf Double field inflation of generalized dilaton-axion models with a new Fayet-Iliopoulos (FI) term
\end{center}

\footnotetext{\hspace*{-0cm}\footnotesize MAN Ping Kwan, Ellgan, E-mail:  ellgan101@akane.waseda.jp}

\begin{center}
\rm MAN Ping Kwan, Ellgan$^{\rm a)\dagger}$
\end{center}

\begin{center}
\begin{footnotesize} \sl
${Department \; of \; Pure \; and \; Applied \; Physics, \; Waseda \; University}^{\rm a)}$ \\
\end{footnotesize}
\end{center}

\section{Abstract}
\noindent We study the inflation dynamics of generalized dilaton-axion models with a new Fayet-Iliopoulos (FI) term. In particular, we find the relationships between the super-potential parameters and the coefficient of natural logarithm of the real part of dilaton-axion fields stored in the K\"ahler potential based on the vacuum conditions at the end of inflation. We also evaluate the feasible initial field values, their corresponding SUSY breaking scales and the iso-curvature parameters. 

\section{Introduction}
\begin{table}[h!]
\begin{center}
\begin{tabular}{ |c|c|c|c| }
\hline
$\text{Slow-roll parameters}$ & $\text{Range(s)}$ & $\text{Spectral indices}$ & $\text{Range(s)}$ \\
\hline
${\epsilon}_{V}$ & $<0.004$ & ${n}_{s} - 1$ & $[-0.0423, -0.0327]$ \\
\hline
${\eta}_{V}$ & $[-0.021, -0.008]$ & $\frac{d {n}_{s}}{d \ln{k}}$ & $[-0.008, 0.012]$ \\
\hline
${\xi}_{V}$ & $[-0.0045, 0.0096]$ & $\frac{d^{2} {n}_{s}}{d \ln{k}^{2}}$ & $[-0.003, 0.023]$ \\
\hline
${H}_{\text{hc}}$ & $< 2.7 \times {10}^{-5} {M}_{\text{pl}}$ & ${V}_{\text{hc}}$ & $< \left( 1.7 \times {10}^{16} \; \text{GeV} \right)^{4}$ \\
\hline
\end{tabular}
\end{center}
\caption{Slow roll potential parameters and spectral indices in Planck 2018}
\label{table:Planck data 2018 slow roll potential parameters and spectral indices}
\end{table}

\noindent Inflation is the vital part for studying the birth of our universe and verifying the consistency of the theory of quantum gravity \cite{{Phys. Rev. D23 (1981), 347-356}, {Phys.Lett. 91B (1980), 99-102}, {Mon. Not. Roy. Astron. Soc. 195 (1981), 467-479}}. The latest observation data are shown in Table \ref{table:Planck data 2018 slow roll potential parameters and spectral indices} \cite{1807.06211}, which are consistent with the slow roll approximation. There are many models satisfying the observation \cite{1303.3787, 1807.06211}, one of which is the most promising model - Starobinsky model (also called $R^2$ inflation model), which can be motivated by, for example, modified gravity theory \cite{{Phys.Lett. 91B (1980), 99-102}, {1303.3787}, {1311.0744}, {1504.01772}, {1603.05537}, {1705.11098}, {1708.08346}} and supergravity theory (SUGRA) \cite{{hep-ph/0004243}, {1011.5945}, {1307.7696}, {1404.3739}}. Since SUGRA describes high energy physics beyond the Standard model (SM), gravity beyond the $\Lambda$CDM cosmological model and the low energy effective theory of superstring theory, it has been adopted to study the primordial dynamics of the universe evolution. In particular, the extension of Starobinsky model in the new-minimal formulation of SUGRA is dual to the standard counterpart coupled to a massive vector multiplet, or a massless vector multiplet and a St\"uckelberg chiral multiplet with the K\"ahler potential in the following form \cite{{1307.1137}, {1309.1085}, {1608.02121}}
\begin{equation}
\frac{K \left( T, \bar{T} \right)}{ {M}_{\text{pl}}^{2} } = - {\rho} \ln{\left( \frac{ T + \bar{T} }{{M}_{\text{pl}}} \right)} + \tilde{\beta} \left( \frac{T + \bar{T}}{{M}_{\text{pl}}} \right), \\
\end{equation}
\noindent where the last term is the FI term of the gauged $R$-symmetry and spontaneous SUSY breaking occurred after the inflation. \cite{1608.02121} showed that this model can be consistent with observations only for ${\rho} = 1, 2$ with some non-perturbative corrections and SUSY is spontaneously broken at the dS vacuum with gravitino mass in the TeV range. Furthermore, \cite{1907.10373} finds that an alternative non-vanishing FI terms \cite{{1712.08601}, {1801.04794}} can allow inflationary models having the following forms of K\"ahler potential $K$
\begin{equation}
\label{Kahler potential without linear terms}
\frac{K \left( T, \bar{T} \right)}{{M}_{\text{pl}}^{2}} = - {\rho} \ln{\left( \frac{T + \overline{T}}{ {M}_{\text{pl}} } \right)}. \\
\end{equation}
\noindent This K\"ahler potential model generalizes various models such as no-scale SUGRA \cite{{Phenomenological SU(1,1) supergravity}, {1503.08867}, {1507.02308}, {1711.11051}, {1809.10114}, {1812.02192}, {1906.10176}} and M theory compactification on a ${G}_{2}$ manifold \cite{{1009.4439},{1010.3173},{1610.04163}}. The model studied in \cite{1907.10373} describes the inflation from the real part of dilaton-axion chiral superfield $T$ with a fixed imaginary part at the dS vacuum. Thus, it is interesting to see how the inflation dynamics changes when the imaginary part also evolves to contribute to the inflation dynamics. That is the main theme of this paper. \\

\vspace{3mm}

\noindent In this paper, we are going to study the inflation dynamics of generalized dilaton-axion models, where both the dilaton and its axionic partner simultaneously contribute to the evolution of inflation. The organization of this paper is as follows. In section \ref{General model}, we give the total potential consisting of $F$ term from the K\"ahler potential given by Eq.(\ref{Kahler potential without linear terms}) and the Polonyi type super-potential, and $D$ term potential from an alternative FI term \cite{{1712.08601}, {1801.04794}}. We give the derivatives of the total potential, the possible ranges of parameters under the constraints of dS vacuum and the corresponding SUSY breaking scales in section \ref{Derivatives of the total potential}. We summarize our results of dS vacuum in section \ref{A short summary} and the equations of motion (E.O.M.s) of the evolution of dilaton and axion. In section \ref{Numerical calculations}, we give the initial conditions and parameters feasible for inflation based on the constraints of Planck observation listed in Table \ref{table:Planck data 2018 slow roll potential parameters and spectral indices} and e-folding number $50$ to $60$. We discuss our results in section \ref{Discussion} and draw conclusions in section \ref{Conclusion}. 

\section{General model}
\label{General model}
\noindent Recall the $F$ term potential in $\mathcal{N}=1$ supergravity (SUGRA) is 
\begin{equation}
\label{F term potential}
{V}_{F} = {e}^{ \frac{K}{ {M}^{2}_{\text{pl}} } } \left( {K}^{I \bar{J}} {D}_{I} W {D}_{\bar{J}} \bar{W} - \frac{3}{ {M}^{2}_{\text{pl}} } |W|^{2} \right). \\
\end{equation}
\noindent where 
\begin{equation}
{K}_{I \bar{J}} \equiv \frac{{\partial}^{2} K}{{\partial} {\Phi}^{I} {\partial} \bar{{\Phi}}^{\bar{J}}}, \quad {D}_{I} W \equiv {\partial}_{I} W + \frac{1}{{M}^{2}_{\text{pl}}} \left( {\partial}_{I} K \right) W = \frac{{\partial} W}{{\partial} {\Phi}^{I}} + \frac{1}{{M}^{2}_{\text{pl}}} \frac{ {\partial} K }{ {\partial} {\Phi}^{I} } W, \\
\end{equation}
\noindent ${\Phi}^{I}$ is complex scalar field for all $I$ from $1$ to the dimension of the field space, the upper bar means the complex conjugate of the corresponding variables/ fields, ${K}^{I \bar{J}}$ is the inverse of the K\"ahler metric ${K}_{I \bar{J}}$ and ${M}_{\text{pl}}$ is the (reduced) Planck mass\footnote{Since the mass dimensions of K\"ahler potential $K$ and super-potential $W$ are $2$ and $3$ respectively, the (F term) potential ${V}_{F}$ has the mass dimension of 4. (i.e. $\left[ K \right] = {M}^{2}_{\text{pl}}$, $\left[ W \right] = {M}^{3}_{\text{pl}}$, $\left[ {V}_{F} \right] = {M}^{4}_{\text{pl}}$) }. Note that the K\"ahler potential is given by 
\begin{equation}
\label{Kahler potential}
\frac{K \left( T, \bar{T} \right)}{ {M}^{2}_{\text{pl}} } = - {\rho} \ln{  \left( \frac{ T + \bar{T} }{ {M}_{\text{pl}} } \right) } , \\
\end{equation}
\noindent while the super-potential is given by
\begin{equation}
W \left( T \right) = {\lambda} + {\mu} T, \\
\end{equation}
\noindent where ${\rho} \in \mathbb{N}$ and  ${\lambda}, \; {\mu} \in \mathbb{C}$ are constants\footnote{The mass dimensions of ${\rho}$, ${\lambda}$, ${\lambda}_{R}$, ${\lambda}_{I}$, ${\mu}$, ${\mu}_{R}$, ${\mu}_{I}$, $T$, ${T}_{R}$ and ${T}_{I}$ are $\left[ {\rho} \right] = 1$, $\left[ {\lambda} \right] = \left[ {\lambda}_{R} \right] = \left[ {\lambda}_{I} \right] = {M}^{3}_{\text{pl}}$, $\left[ {\mu} \right] = \left[ {\mu}_{R} \right] = \left[ {\mu}_{I} \right] = {M}^{2}_{\text{pl}}$, $\left[ T \right] = \left[ {T}_{R} \right] = \left[ {T}_{I} \right] = {M}_{\text{pl}}$. }. Note that the first and second derivatives of the K\"ahler potential are given by
\begin{equation}
{K}_{T} = {K}_{\overline{T}} = \frac{- {\rho} {M}^{2}_{\text{pl}} }{T + \overline{T}}, \quad {K}_{T \overline{T}} = \frac{ {\rho} {M}^{2}_{\text{pl}} }{\left( T + \overline{T} \right)^{2}}, \\
\end{equation}

\noindent The first covariant (K\"ahler) derivative of potential is
\begin{equation}
{D}_{T} W = {\mu} - \frac{ {\rho} }{T + \overline{T}} \left( {\lambda} + {\mu} T \right), \\
\end{equation}
\noindent and the $F$ term potential is
\begin{equation}
{V}_{F} = \frac{-1}{M_{\text{pl}}^2}  \left(\frac{\bar{T}+T}{M_{\text{pl}}}\right)^{-{\rho} }
 \left\{ \bar{\lambda } \left(\mu  \bar{T}+2 \lambda +3 \mu  T\right)+\bar{\mu }
   \left(\bar{T} (3 \lambda +2 \mu  T)+\lambda  T\right)\right\}, \\
   \end{equation}
\noindent or in terms of ${\lambda} = {\lambda}_{R} + i {\lambda}_{I}$, ${\mu} = {\mu}_{R} + i {\mu}_{I}$, ${T} = {T}_{R} + i {T}_{I}$, ${\omega}_{1} = 2 \left( {\lambda}_{R} {\mu}_{R} + {\lambda}_{I} {\mu}_{I} \right)$, ${\omega}_{2} = 2 \left( {\lambda}_{I} {\mu}_{R} - {\lambda}_{R} {\mu}_{I} \right)$, ${\omega}_{1}^{2} + {\omega}_{2}^{2} = 4 \left| {\mu} \right|^{2} \left| {\lambda} \right|^{2}$, ${\omega}_{1} = {\iota}_{1} \left| {\mu} \right|^{2}$, ${\omega}_{2} = {\iota}_{2} \left| {\mu} \right|^{2}$,

\begin{equation}
\begin{split}
{M}_{\text{pl}}^{2} {V}_{F} =&\; \left( \frac{ {M}_{\text{pl}} }{ 2 {T}_{R} } \right)^{\rho} \left\{ \left( {\rho} - 3 \right) \left( \left|{\lambda} \right|^{2} + {\omega}_{2} {T}_{I} + \left| {\mu} \right|^{2} {T}^{2}_{I} \right) + \left( {\rho} - 5 \right) {\omega}_{1} {T}_{R} + \frac{ \left( {\rho}^{2} - 7 {\rho} + 4 \right) }{\rho} \left| {\mu} \right|^{2} {T}^{2}_{R} \right\}. \\
\end{split}
\end{equation}

\noindent For simplicity, we let
\begin{equation}
A \left( {T}_{I} \right) = \frac{1}{ {M}^{2}_{\text{pl}} } \left( \frac{ {M}_{\text{pl}} }{ 2 } \right)^{\rho} \left( {\rho} - 3 \right) \left( \left| {\lambda} \right|^{2} + {\omega}_{2} {T}_{I} + \left| {\mu} \right|^{2} {T}^{2}_{I} \right), \\
\end{equation}
\begin{equation}
\label{B and C}
B = \frac{1}{ {M}^{2}_{\text{pl}} } \left( \frac{ {M}_{\text{pl}} }{ 2 } \right)^{\rho} \left( {\rho} - 5 \right) {\omega}_{1}, \quad C = \frac{1}{ {M}^{2}_{\text{pl}} } \left( \frac{ {M}_{\text{pl}} }{ 2 } \right)^{\rho} \frac{ \left( {\rho}^{2} - 7 \rho + 4 \right) }{\rho} \left| {\mu} \right|^{2}, \\
\end{equation}

\noindent so that the $F$ term potential becomes
\begin{equation}
\label{F term potential in general}
{V}_{F} = A \left( {T}_{I} \right) {T}^{- {\rho}}_{R} + B {T}^{1 - {\rho}}_{R} + C {T}^{2 - {\rho}}_{R}. \\
\end{equation}

\noindent Even though the vacuum of a $F$ term potential is generally AdS, we can obtain the dS vacuum by introducing an abelian vector multiplet with the simplest Fayet-Iliopoulos (FI) type term \cite{1712.08601, 1801.04794} and eliminating the auxiliary field of the vector multiplet, which has a positive contribution \cite{1907.10373}
\begin{equation}
{V}_{D} = \frac{1}{2} g^2 {\xi}^{2}, \\
\end{equation}
\noindent where $g$ is the gauge coupling and $\xi$ is the real FI constant. Hence, in the rest of this paper, we investigate the properties and inflation dynamics of the total potential $V = {V}_{F} + {V}_{D}$. 

\section{Derivatives of the total potential}
\label{Derivatives of the total potential}
\noindent To find our present universe after inflation, we set up some constraints on the derivatives of the $F$ term potential to find the minimum as follows. The first derivatives are
\begin{equation}
\begin{split}
\frac{ {\partial} {V}}{ {\partial} {T}_{R} } =&\; - {\rho} A \left( {T}_{I} \right) {T}^{- {\rho} - 1}_{R} + \left( 1 - {\rho} \right) B {T}^{ - {\rho}}_{R} + \left( 2 - {\rho} \right) C {T}^{1 - {\rho}}_{R}, \\
\end{split}
\end{equation}
\begin{equation}
\begin{split}
\frac{ {\partial} {V}}{ {\partial} {T}_{I} } =&\; \frac{1}{{M}^{2}_{\text{pl}}} \left(\frac{M_{\text{pl}}}{2 T_R}\right)^{ {\rho} } \left( {\rho} - 3 \right) \left( {\omega}_{2} + 2 \left| {\mu} \right|^2 {T}_{I} \right), \\
\end{split}
\end{equation}
\noindent while the second derivatives are
\begin{equation}
\begin{split}
\frac{ {\partial}^{2} {V}}{ {\partial} {T}^{2}_{R} } =&\; \left( {\rho} + 1 \right) \rho A \left( {T}_{I} \right) {T}^{- {\rho} - 2}_{R} + {\rho} \left( {\rho} - 1 \right) B {T}^{ - {\rho} - 1}_{R} + \left( {\rho} - 1 \right) \left( {\rho} - 2 \right) C {T}^{- {\rho}}_{R}, \\
\end{split}
\end{equation}

\begin{equation}
\begin{split}
\frac{ {\partial}^{2} {V}}{ {\partial} {T}_{R} {\partial} {T}_{I} } = \frac{ {\partial}^{2} {V}}{ {\partial} {T}_{I} {\partial} {T}_{R} } =&\; \frac{-2}{M_{\text{pl}}^3} \left(\frac{M_{\text{pl}}}{2 T_R}\right)^{{\rho} + 1} ({\rho} - 3) {\rho} \left( {\omega}_{2} + 2 \left| \mu \right|^2 {T}_{I} \right), \\
\end{split}
\end{equation}

\begin{equation}
\begin{split}
\frac{ {\partial}^{2} {V}}{ {\partial} {T}^{2}_{I} } =&\; \frac{2}{ {M}^{2}_{\text{pl}} } \left(\frac{M_{\text{pl}}}{2 T_R}\right)^{{\rho} } \left( {\rho} - 3 \right) \left| \mu \right|^2. \\
\end{split}
\end{equation}
\noindent The minimum points are the solutions of the following equations

\begin{equation}
\begin{split}
\frac{ {\partial} {V}}{ {\partial} {T}_{R} } =&\; - {T}^{- {\rho} - 1}_{R} \left\{ \rho A \left( {T}_{I} \right) + \left( {\rho} - 1 \right) B {T}_{R} + \left( {\rho} - 2 \right) C {T}^{2}_{R} \right\} = 0 \\
\Rightarrow \; {T}_{R} =&\; \frac{ - \left( \rho - 1 \right) B \pm \sqrt{ \left( \rho - 1 \right)^{2} B^{2} - 4 \rho \left( \rho - 2 \right) A \left( {T}_{I} \right) C } }{2 \left( \rho - 2 \right) C}, \\
\end{split}
\end{equation}

\begin{equation}
\frac{ {\partial} {V}}{ {\partial} {T}_{I} } = \frac{1}{{M}^{2}_{\text{pl}}} \left(\frac{M_{\text{pl}}}{2 T_R}\right)^{{\rho} } ({\rho} -3) \left( {\omega}_{2} + 2 \left| {\mu} \right|^2 {T}_{I} \right) = 0 \quad \Rightarrow \quad {T}_{I} = \frac{ - {\omega}_{2} }{2 \left| {\mu} \right|^{2} } = - \frac{{\iota}_{2}}{2} \quad \text{or} \quad \rho = 3, \\
\end{equation}

\subsection{Suppose $\rho \neq 3$. }
\noindent When ${T}_{I} = \frac{ - {\omega}_{2} }{2 \left| {\mu} \right|^{2} } = {T}_{I0}$, we have
\begin{equation}
A \left( {T}_{I0} \right) = \frac{ \rho - 3 }{ {M}^{2}_{\text{pl}} } \left( \frac{ {M}_{\text{pl}} }{2} \right)^{\rho} \left( \left| {\lambda} \right|^{2} - \frac{ {\omega}^{2}_{2} }{ 4 \left| {\mu} \right|^{2} } \right) = \frac{ \rho - 3 }{ {M}^{2}_{\text{pl}} } \left( \frac{ {M}_{\text{pl}} }{2} \right)^{\rho} \frac{ {\omega}^{2}_{1} }{ 4 \left| {\mu} \right|^{2} }, \\
\end{equation}
\noindent and $B$ and $C$ are given by Eq.(\ref{B and C}). The critical values of ${T}_{R}$ are given by
\begin{equation}
{T}_{R0} = \frac{ \rho \left[ - \left( \rho - 1 \right) \left( \rho - 5 \right) \pm \left( \rho + 1 \right) \right] }{2 \left( \rho - 2 \right) \left( {\rho}^2 - 7 {\rho} + 4 \right) } \frac{ {\omega}_{1} }{ \left| {\mu} \right|^{2} } = \frac{- {\rho} \left( {\rho} - 3 \right) }{ 2 \left( {\rho}^{2} - 7 {\rho} + 4 \right) } \frac{ {\omega}_{1} }{ \left| {\mu} \right|^{2} } \quad \text{or} \quad \frac{ - {\rho} }{2 \left( {\rho} - 2 \right) } \frac{ {\omega}_{1} }{ \left| {\mu} \right|^{2} }. \\
\end{equation}
\noindent Defining 
\begin{equation}
{T}_{R0-} = \frac{- {\rho} \left( {\rho} - 3 \right) {\iota}_{1} }{ 2 \left( {\rho}^{2} - 7 {\rho} + 4 \right) } \quad \text{and} \quad {T}_{R0+} = \quad \frac{ - {\rho} {\iota}_{1} }{2 \left( {\rho} - 2 \right) }, 
\end{equation}
\noindent we know that the positivity of ${T}_{R0}$ give the ranges\footnote{Note that the roots of ${\rho}^{2} - 7 {\rho} + 4 = 0$ are ${\rho} = \frac{1}{2} \left( 7 - \sqrt{33} \right) \approx 0.627719$ or ${\rho} = \frac{1}{2} \left( 7 + \sqrt{33} \right) \approx 6.37228$. } of ${\rho}$ as shown in Table \ref{table:Positivity of TR0}.

\begin{table}[h]
\begin{center}
\begin{tabular}{ |c|c|c|c| }
\hline
Case(s) & ${T}_{R0-} > 0$ & ${T}_{R0+} > 0$ \\
\hline
${\omega}_{1} > 0 \; \left( {\iota}_{1} > 0 \right)$ & $0 < {\rho} < \frac{\left( 7- \sqrt{33} \right)}{2}, \; 3 < {\rho} < \frac{\left( 7+ \sqrt{33} \right)}{2}$ & $0 < {\rho} < 2$ \\
\hline
${\omega}_{1} < 0 \; \left( {\iota}_{1} < 0 \right)$ & ${\rho} < 0, \; \frac{\left( 7- \sqrt{33} \right)}{2} < {\rho} < 3, \; {\rho} > \frac{\left( 7+ \sqrt{33} \right)}{2}$ & ${\rho} < 0, \; {\rho} > 2$ \\
\hline
\end{tabular}
\end{center}
\caption{Possible ranges for satisfying the positivity of ${T}_{R0}$ }
\label{table:Positivity of TR0}
\end{table}

\noindent To check whether the potential is physical at the critical points\footnote{It means the eigenvalues of the mass matrix are all positive. }, 
\noindent the mass matrix evaluated at $\left( {T}_{R0-}, {T}_{I0} \right)$ is 
\begin{equation}
\left. 
\begin{pmatrix}
\frac{ {\partial}^{2} {V}}{ {\partial} {T}^{2}_{R} } & \frac{ {\partial}^{2} {V}}{ {\partial} {T}_{R} {\partial} {T}_{I} } \\
\frac{ {\partial}^{2} {V}}{ {\partial} {T}_{I} {\partial} {T}_{R} } & \frac{ {\partial}^{2} {V}}{ {\partial} {T}_{I} {\partial} {T}_{I} }
\end{pmatrix}
\right|_{ \left( {T}_{R0-}, {T}_{I0} \right) }
= \frac{2 \left| {\mu} \right|^{2} }{ {M}^{2}_{\text{pl}} } \left[ \frac{ \left( {\rho}^{2} - 7 {\rho} + 4 \right) {M}_{\text{pl}} }{ - {\rho} \left( {\rho} - 3 \right) {\iota}_{1} } \right]^{\rho} 
\begin{pmatrix}
\frac{- \left( {\rho} + 1 \right) \left( {\rho}^2 - 7 {\rho} + 4 \right) }{ {\rho} \left( {\rho} - 3 \right) }& 0 \\
0&  \left( {\rho} - 3 \right) \\
\end{pmatrix},
\end{equation}
\noindent which has two positive eigenvalues if $3 < {\rho} < \frac{7 + \sqrt{33} }{2}$. The corresponding square masses of ${T}_{R}$ and ${T}_{I}$ (evaluated at $\left( {T}_{R0-}, {T}_{I0} \right)$) are 

\begin{equation}
\left. {M}_{{T}_{R}}^{2} \right|_{\left( {T}_{R0-}, {T}_{I0} \right)} = \frac{2 \left| {\mu} \right|^{2}}{{M}_{\text{pl}}^{2}} \left[ \frac{\left( {\rho}^{2} - 7 {\rho} + 4 \right) {M}_{\text{pl}} }{ - {\rho} \left( {\rho} - 3 \right) {\iota}_{1} } \right]^{\rho} \frac{ \left( {\rho} + 1 \right) \left( {\rho}^{2} - 7 {\rho} + 4 \right)}{- {\rho} \left( {\rho} - 3 \right)}, \\
\end{equation}

\begin{equation}
\left. {M}_{{T}_{I}}^{2} \right|_{\left( {T}_{R0-}, {T}_{I0} \right)} = \frac{2 \left| {\mu} \right|^{2}}{{M}_{\text{pl}}^{2}} \left[ \frac{\left( {\rho}^{2} - 7 {\rho} + 4 \right) {M}_{\text{pl}} }{ - {\rho} \left( {\rho} - 3 \right) {\iota}_{1} } \right]^{\rho} \left( {\rho} - 3 \right), \\
\end{equation}

\noindent while their ratio is
\begin{equation}
\left. \left( \frac{ {M}_{{T}_{R}} }{ {M}_{{T}_{I}} } \right)^{2} \right|_{\left( {T}_{R0-}, {T}_{I0} \right)} = \frac{ \left( {\rho} + 1 \right) \left( {\rho}^{2} - 7 {\rho} + 4 \right) }{- {\rho} \left( {\rho} - 3 \right)^{2} }. \\
\end{equation}

\noindent The counterpart evaluated $\left( {T}_{R0+}, {T}_{I0} \right)$
\begin{equation}
\left. 
\begin{pmatrix}
\frac{ {\partial}^{2} {V}}{ {\partial} {T}^{2}_{R} } & \frac{ {\partial}^{2} {V}}{ {\partial} {T}_{R} {\partial} {T}_{I} } \\
\frac{ {\partial}^{2} {V}}{ {\partial} {T}_{I} {\partial} {T}_{R} } & \frac{ {\partial}^{2} {V}}{ {\partial} {T}_{I} {\partial} {T}_{I} }
\end{pmatrix}
\right|_{ \left( {T}_{R0+}, {T}_{I0} \right) }
= \frac{2 \left| {\mu} \right|^{2} }{ {M}^{2}_{\text{pl}} } \left[ \frac{\left( {\rho} - 2 \right) {M}_{\text{pl}} }{ - {\rho} {\iota}_{1} } \right]^{\rho} 
\begin{pmatrix}
\frac{ \left( {\rho} + 1 \right) \left( {\rho} - 2 \right) }{ {\rho} }& 0 \\
0&  \left( {\rho} - 3 \right) \\
\end{pmatrix}. 
\end{equation}
\noindent which has two positive eigenvalues if ${\rho} > 3$ and the corresponding square masses of ${T}_{R}$ and ${T}_{I}$ (evaluated at $\left( {T}_{R0+}, {T}_{I0} \right)$) are 

\begin{equation}
\left. {M}_{{T}_{R}}^{2} \right|_{\left( {T}_{R0+}, {T}_{I0} \right)} = \frac{2 \left| {\mu} \right|^{2}}{{M}_{\text{pl}}^{2}} \left[ \frac{\left( {\rho} - 2 \right) {M}_{\text{pl}} }{ - {\rho} {\iota}_{1} } \right]^{\rho} \frac{ \left( {\rho} + 1 \right) \left( {\rho} - 2 \right)}{ {\rho} }, \\
\end{equation}

\begin{equation}
\left. {M}_{{T}_{I}}^{2} \right|_{\left( {T}_{R0+}, {T}_{I0} \right)} = \frac{2 \left| {\mu} \right|^{2}}{{M}_{\text{pl}}^{2}} \left[ \frac{\left( {\rho} - 2 \right) {M}_{\text{pl}} }{ - {\rho} {\iota}_{1} } \right]^{{\rho}} \left( {{\rho}} - 3 \right), \\
\end{equation}

\noindent while their ratio is 
\begin{equation}
\left. \left( \frac{ {M}_{{T}_{R}} }{ {M}_{{T}_{I}} } \right)^{2} \right|_{\left( {T}_{R0+}, {T}_{I0} \right)} = \frac{ \left( {{\rho}} + 1 \right) \left( {{\rho}} - 2 \right) }{ {{\rho}} \left( {{\rho}} - 3 \right) }. \\
\end{equation}

\noindent Combining with the ranges given by the positivity of ${T}_{R0}$, we can finalize the possible ranges to produce a dS vacuum as shown in Table \ref{table:Positivity of TR0 and positive eigenvalues of the mass matrix}. 
\begin{table}[h]
\begin{center}
\begin{tabular}{ |c|c|c|c| }
\hline
Case(s) & ${T}_{R0-} > 0, \; {\lambda}_{1,2} > 0$ & ${T}_{R0+} > 0, \; {\lambda}_{1,2} > 0$ \\
\hline
${\omega}_{1} > 0 \; \left( {\iota}_{1} > 0 \right)$ & $3 < {{\rho}} < \frac{7 + \sqrt{33} }{2}$ & $\times$ \\
\hline
${\omega}_{1} < 0 \; \left( {\iota}_{1} < 0 \right)$ & $\times$ & ${{\rho}} > 3$ \\
\hline
\end{tabular}
\end{center}
\caption{Possible ranges for satisfying the positivity of ${T}_{R0}$ and positive eigenvalues of the mass matrix ${\lambda}_{1}, {\lambda}_{2}$}
\label{table:Positivity of TR0 and positive eigenvalues of the mass matrix}
\end{table}
\noindent Now, based on the physically feasible ranges, we can obtain the corresponding gravitino mass and the $F$ term and $D$ term SUSY breaking scales.

\subsubsection{Suppose ${\omega}_{1} > 0$, $3 < {{\rho}} < \frac{7 + \sqrt{33} }{2}$. }
\noindent The gravitino mass evaluated at $\left( {T}_{R0-}, {T}_{I0} \right)$ is 
\begin{equation}
\begin{split}
{M}_{3/2-}\equiv&\; \left. \frac{ {e}^{ \frac{K}{2 {M}_{\text{pl}}^{2} } } }{ {M}_{\text{pl}}^{2} } \left| W \left( {T} \right) \right| \right|_{-} = \frac{1}{ {M}_{\text{pl}}^{2} } \left( \frac{ {M}_{\text{pl}} ({\rho} - 2) \left({\rho}^{2} - 7 {\rho} +4 \right) }{- \left( {\rho} - 3 \right) {\rho} {\iota}_{1} } \right)^{{\rho} /2} \\
   &\; \times \left| \frac{ \left( {\rho}^{3} - 10 {\rho}^{2} + 21{\rho} - 8 \right) {\iota}_{1} {\mu} }{2 \left( {\rho} - 2 \right) \left( {\rho}^{2} - 7 {\rho} + 4 \right) } \right|, \\
\end{split}
\end{equation}
\noindent which forms the following ratios with 

\begin{equation}
\left. \left( \frac{ {M}_{{T}_{R}} }{ {M}_{3/2} } \right)^{2} \right|_{\left( {T}_{R0-}, {T}_{I0} \right)} = \frac{- 8 \left( {\rho} + 1 \right) \left( {\rho}^{2} - 7 {\rho} + 4 \right)^{3} }{ \left( {\rho} - 2 \right)^{ \left( {\rho} - 2 \right) } {\rho} \left( {\rho} - 3 \right) \left( {\rho}^{3} - 10 {\rho}^{2} + 21{\rho} - 8 \right)^{2} } \left( \frac{ {M}_{\text{pl}} }{ {\iota}_{1} } \right)^{2}, \\
\end{equation}


\noindent The $F$ term SUSY breaking scale evaluated at $\left( {T}_{R0-}, {T}_{I0} \right)$ is

\begin{equation}
\begin{split}
\left. {F}_{T} \right|_{-} \equiv \left. {e}^{ \frac{K}{2 {M}_{\text{pl}}^{2} } } \left| {D}_{T} W \left( T \right) \right| \right|_{-} =&\; \left( \frac{ {M}_{\text{pl}} ({\rho} - 2) \left({\rho}^2 - 7 {\rho} + 4 \right) }{- ({\rho} -3) {\rho}  {\iota}_{1} } \right)^{{\rho} /2} \\
&\; \times \left| \frac{ \left( {\rho} - 7 \right) \left( {\rho} - 2 \right) \left( {\rho} - 1 \right) {\mu} }{2 \left( {\rho} - 3 \right)} \right|. \\
\end{split}
\end{equation}
\noindent while the $D$ term SUSY breaking scale evaluated at $\left( {T}_{R0-}, {T}_{I0} \right)$ is
\begin{equation}
\begin{split}
\left. \frac{1}{2} g^2 {\xi}^{2} \right|_{-} \equiv&\; \frac{ {\iota}_{1} \left| {\mu} \right|^{2} }{{\rho} {M}_{\text{pl}} } \left(\frac{\left({\rho} ^2-7 {\rho} +4\right) {M}_{\text{pl}}}{- ({\rho} -3) {\rho}  {\iota}_{1} }\right)^{{\rho} -1}. \\
\end{split}
\end{equation}

\subsubsection{Suppose ${\omega}_{1} < 0$, ${\rho} > 3$. }
\noindent The gravitino mass evaluated at $\left( {T}_{R0+}, {T}_{I0} \right)$ is
\begin{equation}
\begin{split}
{M}_{3/2+}\equiv&\; \left. \frac{ {e}^{ \frac{K}{2 {M}_{\text{pl}}^{2} } } }{ {M}_{\text{pl}}^{2} } \left| W \left( {T} \right) \right| \right|_{+} = \frac{1}{ {M}_{\text{pl}}^{2} } \left( \frac{({\rho} - 2) {M}_{\text{pl}} }{- {\rho}  {\iota}_{1} } \right)^{{\rho} /2} \times \left| \frac{- {\iota}_{1} {\mu} }{\left( {\rho} - 2 \right) } \right|, \\
\end{split}
\end{equation}
\noindent and the $F$ term SUSY breaking scale evaluated at $\left( {T}_{R0+}, {T}_{I0} \right)$ is

\begin{equation}
\left. {F}_{T} \right|_{+} \equiv \left. {e}^{ \frac{K}{2 {M}_{\text{pl}}^{2} } } \left| {D}_{T} W \left( T \right) \right| \right|_{+} = \left( \frac{({\rho} - 2) {M}_{\text{pl}} }{- {\rho}  {\iota}_{1} } \right)^{{\rho} /2} \times \left| 0 \right| = 0. \\
\end{equation}
\noindent while the $D$ term SUSY breaking scale evaluated at $\left( {T}_{R0+}, {T}_{I0} \right)$ is
\begin{equation}
\begin{split}
\left. \frac{1}{2} g^2 {\xi}^{2} \right|_{+} \equiv&\; \frac{ 3 {\iota}_{1}^{2} \left| {\mu} \right|^{2} }{ {M}_{\text{pl}}^{2} \left( {\rho} - 2 \right)^{2}} \left( \frac{\left( {\rho} - 2 \right) {M}_{\text{pl}} }{- {\rho} {\iota}_{1} } \right)^{\rho}. \\
\end{split}
\end{equation}

\subsection{Suppose ${\rho} = 3$. }
\noindent The total potential becomes
\begin{equation}
{V} = -\frac{M_{\text{pl}} }{3 \left(\bar{T}+T\right)^2} \left[ 3 \mu  \bar{\lambda} + \bar{\mu} \left(2 \mu \bar{T}+3 \lambda +2 \mu  T\right) \right] + \frac{1}{2} g^2 {\xi}^{2}, \\
\end{equation}
\noindent or in terms of ${\lambda} = {\lambda}_{R} + i {\lambda}_{I}$, ${\mu} = {\mu}_{R} + i {\mu}_{I}$, ${T} = {T}_{R} + i {T}_{I}$, ${\omega}_{1} = 2 \left( {\lambda}_{R} {\mu}_{R} + {\lambda}_{I} {\mu}_{I} \right)$, ${\omega}_{2} = 2 \left( {\lambda}_{I} {\mu}_{R} - {\lambda}_{R} {\mu}_{I} \right)$, 
\begin{equation}
{V} \left( {T}_{R} \right) = \frac{- M_{\text{pl}} }{8 {T}_{R}^3} \left( \frac{8}{3} \left| {\mu} \right|^2 {T}_{R}^2 + 2 {T}_{R} {\omega}_{1} \right) + \frac{1}{2} g^2 {\xi}^{2}, \\
\end{equation}
\noindent which is independent of the imaginary part of $T$, ${T}_{I}$. This model does not form a double field inflation model since the second derivative of ${V}$ with respect to ${T}_{I}$, $\frac{ {\partial}^{2} {V} }{ {\partial} {T}_{I}^{2} }$ is equal to zero, but it can be considered as a single field model if we fix ${T}_{I}$ to some real values. To consider it as a single field model and study the properties of minimum  point, we first find the first and second derivatives

\begin{equation}
\frac{{\partial} {V} }{{\partial} {T}_{R} } = \frac{M_{\text{pl}} }{6 {T}_{R}^{3} } \left(2 \left| {\mu} \right|^{2} {T}_{R} + 3 {\omega}_{1} \right)
\end{equation}
\begin{equation}
\frac{{\partial}^{2} {V} }{{\partial} {T}_{R}^{2} } = \frac{- M_{\text{pl}} }{6 {T}_{R}^{4} } \left(4 \left| {\mu} \right|^{2} {T}_{R} + 9 {\omega}_{1} \right)
\end{equation}

\noindent Finding the minimum point by taking $\frac{{\partial} {V} }{{\partial} {T}_{R} } = 0$, we have
\begin{equation}
\frac{{\partial} {V} }{{\partial} {T}_{R} } = 0 \quad \Rightarrow \quad {T}_{R} = \frac{- 3 {\omega}_{1} }{ 2 \left| {\mu} \right|^{2} } = \frac{-3 {\iota}_{1}}{2} =: {T}_{R0}. \\
\end{equation}
\noindent The second derivative evaluated at ${T}_{R} = {T}_{R0}$ is
\begin{equation}
\left. \frac{{\partial}^{2} {V} }{{\partial} {T}_{R}^{2} } \right|_{ {T}_{R} = {T}_{R0} } = - \frac{8 \left| {\mu} \right|^{2} {M}_{\text{pl}} }{81 {\iota}_{1}^{3} }. 
\end{equation}
\noindent The positivity of ${T}_{R}$ and $\left. \frac{{\partial}^{2} {V} }{{\partial} {T}_{R}^{2} } \right|_{ {T}_{R} = {T}_{R0} }$ requires ${\omega}_{1} < 0$. The gravitino mass evaluated at ${T}_{R} = {T}_{R0}$ is
\begin{equation}
\begin{split}
{M}_{3/2} \equiv&\; \left. \frac{ {e}^{ \frac{K}{2 {M}_{\text{pl}}^{2} } } }{ {M}_{\text{pl}}^{2} } \left| W \left( {T} \right) \right| \right|_{ {T}_{R} = {T}_{R0} } = \frac{1}{M_{\text{pl}}^2} \left(\frac{ {M}_{\text{pl}} }{-3 {\iota}_{1} } \right)^{3/2} \times \left| \frac{- 2 {\omega}_{1} + i \left( {\omega}_{2} + 2 \left| {\mu} \right|^{2} {T}_{I} \right) }{ 2 \bar{\mu} } \right|. \\
\end{split}
\end{equation}
\noindent The $F$ term SUSY breaking scale (evaluated at ${T}_{R} = {T}_{R0}$) is 
\begin{equation}
\begin{split}
\left. {F}_{T} \right|_{0} \equiv&\; \left. {e}^{ \frac{K}{2 {M}_{\text{pl}}^{2} } } \left| {D}_{T} W \left( T \right) \right| \right|_{{T}_{R} = {T}_{R0}} = \left(\frac{ {M}_{\text{pl}} }{-3 {\iota}_{1} } \right)^{3/2} \times \left| \frac{i {\mu} }{2 {\omega}_{1} } \left( {\omega}_{2} + 2 \left| {\mu} \right|^{2} {T}_{I} \right) \right|, \\
\end{split}
\end{equation}
\noindent while the $D$ term SUSY breaking scale (evaluated at ${T}_{R} = {T}_{R0}$) is
\begin{equation}
\left. \frac{1}{2} g^2 {\xi}^{2} \right|_{0} = - \frac{{M}_{\text{pl}} \left| {\mu} \right|^{4}}{9 {\omega}_{1}} = - \frac{{M}_{\text{pl}} \left| {\mu} \right|^{2}}{9 {\iota}_{1}}. \\
\end{equation}

\section{A short summary of total potentials and required tools for inflation analysis}
\label{A short summary}
\subsection{Suppose ${\rho} \neq 3$. }
\subsubsection{Total potentials}
\noindent By using Eq.(\ref{F term potential}), (27), (30) and Table 2, the total potentials (normalized by $\left| {\mu} \right|^{2}$) are given by $V = {V}_{F} + {V}_{D}$, where
\begin{equation}
\label{F term of total potential}
\frac{{V}_{F} \left( {T}_{R}, {T}_{I} \right)}{\left| {\mu} \right|^{2}} = \frac{1}{{M}_{\text{pl}}^{2}} \left( \frac{{M}_{\text{pl}}}{2 {T}_{R}} \right)^{\rho} \left\{ \left( {\rho} - 3 \right) \left[ \frac{\left( {\iota}_{1}^{2} + {\iota}_{2}^{2} \right)}{4} + {\iota}_{2} {T}_{I} + {T}_{I}^{2} \right] + \left( {\rho} - 5 \right) {\iota}_{1} {T}_{R} + \frac{\left( {\rho}^{2} - 7 {\rho} + 4 \right)}{\rho} {T}_{R}^{2} \right\}, \\
\end{equation}
\noindent and 
\begin{equation}
\label{D term of total potential when omega1 greater than zero}
\frac{{V}_{D}}{\left| {\mu} \right|^{2}} = \frac{g^2 {\xi}^{2}}{2 \left| {\mu} \right|^{2}} =
\frac{ {\iota}_{1} }{{\rho} {M}_{\text{pl}} } \left(\frac{\left( {\rho}^{2} - 7 {\rho} + 4 \right) {M}_{\text{pl}}}{- \left( {\rho} - 3 \right) {\rho}  {\iota}_{1} }\right)^{{\rho} -1}, \\
\end{equation}
\noindent if ${\omega}_{1} > 0$ and $3 < {\rho} < \frac{7 + \sqrt{33}}{2}$, or 
\begin{equation}
\label{D term of total potential when omega1 smaller than zero}
\frac{{V}_{D}}{\left| {\mu} \right|^{2}} = \frac{g^2 {\xi}^{2}}{2 \left| {\mu} \right|^{2}} = \frac{ 3 {\iota}_{1}^{2} }{ {M}_{\text{pl}}^{2} \left( {\rho} - 2 \right)^{2}} \left( \frac{\left( {\rho} - 2 \right) {M}_{\text{pl}} }{- {\rho} {\iota}_{1} } \right)^{\rho}. \\
\end{equation}
\noindent if ${\omega}_{1} < 0$ and ${\rho} > 3$. 

\subsubsection{SUGRA Lagrangian setup}
\noindent We are going to find the corresponding adiabatic and iso-curvature perturbation for double field models. During inflation, since the non-zero vacuum expectation values of fermions break the Lorentz symmetry, which is not physical during inflation, we assume that only bosons contribute to the background of inflation dynamics. Hence, we consider the bosonic part of the Lagrangian, which is given by \cite{gianluca_calcagni_2017}

\begin{equation}
\label{SUGRA Lagrangian}
\mathcal{L} = \sqrt{-g} \mathcal{L}_{\text{SUGRA}} = \sqrt{-g} \left[ \frac{{M}^{2}_{\text{pl}}}{2} R - {K}_{i \bar{j}} {\triangledown}_{\mu} {\phi}^{i} {\triangledown}^{\mu} {\bar{\phi}}^{\bar{j}} - V \right], \\
\end{equation}
\noindent where $V = {V}_{F} + {V}_{D}$ while ${V}_{F}$ and ${V}_{D}$ are $F$ term potential and $D$ term potential respectively. The kinetic term in $\mathcal{L}_{\text{SUGRA}}$ can be written as
\begin{equation}
{K}_{i \bar{j}} {\triangledown}_{\mu} {\phi}^{i} {\triangledown}^{\mu} {\bar{\phi}}^{\bar{j}} = \frac{ {\rho} {M}^{2}_{\text{pl}} }{\left( T + \bar{T} \right)^{2}} {\triangledown}_{\mu} T {\triangledown}^{\mu} {\bar{T}} = \frac{ {\rho} {M}^{2}_{\text{pl}} }{4 {T}_{R}^2} \left( {\triangledown}_{\mu} {T}_{R} {\triangledown}^{\mu} {T}_{R} + {\triangledown}_{\mu} {T}_{I} {\triangledown}^{\mu} {T}_{I} \right). \\
\end{equation}
\noindent If we treat $\mathcal{L}$ in Eq.(\ref{SUGRA Lagrangian}) in the Jordan frame, by Eq.(\ref{Field space metric transformation}), we obtain
\begin{equation}
f \left( {\phi}^{I} \right) = f \left( {\phi}^{1}, {\phi}^{2} \right) = \frac{1}{2} {M}^{2}_{\text{pl}}, \quad {\phi}^{1} = {T}_{R}, \quad {\phi}^{2} = {T}_{I}, \quad \tilde{\mathcal{G}}_{IJ} = \frac{ {\rho} {M}^{2}_{\text{pl}} }{2 {T}_{R}^2} {\delta}_{IJ}, \\
\end{equation}
\noindent for all $I, J \in \left\{ 1, 2 \right\}$. Since $f \left( {\phi}^{I} \right)$ is a constant, the metric in the field space in the Jordan frame $\tilde{\mathcal{G}}_{IJ}$ is the same as that in the Einstein frame ${\mathcal{G}}_{IJ}$. Thus, 
The conformal factor becomes
\begin{equation}
{\Omega}^{2} = 1, \\
\end{equation}
\noindent and the metric of the field space in the Einstein frame is
\begin{equation}
{\mathcal{G}}_{IJ} = \frac{ {\rho} {M}^{2}_{\text{pl}} }{2 {T}_{R}^{2} } {\delta}_{IJ}. \\
\end{equation}

\noindent We expand the fields to the first order around its classical background values
\begin{equation}
{T}_{R} \left( {x}^{\mu} \right) = {T}_{Rb} \left( t \right) + {\delta} {T}_{R} \left( {x}^{\mu} \right), \quad \text{and} \quad {T}_{I} \left( {x}^{\mu} \right) = {T}_{Ib} \left( t \right) + {\delta} {T}_{I} \left( {x}^{\mu} \right), \\
\end{equation}
\noindent The norm of the velocity vector is given by the background components of the fields as follows
\begin{equation}
\dot{\sigma}^{2} = \mathcal{G}_{IJ} \dot{\varphi}^{I} \dot{\varphi}^{J} = \frac{ {\rho} {M}^{2}_{\text{pl}} }{2 {T}_{Rb}^{2} } \left( \dot{T}_{Rb}^{2} + \dot{T}_{Ib}^{2} \right) \quad \Rightarrow \quad \dot{\sigma} = \frac{ {M}_{\text{pl}} }{ \left| {T}_{Rb} \right| } \sqrt{ \frac{\rho}{2} } \sqrt{ \dot{T}_{Rb}^{2} + \dot{T}_{Ib}^{2} }, \\
\end{equation}
\noindent where the norm $\dot{\sigma}$ is defined to be positive. Since the background components of fields depend on cosmic time $t$ only, we can easily obtain the Laplacian as
\begin{equation}
\square \phi = - \left( \ddot{\phi} + 3 H \dot{\phi} \right), \quad \forall {\phi} \in \left\{ {T}_{Rb}, {T}_{Ib} \right\}, \\
\end{equation}
\noindent and hence the equations of motion (E.O.M.)s of ${T}_{Rb}$ and ${T}_{Ib}$ are given by
\begin{equation}
\label{E.O.M.s}
\begin{split}
- \left( \ddot{T}_{Rb} + 3 H \dot{T}_{Rb} \right) - \left( \frac{-1}{ {T}_{Rb} } \dot{T}_{Rb}^{2} + \frac{1}{ {T}_{Rb} } \dot{T}_{Ib}^{2} \right) - \frac{2 {T}_{Rb}^{2} }{ {\rho} {M}_{\text{pl}}^{2} } \left. {V}_{ {T}_{R} } \right|_{b} =&\; 0, \\
- \left( \ddot{T}_{Ib} + 3 H \dot{T}_{Ib} \right) - \left( \frac{-1}{ {T}_{Rb} } \dot{T}_{Rb} \dot{T}_{Ib} - \frac{1}{ {T}_{Rb} } \dot{T}_{Rb} \dot{T}_{Ib} \right) - \frac{2 {T}_{Rb}^{2} }{ {\rho} {M}_{\text{pl}}^{2} } \left. {V}_{ {T}_{I} } \right|_{b} =&\; 0, \\
\end{split}
\end{equation}
\noindent where $\left. {V}_{ {T}_{R} } \right|_{b}$ and $\left. {V}_{ {T}_{I} } \right|_{b}$ are the first derivatives of the potential with respect to ${T}_{R}$ and ${T}_{I}$ evaluated at the background respectively and the dot over ${T}_{Rb}$ and ${T}_{Ib}$ means the derivatives of the corresponding background fields with respect to cosmic time $t$.

\subsection{Suppose ${\rho} = 3$. }
\subsubsection{Total potentials and slow-roll parameters}
\noindent Note that the total potential (normalized by $\left| {\mu} \right|^{2}$) are given by
\begin{equation}
\label{Total potential at alpha 3}
\frac{{V} \left( {T}_{R} \right)}{\left| {\mu} \right|^{2}} = \frac{- M_{\text{pl}} }{8 {T}_{R}^3} \left( \frac{8}{3} {T}_{R}^2 + 2 {T}_{R} {\iota}_{1} \right) - \frac{{M}_{\text{pl}}}{9 {\iota}_{1}}, \\
\end{equation}
\noindent which is independent of ${T}_{I}$. Hence, 
\begin{equation}
\frac{{\partial} V}{ {\partial}{T}_{I} } = \frac{{\partial}^{2} V}{{\partial}{T}_{R} {\partial}{T}_{I} } = \frac{{\partial}^{2} V}{{\partial}{T}_{I} {\partial}{T}_{R} } = \frac{{\partial}^{2} V}{{\partial}{T}_{I}^{2}} = 0, \quad \forall \left( {T}_{R}, {T}_{I} \right) \in \mathbb{R}^{2}. 
\end{equation}

\subsubsection{Equations of motion}
\noindent The bosonic part of Lagrangian and the kinetic terms of ${\rho} = 3$ case are given by Eq.(\ref{E.O.M.s}). Hence, the E.O.M.s of ${T}_{R}$ and ${T}_{I}$ become 
\begin{equation}
\label{EOMs for alpha = 3}
\begin{split}
- \left( \ddot{T}_{Rb} + 3 H \dot{T}_{Rb} \right) - \left( \frac{-1}{ {T}_{Rb} } \dot{T}_{Rb}^{2} + \frac{1}{ {T}_{Rb} } \dot{T}_{Ib}^{2} \right) - \frac{2 {T}_{Rb}^{2} }{ {\rho} {M}_{\text{pl}}^{2} } \left. {V}_{ {T}_{R} } \right|_{b} =&\; 0, \\
- \left( \ddot{T}_{Ib} + 3 H \dot{T}_{Ib} \right) - \left( \frac{-1}{ {T}_{Rb} } \dot{T}_{Rb} \dot{T}_{Ib} - \frac{1}{ {T}_{Rb} } \dot{T}_{Rb} \dot{T}_{Ib} \right) =&\; 0. \\
\end{split}
\end{equation}
\noindent We will adopt Eq. (\ref{EOMs for alpha = 3}) to evaluate their corresponding evolutions.

\section{Numerical calculations}
\label{Numerical calculations}
\subsection{Suppose ${\rho} \neq 3$. }
\begin{table}[h!]
\begin{center}
\begin{tabular}{ |c|c|c|c|c|c|c|c|c| }
\hline
$\left| {\mu} \right|/ {M}_{\text{pl}}^{2}$ & ${N}_{\text{hc}}$ & ${{T}_{R}}'(N=0)$ & ${{T}_{I}}'(N=0)$ \\
\hline
$3 \times {10}^{-4}$ & $0$ & ${10}^{-5}$ & ${10}^{-5}$ \\
\hline
\end{tabular}
\end{center}
\caption{Fixed values for numerical calculations}
\label{table: Fixed values for numerical calculations}
\end{table}

\noindent We are going to show the feasible parameters for inflation, and the corresponding ${\beta}_{\text{iso}}$ and $\cos{\left( \Delta \right)}$ are evaluated based on that parameters. Recall ${\beta}_{\text{iso}}$ and $\cos{\left( \Delta \right)}$ are given by
\begin{equation}
\label{betaiso from Nhc to Nend}
{\beta}_{\text{iso}} = \frac{{T}_{\mathcal{SS}} \left( {N}_{\text{hc}}, \; {N}_{\text{end}} \right)^{2}}{1 + {T}_{\mathcal{SS}} \left( {N}_{\text{hc}}, \; {N}_{\text{end}} \right)^{2} + {T}_{\mathcal{RS}} \left( {N}_{\text{hc}}, \; {N}_{\text{end}} \right)^{2}}, \\
\end{equation}

\begin{equation}
\label{cosine Delta from Nhc to Nend}
\cos{\left( \Delta \right)} = \frac{{T}_{\mathcal{RS}} \left( {N}_{\text{hc}}, \; {N}_{\text{end}} \right)^{2}}{ \sqrt{1 + {T}_{\mathcal{RS}} \left( {N}_{\text{hc}}, \; {N}_{\text{end}} \right)^{2}} }
\end{equation}

\noindent First, we find the initial conditions based on the constraints of ${\epsilon}_{V}$, ${\eta}_{V}$, ${n}_{s}$, ${V}_{\text{hc}}$ listed in Table \ref{table:Planck data 2018 slow roll potential parameters and spectral indices} and the initial conditions listed on Table \ref{table: Fixed values for numerical calculations}, and extract some points for path evolution. Next, we set some fixed parameters as shown in Table \ref{table: Fixed values for numerical calculations}, while various sets of parameters are listed in Table \ref{table: Feasible parameter sets for inflation for positive omega for alpha = 4} for ${\rho} = 4$, Table \ref{table: Feasible parameter sets for inflation for positive omega for alpha = 5} for ${\rho} = 5$ and Table \ref{table: Feasible parameter sets for inflation for positive omega for alpha = 6} for ${\rho} = 6$ respectively.

\subsubsection{Suppose ${\omega}_{1}>0, \; 3 < {\rho} < \frac{7 + \sqrt{33}}{2}$.} 
\begin{figure}[h!]
\centering
\includegraphics[width=70mm, height=55mm]{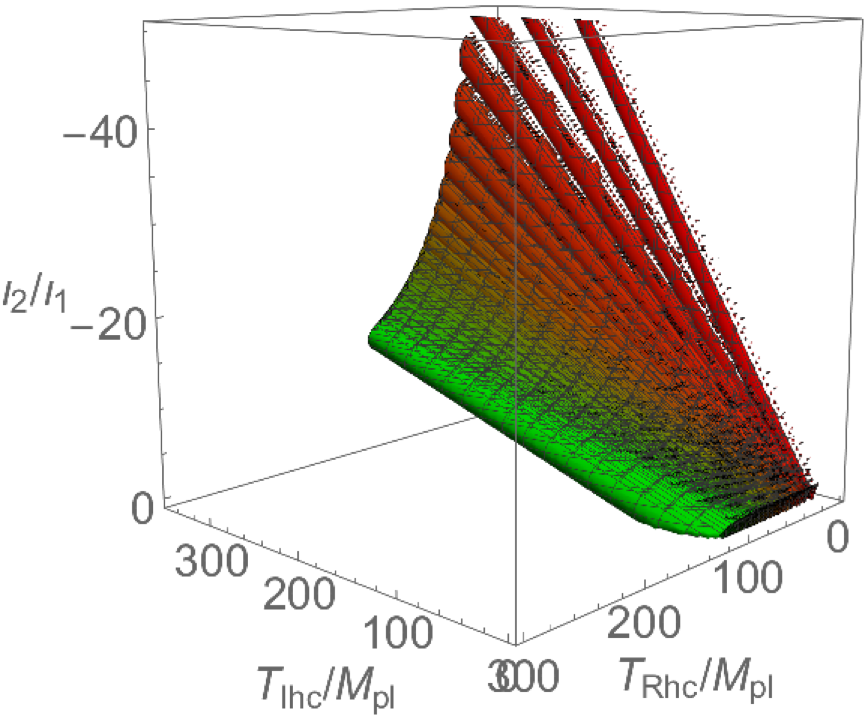} 
\includegraphics[width=70mm, height=55mm]{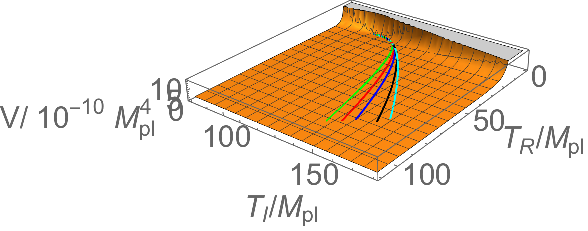} 
\includegraphics[width=70mm, height=55mm]{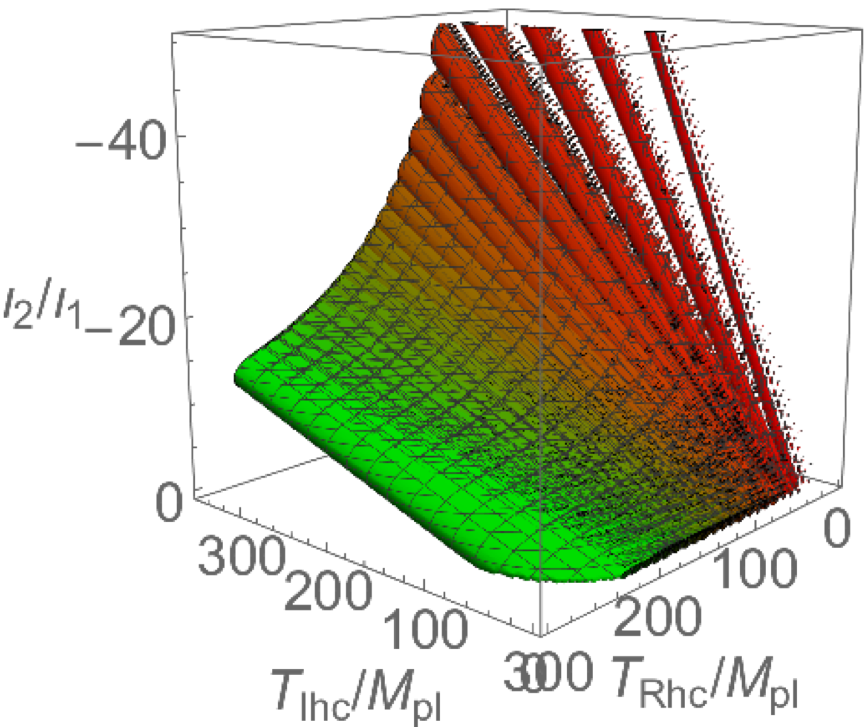} 
\includegraphics[width=70mm, height=55mm]{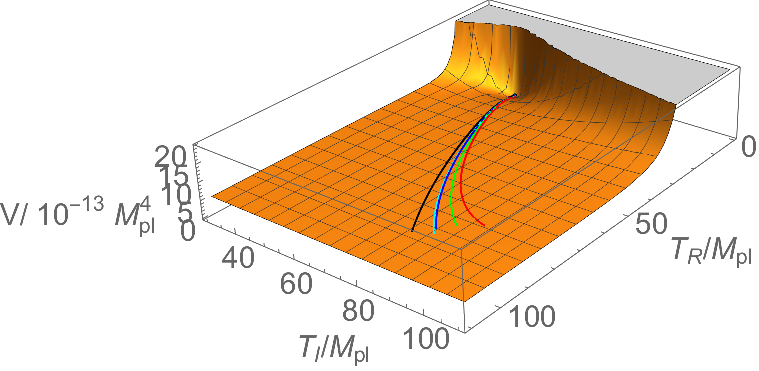} 
\includegraphics[width=70mm, height=55mm]{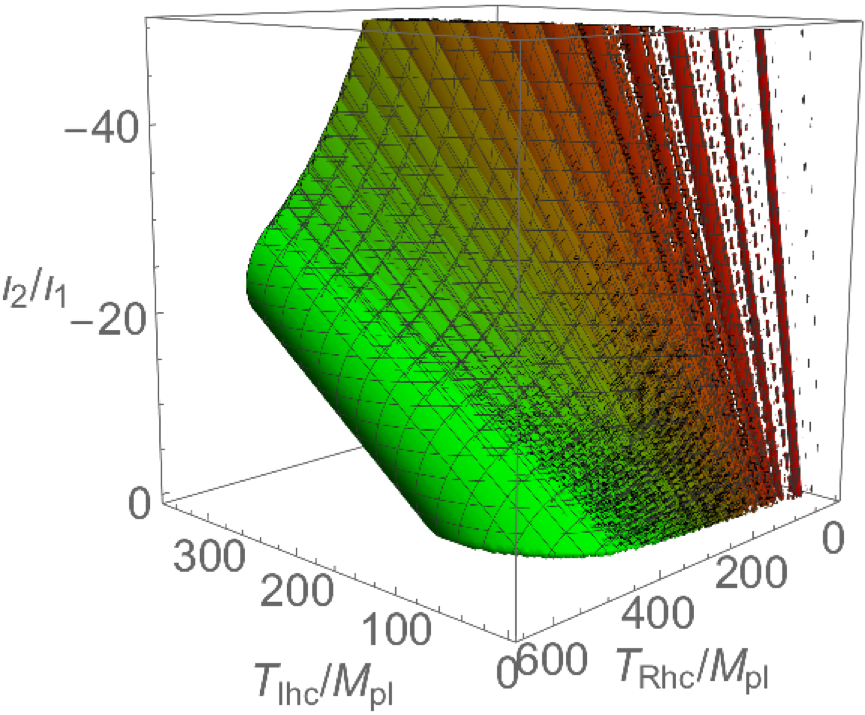} 
\includegraphics[width=70mm, height=55mm]{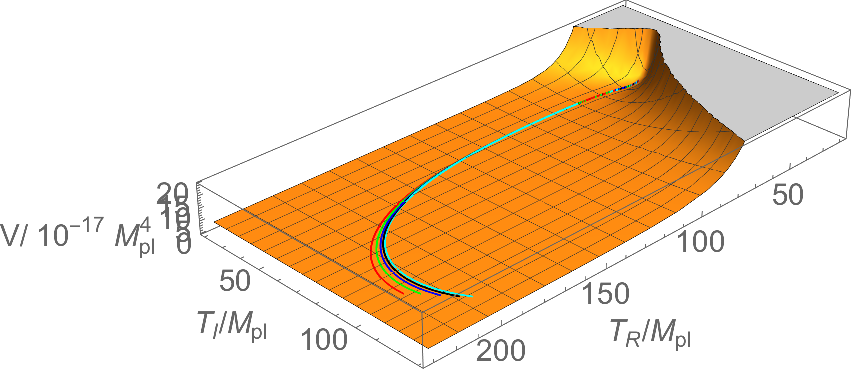} 
\caption{Left: Regions of possible values of ${T}_{R \text{hc}}$, ${T}_{I \text{hc}}$ ${\iota}_{2}/{\iota}_{1}$ for inflation based on the constraints listed in Table \ref{table:Planck data 2018 slow roll potential parameters and spectral indices} at various integers ${\rho}$, a fixed $\left| {\mu} \right| = 3 \times {10}^{-4} {M}_{\text{pl}}^{2}$ and ${\omega}_{1} > 0$ $\left( {\iota}_{1} > 0 \right)$. Each "C" shaped tube moves from red color to green color as ${\iota}_{1}$ increases. Please pay attention to the direction of magnitude of each axis in each graph. $1$st row: ${\rho} = 4$, from ${\iota}_{1} = 9 {M}_{\text{pl}}$ (red) to ${\iota}_{1} = 49 {M}_{\text{pl}}$ (green) with an increment of $2 {M}_{\text{pl}}$; $2$nd row: ${\rho} = 5$, from ${\iota}_{1} = 6 {M}_{\text{pl}}$ (red) to ${\iota}_{1} = 46 {M}_{\text{pl}}$ (green) with an increment of $2 {M}_{\text{pl}}$; $3$rd row: ${\rho} = 6$, from ${\iota}_{1} = 1 {M}_{\text{pl}}$ (red) to ${\iota}_{1} = 21 {M}_{\text{pl}}$ (green) with an increment of $1 {M}_{\text{pl}}$. (These regions are restricted by Table \ref{table:Planck data 2018 slow roll potential parameters and spectral indices} only without the e-folding constraints $50 \leq {N}_{\text{end}} - {N}_{\text{hc}} \leq 60$. ) Right: The trajectories of background fields on the potential surface. $1$st row: ${\rho} = 4$, $2$nd row: ${\rho} = 5$, $3$rd row: ${\rho} = 6$. Various colors of the paths correspond to the parameter sets lised on Table \ref{table: Feasible parameter sets for inflation for positive omega for alpha = 4}, \ref{table: Feasible parameter sets for inflation for positive omega for alpha = 5} and \ref{table: Feasible parameter sets for inflation for positive omega for alpha = 6} respectively. }
\label{fig: Alpha 456 and omega1 greater than 0}
\end{figure}

\begin{figure}[h!]
\centering
\includegraphics[width=75mm, height=75mm]{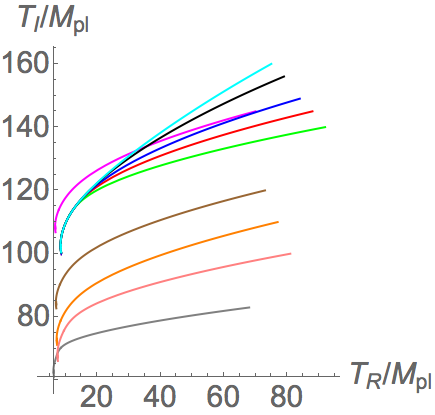} 
\includegraphics[width=40mm, height=75mm]{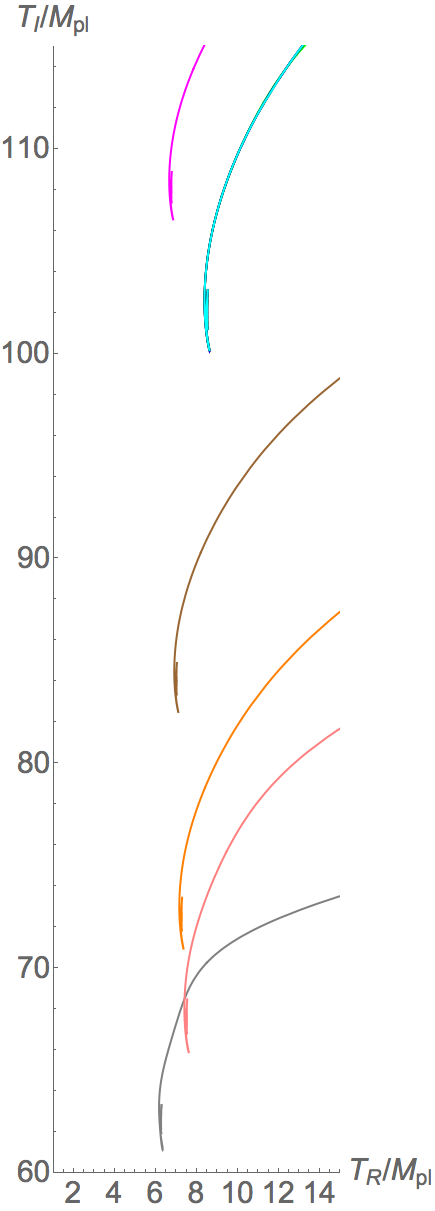} 
\includegraphics[width=75mm, height=75mm]{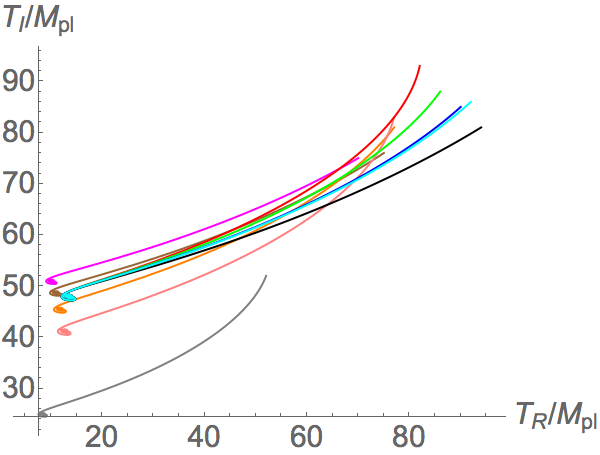} 
\includegraphics[width=40mm, height=75mm]{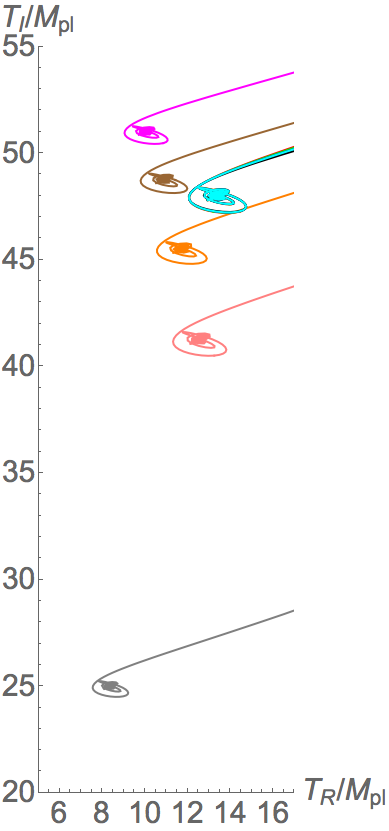} 
\includegraphics[width=75mm, height=65mm]{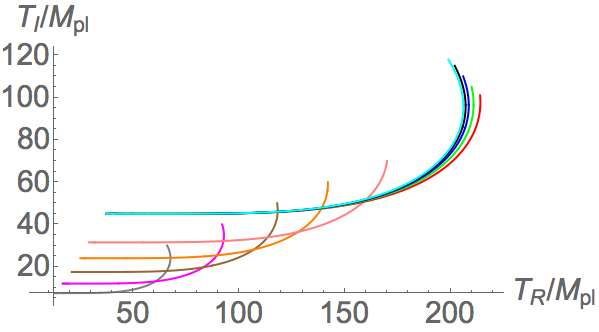} 
\includegraphics[width=52.5mm, height=55mm]{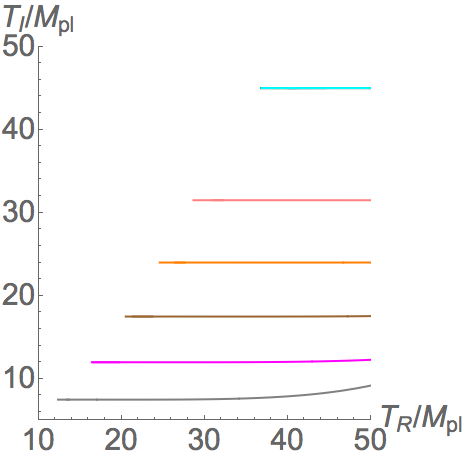} 
\caption{Evolution paths of inflation dynamics at ${\omega}_{1} > 0 \; \left( {\iota}_{1} > 0 \right)$. $1$st row: ${\rho} = 4$, $2$nd row: ${\rho} = 5$, $3$rd row: ${\rho} = 6$. Various colors refer to various parameter sets listed on Table \ref{table: Feasible parameter sets for inflation for positive omega for alpha = 4}, \ref{table: Feasible parameter sets for inflation for positive omega for alpha = 5} and \ref{table: Feasible parameter sets for inflation for positive omega for alpha = 6} correspondingly. }
\label{fig: evolution paths at omega1 greater than 0}
\end{figure}

\begin{figure}[h!]
\centering
\includegraphics[width=72.5mm, height=55mm]{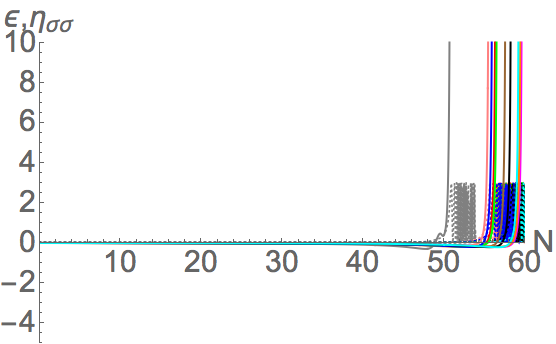} 
\includegraphics[width=72.5mm, height=55mm]{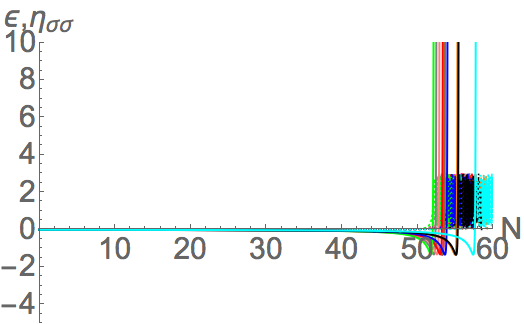} 
\includegraphics[width=72.5mm, height=55mm]{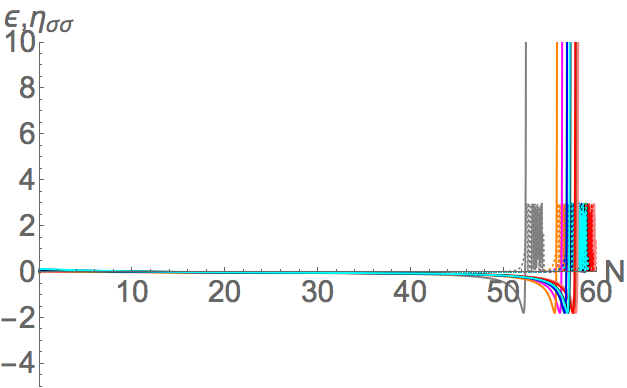} 
\caption{Evolutions of slow-roll parameters ${\epsilon}$ and ${\eta}_{\sigma \sigma}$ at ${\omega}_{1} > 0 \; \left( {\iota}_{1} > 0 \right)$. $1$st row: ${\rho} = 4$, $2$nd row: ${\rho} = 5$, $3$rd row: ${\rho} = 6$. Various colors refer to various parameter sets listed on Table \ref{table: Feasible parameter sets for inflation for positive omega for alpha = 4}, \ref{table: Feasible parameter sets for inflation for positive omega for alpha = 5} and \ref{table: Feasible parameter sets for inflation for positive omega for alpha = 6} correspondingly. Dotted lines represent the evolutions of $\epsilon$ while the solid lines represent the counterpart of ${\eta}_{{\sigma}{\sigma}}$. }
\label{fig: Epsilon EtaSigmaSigma at omega1 greater than 0}
\end{figure}

\begin{figure}[h!]
\centering
\includegraphics[width=72.5mm, height=55mm]{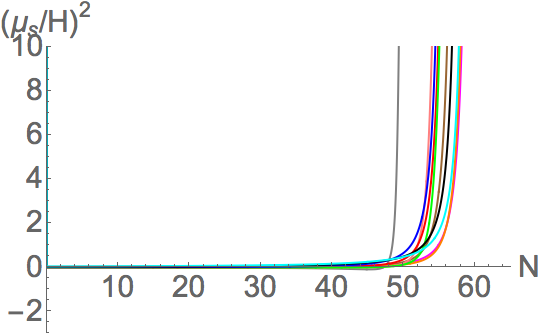} 
\includegraphics[width=72.5mm, height=55mm]{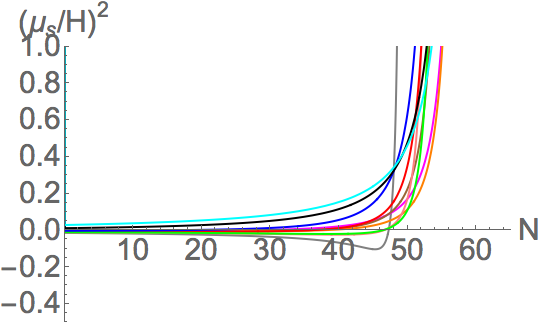} 
\includegraphics[width=72.5mm, height=55mm]{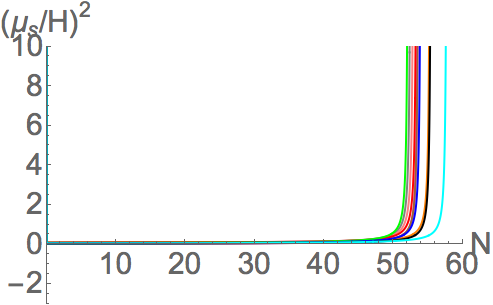} 
\includegraphics[width=72.5mm, height=55mm]{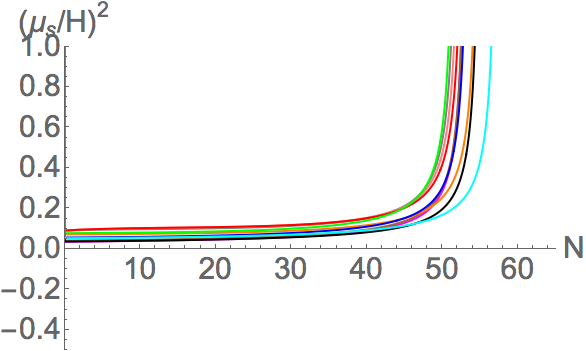} 
\includegraphics[width=72.5mm, height=55mm]{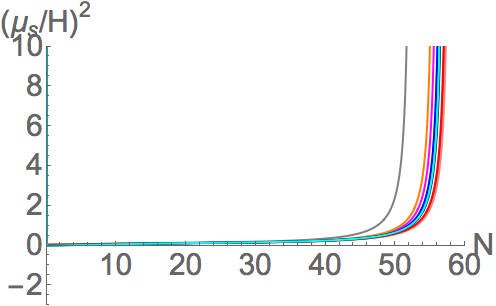} 
\includegraphics[width=72.5mm, height=55mm]{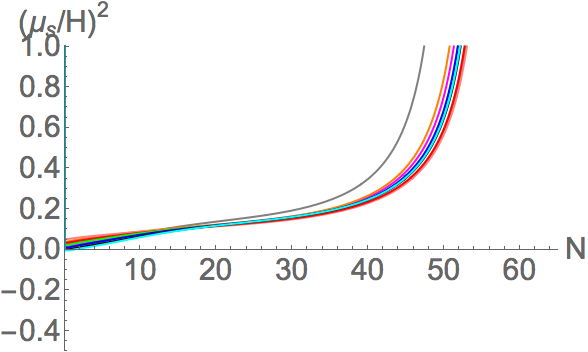} 
\caption{Square of effective mass per Hubble parameter $\left( {\mu}_{s}/H \right)^{2}$ at ${\omega}_{1} > 0 \; \left( {\iota}_{1} > 0 \right)$. $1$st row: ${\rho} = 4$, $2$nd row: ${\rho} = 5$, $3$rd row: ${\rho} = 6$. The graphs on the right hand side (R.H.S.) are the close shots of that on the left hand side (L.H.S.) near zero. Various colors refer to various parameter sets listed on Table \ref{table: Feasible parameter sets for inflation for positive omega for alpha = 4}, \ref{table: Feasible parameter sets for inflation for positive omega for alpha = 5} and \ref{table: Feasible parameter sets for inflation for positive omega for alpha = 6} correspondingly. }
\label{fig: Square of effective mass per Hubble parameter at omega1 greater than 0}
\end{figure}

\begin{figure}[h!]
\centering
\includegraphics[width=72.5mm, height=55mm]{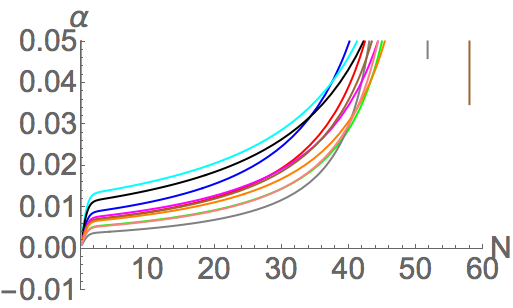} 
\includegraphics[width=72.5mm, height=55mm]{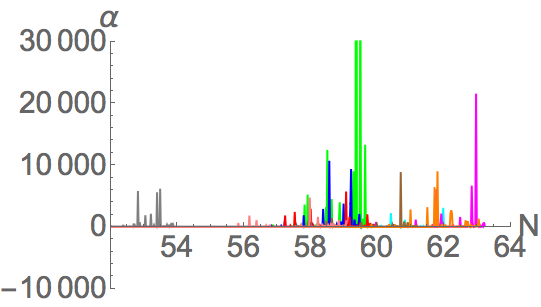} 
\includegraphics[width=72.5mm, height=55mm]{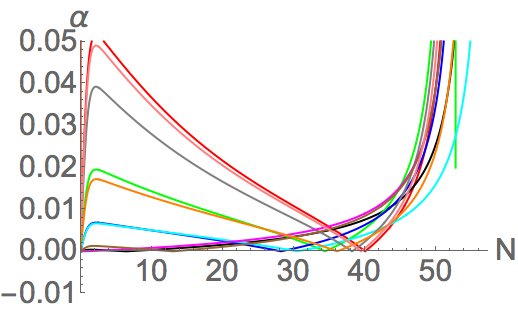} 
\includegraphics[width=72.5mm, height=55mm]{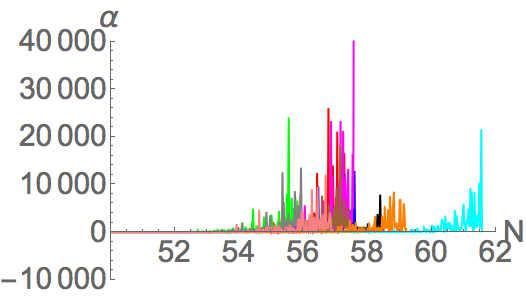} 
\includegraphics[width=72.5mm, height=55mm]{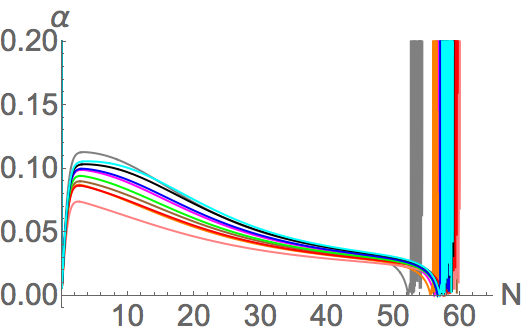} 
\includegraphics[width=72.5mm, height=55mm]{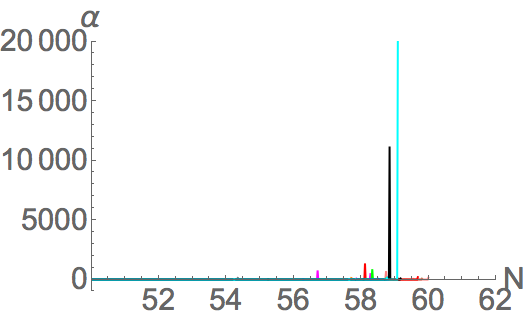} 
\caption{2 times of turn rate per Hubble parameter ${\alpha} = 2 {\omega}/H$ at ${\omega}_{1} > 0 \; \left( {\iota}_{1} > 0 \right)$. $1$st row: ${\rho} = 4$, $2$nd row: ${\rho} = 5$, $3$rd row: ${\rho} = 6$. Left graphs show the overall evolutions, while right graphs focus on the changes at the end of inflation. Various colors refer to various parameter sets listed on Table \ref{table: Feasible parameter sets for inflation for negative omega for alpha = 4}, \ref{table: Feasible parameter sets for inflation for negative omega for alpha = 5} and \ref{table: Feasible parameter sets for inflation for negative omega for alpha = 6} correspondingly. }
\label{fig: 2 times of turn rate per Hubble parameter at omega1 greater than 0}
\end{figure}

\begin{figure}[h!]
\centering
\includegraphics[width=72.5mm, height=55mm]{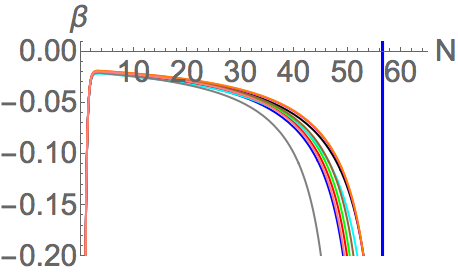} 
\includegraphics[width=72.5mm, height=55mm]{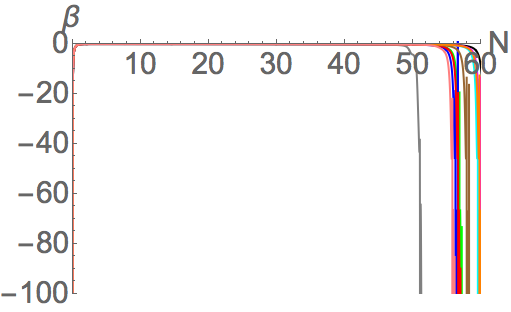} 
\includegraphics[width=72.5mm, height=55mm]{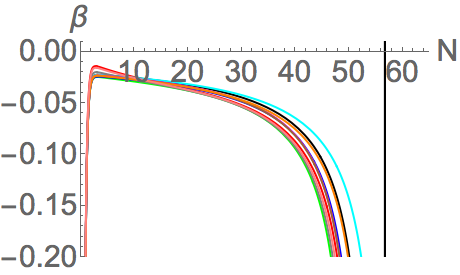} 
\includegraphics[width=72.5mm, height=55mm]{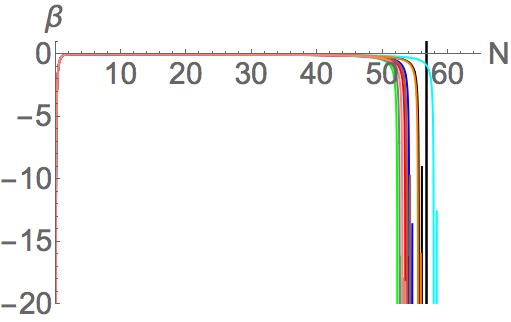} 
\includegraphics[width=72.5mm, height=55mm]{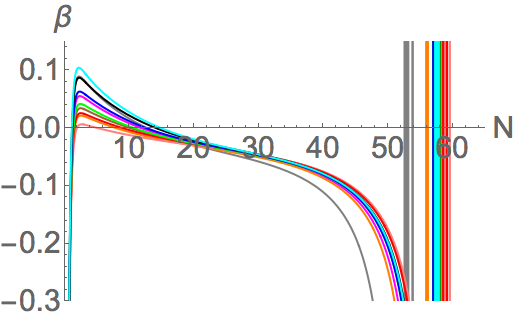} 
\includegraphics[width=72.5mm, height=55mm]{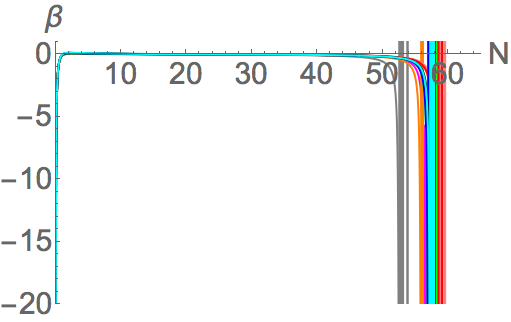} 
\caption{$\beta$ at ${\omega}_{1} > 0 \; \left( {\iota}_{1} > 0 \right)$. $1$st row: ${\rho} = 4$, $2$nd row: ${\rho} = 5$, $3$rd row: ${\rho} = 6$. The graphs on the right hand side (R.H.S.) are the close shots of that on the left hand side (L.H.S.). Various colors refer to various parameter sets listed on Table \ref{table: Feasible parameter sets for inflation for negative omega for alpha = 4}, \ref{table: Feasible parameter sets for inflation for negative omega for alpha = 5} and \ref{table: Feasible parameter sets for inflation for negative omega for alpha = 6} correspondingly. }
\label{fig: Beta at omega1 greater than 0}
\end{figure}

\noindent In Figure \ref{fig: Alpha 456 and omega1 greater than 0}, we show the possible positive values of ${T}_{R \text{hc}}$\footnote{In this paper, we assume the first horizon crossing is the start of inflation. The subscript "hc" means "horizon crossing". }, ${T}_{I \text{hc}}$ and ${\iota}_{2}/ {\iota}_{1}$ based on the Planck observation constraints listed in Table \ref{table:Planck data 2018 slow roll potential parameters and spectral indices} at various ${\rho} = 4, 5, 6$. The "C" shape tube regions gradually change the color from red to green as ${\iota}_{1}$ gradually increases with a constant increment (For details, please refer to the description of Figure \ref{fig: Alpha 456 and omega1 greater than 0}). We extract some of the possible parameters satisfying the e-folding constraints $50 \leq {N}_{\text{end}} - {N}_{\text{hc}} \leq 60$ as listed in Table \ref{table: Feasible parameter sets for inflation for positive omega for alpha = 4}, \ref{table: Feasible parameter sets for inflation for positive omega for alpha = 5} and \ref{table: Feasible parameter sets for inflation for positive omega for alpha = 6}, and their corresponding evolution paths are plotted in Figure \ref{fig: evolution paths at omega1 greater than 0}. The colors correspond to the counterparts of the paths shown in Figure \ref{fig: evolution paths at omega1 greater than 0}. One can see that for ${\rho} = 4$, the background fields move towards their corresponding minimum point (dS vacuum point)\footnote{One can find their corresponding minimum point $\left( {T}_{R \text{end}}, {T}_{I \text{end}} \right)$ in Table \ref{table: Feasible parameter sets for inflation for positive omega for alpha = 4}, \ref{table: Feasible parameter sets for inflation for positive omega for alpha = 5} and \ref{table: Feasible parameter sets for inflation for positive omega for alpha = 6}. Here "dS" means "dS vacuum". } in a nearly straight line, turn to the minimum point with a larger turn rate and then oscillate around the minimum point. As ${\rho}$ increases gradually up to $6$, the background fields greatly turn in the initial stage, move towards their corresponding minimum point and finally oscillate around the minimum point. \\
\vspace{3mm}

\noindent Next, in Figure \ref{fig: Epsilon EtaSigmaSigma at omega1 greater than 0}, we can see the evolutions of $\epsilon$ and ${\eta}_{{\sigma}{\sigma}}$ of all the listed parameter sets in Table \ref{table: Feasible parameter sets for inflation for positive omega for alpha = 4}, \ref{table: Feasible parameter sets for inflation for positive omega for alpha = 5} and \ref{table: Feasible parameter sets for inflation for positive omega for alpha = 6}. Basically, starting from very small positive values, all the $\epsilon$ (dotted) lines evolve and keep at a close-to-zero value, surge to 3 after $50$ e-folding and then oscillate between $0$ and $3$. Also, ${\eta}_{{\sigma}{\sigma}}$ (solid) lines start from a value very close to zero, evolve around zero and then sharply increase at the end of inflation. That means the background fields roll slowly initially and then run at an increasing rate, which is consistent with the mechanism of inflation. In Figure \ref{fig: Square of effective mass per Hubble parameter at omega1 greater than 0}, after dropping sharply from a large positive value to a small positive value, the effective mass square of the entropy perturbation $\left( {\mu}_{s}/H \right)^{2}$, which is given by Eq.(\ref{Effective mass squared of entropy perturbations}), remains light and physical. This shows that the trajectories roll from the plateau. Since ${\mu}_{s}^{2}$ is related to the curvature of the potential in the direction orthogonal to the trajectory, light mass square means the curvature orthogonal to the trajectory is small at the initial stage. When the background fields roll off the ridge and reach $50$ e-folding, the entropy mass square largely grows because the curvature orthogonal to the trajectory is large and the turn rate becomes significant as the background field oscillates around their corresponding minimum points.

\begin{sidewaystable}[h!]
\begin{center}
\begin{tabular}{ |c|c|c|c|c|c|c|c|c|c|c|c|c|c| }
\hline
$\text{Color}$ & ${\iota}_{1}$ & ${\iota}_{2}/ {\iota}_{1}$ & ${M}_{3/2-}/ {M}_{\text{pl}}$ & $\left. {F}_{T} \right|_{-} / {M}_{\text{pl}}$ & ${T}_{R \text{end}}$ & ${T}_{I \text{end}}$ & ${T}_{R \text{hc}}$ & ${T}_{I \text{hc}}$ & ${N}_{\text{end}}$ & ${N}_{\text{stop}}$ & ${\beta}_{\text{iso}}$ & $\cos{\left( \Delta \right)}$ \\
\hline
$\text{Gray}$ & $25$ & $-5$ & $1.2 \times {10}^{-4}$ & $6.912 \times {10}^{-5}$ & $6.25$ & $62.5$ & $68$ & $83$ & $50.4485$ & $53.8960$ & $1.87873 \times {10}^{-39}$ & $2.2983 \times {10}^{-5}$ \\
\hline
$\text{Magenta}$ & $27$ & $-8$ & $1.1111 \times {10}^{-4}$ & $5.9259 \times {10}^{-5}$ & $6.75$ & $108$ & $70$ & $145$ & $59.3514$ & $63.2240$ & $5.48281 \times {10}^{-40}$ & $2.86372 \times {10}^{-5}$ \\
\hline
$\text{Brown}$ & $28$ & $-6$ & $1.0714 \times {10}^{-4}$ & $5.5102 \times {10}^{-5}$ & $7$ & $84$ & $73$ & $120$ & $57.3194$ & $61.1881$ & $5.32875 \times {10}^{-39}$ & $2.58454 \times {10}^{-5}$ \\
\hline
$\text{Orange}$ & $29$ & $-5$ & $1.0344 \times {10}^{-4}$ & $5.1367 \times {10}^{-5}$ & $7.25$ & $72.5$ & $77$ & $110$ & $59.1943$ & $63.0620$ & $6.85479 \times {10}^{-38}$ & $2.52503 \times {10}^{-5}$ \\
\hline
$\text{Pink}$ & $30$ & $-4.5$ & $5 \times {10}^{-4}$ & $4.8 \times {10}^{-5}$ & $7.5$ & $67.5$ & $81$ & $100$ & $55.2191$ & $59.0726$ & $1.11729 \times {10}^{-39}$ & $2.15684 \times {10}^{-5}$ \\
\hline
$\text{Red}$ & $34$ & $-6$ & $4.41176 \times {10}^{-4}$ & $3.73702 \times {10}^{-5}$ & $8.5$ & $102$ & $88$ & $145$ & $56.0756$ & $59.9293$ & $5.14412 \times {10}^{-37}$ & $2.08999 \times {10}^{-5}$ \\
\hline
$\text{Green}$ & $34$ & $-6$ & $4.41176 \times {10}^{-4}$ & $3.73702 \times {10}^{-5}$ & $8.5$ & $102$ & $92$ & $140$ & $56.2624$ & $59.9871$ & $1.35243 \times {10}^{-42}$ & $1.94259 \times {10}^{-5}$ \\
\hline
$\text{Blue}$ & $34$ & $-6$ & $4.41176 \times {10}^{-4}$ & $3.73702 \times {10}^{-5}$ & $8.5$ & $102$ & $84$ & $149$ & $55.6707$ & $59.5427$ & $4.45289 \times {10}^{-40}$ & $2.29009 \times {10}^{-5}$ \\
\hline
$\text{Black}$ & $34$ & $-6$ & $4.41176 \times {10}^{-4}$ & $3.73702 \times {10}^{-5}$ & $8.5$ & $102$ & $79$ & $156$ & $57.9958$ & $61.0403$ & $4.68586 \times {10}^{-43}$ & $3.08396 \times {10}^{-5}$ \\
\hline
$\text{Cyan}$ & $34$ & $-6$ & $4.41176 \times {10}^{-4}$ & $3.73702 \times {10}^{-5}$ & $8.5$ & $102$ & $75$ & $160$ & $58.9596$ & $62.0001$ & $2.30095 \times {10}^{-41}$ & $4.50112 \times {10}^{-5}$ \\
\hline
\end{tabular}
\end{center}
\caption{Feasible parameter sets for inflation for ${\omega}_{1}>0$ (or ${\iota}_{1} > 0$) and ${\rho} = 4$. Recall the unit of each variable is the following. $\left[ {\iota}_{1} \right] = {M}_{\text{pl}}$, $\left[ {M}_{3/2-} \right] = {M}_{\text{pl}}$, $\left[ \left. {F}_{T} \right|_{-} \right] = {M}_{\text{pl}}^{2}$, $\left[ {T}_{R \text{end}} \right] = \left[ {T}_{I \text{end}} \right] = \left[ {T}_{R \text{hc}} \right] = \left[ {T}_{I \text{hc}} \right] = {M}_{\text{pl}}$. ${N}_{\text{end}}$ is the e-folding number when ${\epsilon} = 1$, which represents the end of inflation. ${N}_{\text{stop}}$ means the e-folding number that we stop the numerical calculation. }
\label{table: Feasible parameter sets for inflation for positive omega for alpha = 4}

\begin{center}
\begin{tabular}{ |c|c|c|c|c|c|c|c|c|c|c|c|c|c| }
\hline
$\text{Color}$ & ${\iota}_{1}$ & ${\iota}_{2}/ {\iota}_{1}$ & ${M}_{3/2-}/ {10}^{-5} {M}_{\text{pl}}$ & $\left. {F}_{T} \right|_{-}/ {10}^{-5} {M}_{\text{pl}}^{2}$ & ${T}_{R \text{end}}$ & ${T}_{I \text{end}}$ & ${T}_{R \text{hc}}$ & ${T}_{I \text{hc}}$ & ${N}_{\text{end}}$ & ${N}_{\text{stop}}$ & ${\beta}_{\text{iso}}$ & $\cos{\left( \Delta \right)}$ \\
\hline
$\text{Gray}$ & $10$ & $-5$ & $19.7028$ & $2.47431$ & $8.3333$ & $25$ & $52$ & $52$ & $52.3575$ & $55.9694$ & $2.66698 \times {10}^{-36}$ & $5.87517 \times {10}^{-5}$ \\
\hline
$\text{Magenta}$ & $12$ & $-8.5$ & $14.9884$ & $1.56856$ & $10$ & $51$ & $70$ & $75$ & $53.4696$ & $57.5777$ & $2.34071 \times {10}^{-38}$ & $7.95594 \times {10}^{-5}$ \\
\hline
$\text{Brown}$ & $13$ & $-7.5$ & $13.2927$ & $1.28409$ & $10.8333$ & $48.75$ & $75$ & $76$ & $53.4769$ & $57.4120$ & $3.60886 \times {10}^{-38}$ & $9.36358 \times {10}^{-5}$ \\
\hline
$\text{Orange}$ & $14$ & $-6.5$ & $11.8942$ & $1.0669$ & $11.6667$ & $45.5$ & $77$ & $81$ & $55.1485$ & $59.2092$ & $1.17279 \times {10}^{-37}$ & $1.31272 \times {10}^{-4}$ \\
\hline
$\text{Pink}$ & $15$ & $-5.5$ & $10.7249$ & $0.897895$ & $12.5$ & $41.25$ & $77$ & $83$ & $52.765$ & $56.8757$ & $4.02109 \times {10}^{-35}$ & $3.01275 \times {10}^{-5}$ \\
\hline
$\text{Red}$ & $16$ & $-6$ & $9.73528$ & $0.764106$ & $13.3333$ & $48$ & $82$ & $93$ & $53.206$ & $57.3168$ & $2.04007 \times {10}^{-36}$ & $2.65296 \times {10}^{-5}$ \\
\hline
$\text{Green}$ & $16$ & $-6$ & $9.73528$ & $0.764106$ & $13.3333$ & $48$ & $86$ & $88$ & $51.9849$ & $55.8550$ & $5.58572 \times {10}^{-37}$ & $9.72256 \times {10}^{-5}$ \\
\hline
$\text{Blue}$ & $16$ & $-6$ & $9.73528$ & $0.764106$ & $13.3333$ & $48$ & $90$ & $85$ & $53.8209$ & $57.6154$ & $2.09012 \times {10}^{-38}$ & $2.90824 \times {10}^{-4}$ \\
\hline
$\text{Black}$ & $16$ & $-6$ & $9.73528$ & $0.764106$ & $13.3333$ & $48$ & $94$ & $81$ & $55.2881$ & $58.4560$ & $7.36786 \times {10}^{-37}$ & $6.65673 \times {10}^{-5}$ \\
\hline
$\text{Cyan}$ & $16$ & $-6$ & $9.73528$ & $0.764106$ & $13.3333$ & $48$ & $92$ & $86$ & $57.56$ & $61.5649$ & $2.27647 \times {10}^{-39}$ & $3.40788 \times {10}^{-4}$ \\
\hline
\end{tabular}
\end{center}
\caption{Feasible parameter sets for inflation for ${\omega}_{1}>0$ (or ${\iota}_{1} > 0$) and ${\rho} = 5$. Recall the unit of each variable is the following. $\left[ {\iota}_{1} \right] = {M}_{\text{pl}}$, $\left[ {M}_{3/2-} \right] = {M}_{\text{pl}}$, $\left[ \left. {F}_{T} \right|_{-} \right] = {M}_{\text{pl}}^{2}$, $\left[ {T}_{R \text{end}} \right] = \left[ {T}_{I \text{end}} \right] = \left[ {T}_{R \text{hc}} \right] = \left[ {T}_{I \text{hc}} \right] = {M}_{\text{pl}}$. ${N}_{\text{end}}$ is the e-folding number when ${\epsilon} = 1$, which represents the end of inflation. ${N}_{\text{stop}}$ means the e-folding number that we stop the numerical calculation. }
\label{table: Feasible parameter sets for inflation for positive omega for alpha = 5}
\end{sidewaystable}

\begin{sidewaystable}[h!]
\begin{center}
\begin{tabular}{ |c|c|c|c|c|c|c|c|c|c|c|c|c|c| }
\hline
$\text{Color}$ & ${\iota}_{1}$ & ${\iota}_{2}/ {\iota}_{1}$ & ${M}_{3/2-}/ {10}^{-5} {M}_{\text{pl}}$ & $\left. {F}_{T} \right|_{-}/ {10}^{-5} {M}_{\text{pl}}^{2}$ & ${T}_{R \text{end}}$ & ${T}_{I \text{end}}$ & ${T}_{R \text{hc}}$ & ${T}_{I \text{hc}}$ & ${N}_{\text{end}}$ & ${N}_{\text{stop}}$ & ${\beta}_{\text{iso}}$ & $\cos{\left( \Delta \right)}$ \\
\hline
$\text{Gray}$ & $3$ & $-5$ & $5.0114$ & $0.3251$ & $13.5$ & $7.5$ & $66$ & $30$ & $52.2627$ & $54.3395$ & $2.39368 \times {10}^{-37}$ & $8.4469 \times {10}^{-6}$ \\
\hline
$\text{Magenta}$ & $4$ & $-6$ & $2.81893$ & $0.1372$ & $18$ & $12$ & $92$ & $40$ & $56.1668$ & $58.2761$ & $6.7651 \times {10}^{-37}$ & $9.20973 \times {10}^{-6}$ \\
\hline
$\text{Brown}$ & $5$ & $-7$ & $1.8041$ & $0.07023$ & $22.5$ & $17.5$ & $118$ & $50$ & $57.7282$ & $59.8035$ & $5.1486 \times {10}^{-38}$ & $9.9622 \times {10}^{-6}$ \\
\hline
$\text{Orange}$ & $6$ & $-8$ & $1.2529$ & $0.04064$ & $27$ & $24$ & $142$ & $60$ & $55.6072$ & $57.7068$ & $5.51104 \times {10}^{-39}$ & $1.01895 \times {10}^{-5}$ \\
\hline
$\text{Pink}$ & $7$ & $-9$ & $0.920467$ & $0.0255952$ & $31.5$ & $31.5$ & $170$ & $70$ & $57.8876$ & $59.9947$ & $3.22952 \times {10}^{-40}$ & $1.2694 \times {10}^{-5}$ \\
\hline
$\text{Red}$ & $9$ & $-10$ & $0.556826$ & $0.0120427$ & $40.5$ & $45$ & $214$ & $101$ & $57.5722$ & $59.6827$ & $2.29669 \times {10}^{-41}$ & $6.38619 \times {10}^{-6}$ \\
\hline
$\text{Green}$ & $9$ & $-10$ & $0.556826$ & $0.0120427$ & $40.5$ & $45$ & $210$ & $105$ & $56.6762$ & $58.7541$ & $4.01603 \times {10}^{-37}$ & $4.91618 \times {10}^{-6}$ \\
\hline
$\text{Blue}$ & $9$ & $-10$ & $0.556826$ & $0.0120427$ & $40.5$ & $45$ & $206$ & $110$ & $56.6968$ & $58.8117$ & $7.46009 \times {10}^{-38}$ & $3.73375 \times {10}^{-6}$ \\
\hline
$\text{Black}$ & $9$ & $-10$ & $0.556826$ & $0.0120427$ & $40.5$ & $45$ & $202$ & $115$ & $57.0605$ & $59.1691$ & $3.21796 \times {10}^{-39}$ & $2.89628 \times {10}^{-6}$ \\
\hline
$\text{Cyan}$ & $9$ & $-10$ & $0.556826$ & $0.0120427$ & $40.5$ & $45$ & $199$ & $118$ & $56.9827$ & $59.0902$ & $5.66193 \times {10}^{-39}$ & $2.46863 \times {10}^{-6}$ \\
\hline
\end{tabular}
\end{center}
\caption{Feasible parameter sets for inflation for ${\omega}_{1} > 0$ (or ${\iota}_{1} > 0$) and ${\rho} = 6$. Recall the unit of each variable is the following. $\left[ {\iota}_{1} \right] = {M}_{\text{pl}}$, $\left[ {M}_{3/2-} \right] = {M}_{\text{pl}}$, $\left[ \left. {F}_{T} \right|_{-} \right] = {M}_{\text{pl}}^{2}$, $\left[ {T}_{R \text{end}} \right] = \left[ {T}_{I \text{end}} \right] = \left[ {T}_{R \text{hc}} \right] = \left[ {T}_{I \text{hc}} \right] = {M}_{\text{pl}}$. ${N}_{\text{end}}$ is the e-folding number when ${\epsilon} = 1$, which represents the end of inflation. ${N}_{\text{stop}}$ means the e-folding number that we stop the numerical calculation. }
\label{table: Feasible parameter sets for inflation for positive omega for alpha = 6}

\begin{center}
\begin{tabular}{ |c|c|c|c|c|c|c|c|c|c|c|c|c|c| }
\hline
$\text{Color}$ & ${\iota}_{1}$ & ${\iota}_{2}/ {\iota}_{1}$ & ${M}_{3/2+}/ {10}^{-6} {M}_{\text{pl}}$ & ${T}_{R \text{end}}$ & ${T}_{I \text{end}}$ & ${T}_{R \text{hc}}$ & ${T}_{I \text{hc}}$ & ${N}_{\text{end}}$ & ${N}_{\text{stop}}$ & ${\beta}_{\text{iso}}$ & $\cos{\left( \Delta \right)}$ \\
\hline
$\text{Red}$ & $-14$ & $11$ & $2.67857$ & $14$ & $77$ & $185$ & $180$ & $58.3165$ & $62.6594$ & $9.9157 \times {10}^{-40}$ & $1.31875 \times {10}^{-5}$ \\
\hline
$\text{Green}$ & $-14$ & $11$ & $2.67857$ & $14$ & $77$ & $190$ & $175$ & $58.1038$ & $62.4462$ & $4.82523 \times {10}^{-39}$ & $1.14167 \times {10}^{-5}$ \\
\hline
$\text{Blue}$ & $-14$ & $11$ & $2.67857$ & $14$ & $77$ & $195$ & $170$ & $58.0405$ & $61.3501$ & $3.07253 \times {10}^{-40}$ & $1.02649 \times {10}^{-5}$ \\
\hline
$\text{Black}$ & $-14$ & $11$ & $2.67857$ & $14$ & $77$ & $200$ & $165$ & $58.1267$ & $61.4364$ & $1.49867 \times {10}^{-40}$ & $9.46949 \times {10}^{-6}$ \\
\hline
$\text{Cyan}$ & $-14$ & $11$ & $2.67857$ & $14$ & $77$ & $205$ & $160$ & $58.3628$ & $61.6725$ & $6.49472 \times {10}^{-38}$ & $8.89825 \times {10}^{-6}$ \\
\hline 
$\text{Gray}$ & $-15$ & $10$ & $2.5$ & $150$ & $75$ & $220$ & $160$ & $57.2553$ & $60.5650$ & $7.23189 \times {10}^{-35}$ & $7.98833 \times {10}^{-6}$ \\
\hline
$\text{Magenta}$ & $-16$ & $9$ & $2.34375$ & $144$ & $72$ & $233$ & $160$ & $55.8962$ & $59.2060$ & $2.04449 \times {10}^{-35}$ & $7.30653 \times {10}^{-6}$ \\
\hline
$\text{Brown}$ & $-17$ & $8$ & $2.20588$ & $136$ & $68$ & $250$ & $160$ & $56.3214$ & $59.6312$ & $2.21076 \times {10}^{-39}$ & $6.80653 \times {10}^{-6}$ \\
\hline
$\text{Orange}$ & $-18$ & $7$ & $2.08333$ & $126$ & $63$ & $265$ & $160$ & $56.3138$ & $59.6237$ & $3.33083 \times {10}^{-36}$ & $6.41033 \times {10}^{-6}$ \\
\hline
$\text{Pink}$ & $-19$ & $6$ & $1.97368$ & $114$ & $57$ & $280$ & $160$ & $56.5409$ & $59.8508$ & $4.60472 \times {10}^{-36}$ & $6.10035 \times {10}^{-6}$ \\
\hline
\end{tabular}
\end{center}
\caption{Feasible parameter sets for inflation for ${\omega}_{1} < 0$ (or ${\iota}_{1} < 0$) and ${\rho} = 4$. Note that the $F$ term SUSY breaking scale is $0$ for all real values of parameters and fields. Recall the unit of each variable is the following. $\left[ {\iota}_{1} \right] = {M}_{\text{pl}}$, $\left[ {M}_{3/2+} \right] = {M}_{\text{pl}}$, $\left[ {T}_{R \text{end}} \right] = \left[ {T}_{I \text{end}} \right] = \left[ {T}_{R \text{hc}} \right] = \left[ {T}_{I \text{hc}} \right] = {M}_{\text{pl}}$. ${N}_{\text{end}}$ is the e-folding number when ${\epsilon} = 1$, which represents the end of inflation. ${N}_{\text{stop}}$ means the e-folding number that we stop the numerical calculation. }
\label{table: Feasible parameter sets for inflation for negative omega for alpha = 4}
\end{sidewaystable}

\subsubsection{Suppose ${\omega}_{1}<0, \; {\rho} > 3$.} 

\begin{figure}[h!]
\centering
\includegraphics[width=70mm, height=55mm]{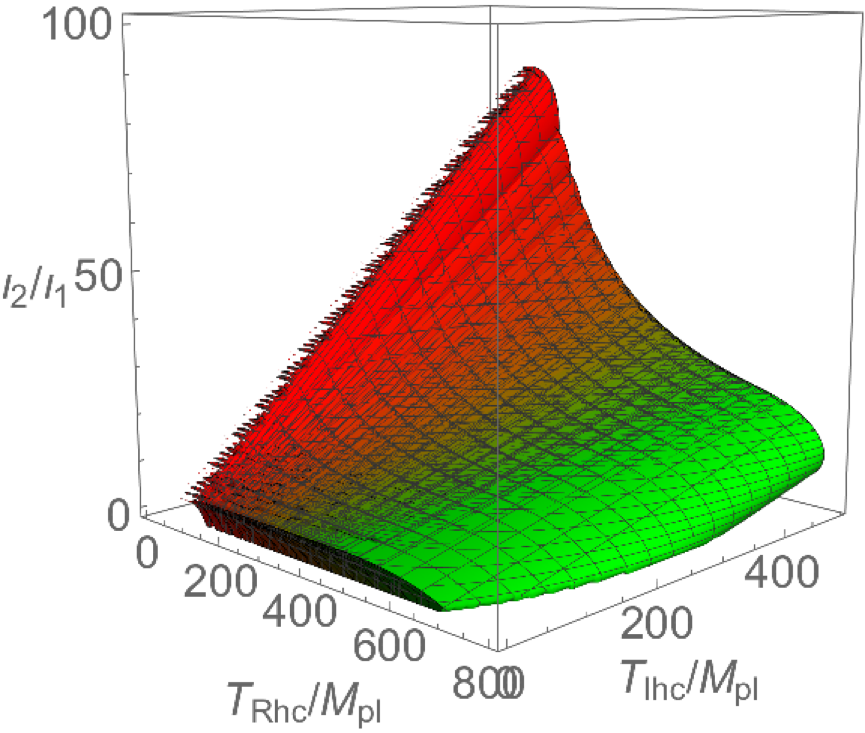} 
\includegraphics[width=70mm, height=55mm]{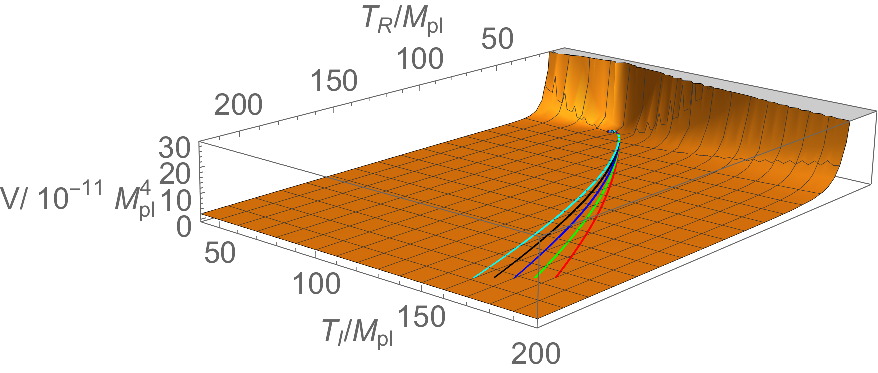} 
\includegraphics[width=70mm, height=55mm]{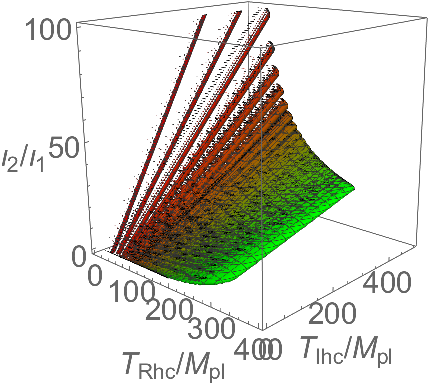} 
\includegraphics[width=70mm, height=55mm]{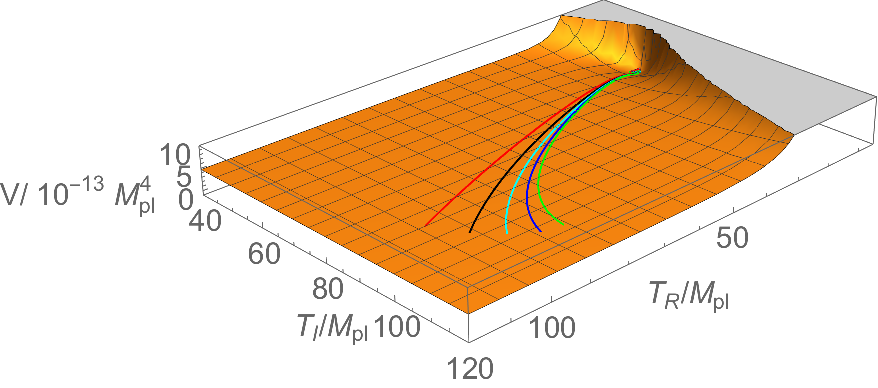} 
\includegraphics[width=70mm, height=55mm]{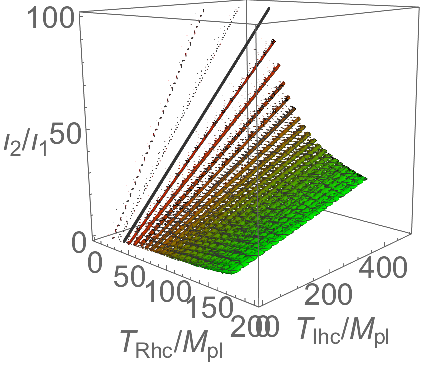} 
\includegraphics[width=70mm, height=55mm]{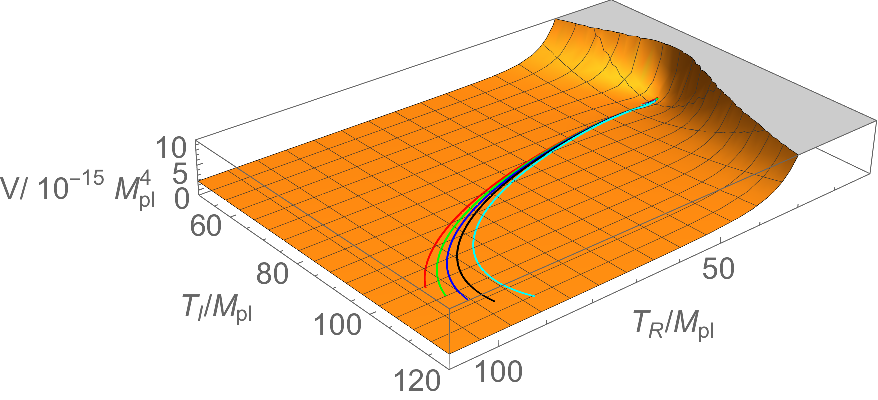} 
\caption{Left: Regions of possible values of ${T}_{R \text{hc}}$, ${T}_{I \text{hc}}$ ${\iota}_{2}/{\iota}_{1}$ for inflation based on the constraints listed in Table \ref{table:Planck data 2018 slow roll potential parameters and spectral indices} at various integers ${\rho}$, a fixed $\left| {\mu} \right| = 3 \times {10}^{-4} {M}_{\text{pl}}^{2}$ and ${\omega}_{1} < 0$ $\left( {\iota}_{1} < 0 \right)$. Each "C" like cylinder moves from red color to green color as ${\iota}_{1}$ increases. Please pay attention to the direction of magnitude of each axis in each graph. $1$st row: ${\rho} = 4$, from ${\iota}_{1} = - 12 {M}_{\text{pl}}$ (red) to ${\iota}_{1} = -52 {M}_{\text{pl}}$ (green) with an increment of $2 {M}_{\text{pl}}$; $2$nd row: ${\rho} = 5$, from ${\iota}_{1} = - 6 {M}_{\text{pl}}$ (red) to ${\iota}_{1} = - 46 {M}_{\text{pl}}$ (green) with an increment of $2 {M}_{\text{pl}}$; $3$rd row: ${\rho} = 6$, from ${\iota}_{1} = - 6 {M}_{\text{pl}}$ (red) to ${\iota}_{1} = - 46 {M}_{\text{pl}}$ (green) with an increment of $2 {M}_{\text{pl}}$. (These regions are restricted by Table \ref{table:Planck data 2018 slow roll potential parameters and spectral indices} only without the e-folding constraints $50 \leq {N}_{\text{end}} - {N}_{\text{hc}} \leq 60$. ) Right: The trajectories of background fields on the potential surface. $1$st row: ${\rho} = 4$, $2$nd row: ${\rho} = 5$, $3$rd row: ${\rho} = 6$. Various colors of the paths correspond to the parameter sets lised on Table \ref{table: Feasible parameter sets for inflation for negative omega for alpha = 4}, \ref{table: Feasible parameter sets for inflation for negative omega for alpha = 5} and \ref{table: Feasible parameter sets for inflation for negative omega for alpha = 6} respectively. }
\label{fig: Alpha 456 and omega1 smaller than 0}
\end{figure}

\begin{figure}[h!]
\centering
\includegraphics[width=80mm, height=60mm]{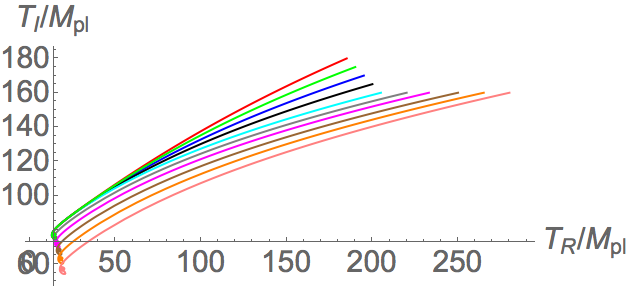} 
\includegraphics[width=45mm, height=60mm]{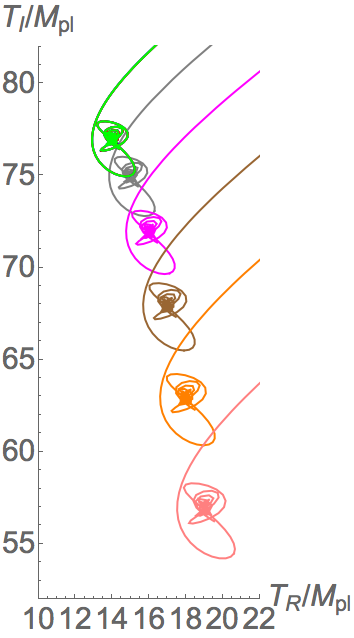} 
\includegraphics[width=80mm, height=60mm]{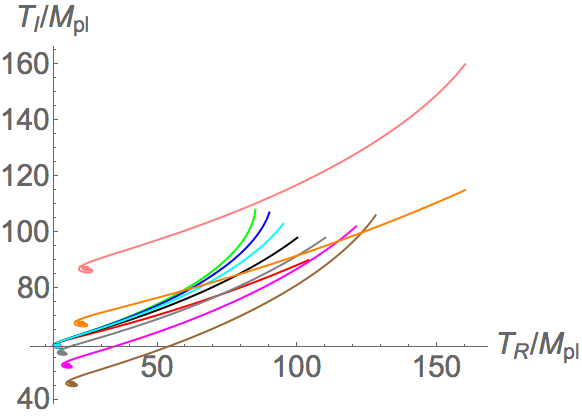} 
\includegraphics[width=45mm, height=60mm]{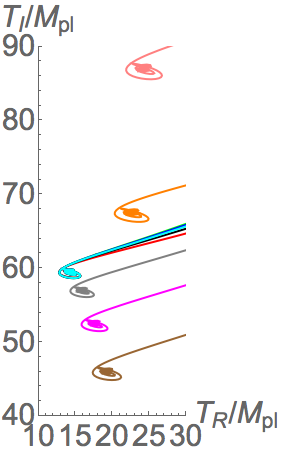} 
\includegraphics[width=80mm, height=60mm]{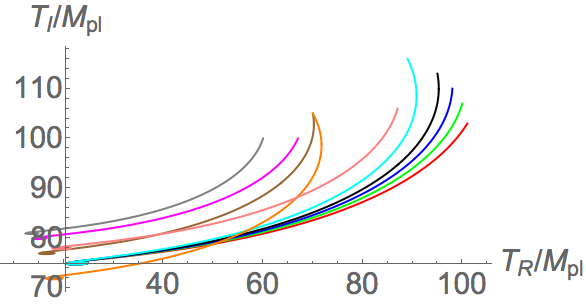} 
\includegraphics[width=45mm, height=60mm]{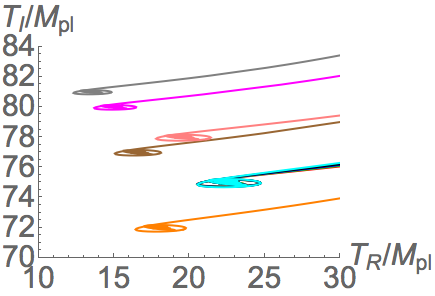} 
\caption{Evolution paths of inflation dynamics at ${\omega}_{1} < 0 \; \left( {\iota}_{1} < 0 \right)$. $1$st row: ${\rho} = 4$, $2$nd row: ${\rho} = 5$, $3$rd row: ${\rho} = 6$. Various colors refer to various parameter sets listed on Table \ref{table: Feasible parameter sets for inflation for negative omega for alpha = 4}, \ref{table: Feasible parameter sets for inflation for negative omega for alpha = 5} and \ref{table: Feasible parameter sets for inflation for negative omega for alpha = 6} correspondingly. }
\label{fig: evolution paths at omega1 smaller than 0}
\end{figure}

\begin{figure}[h!]
\centering
\includegraphics[width=72.5mm, height=55mm]{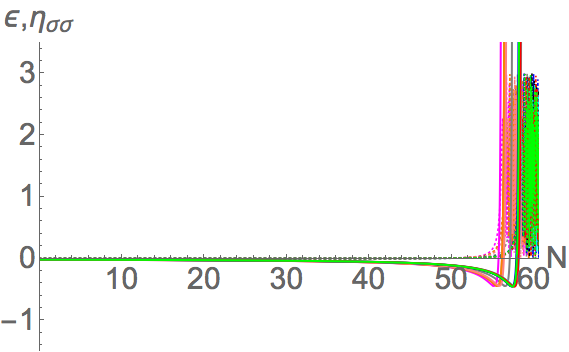} 
\includegraphics[width=72.5mm, height=55mm]{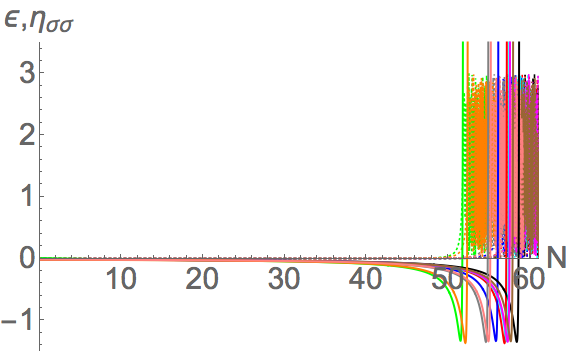} 
\includegraphics[width=72.5mm, height=55mm]{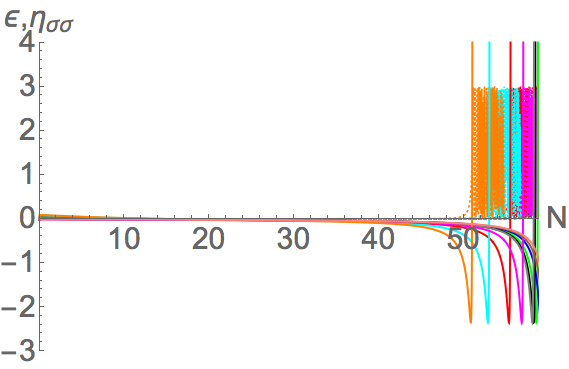} 
\caption{Evolutions of slow-roll parameters at ${\omega}_{1} < 0 \; \left( {\iota}_{1} < 0 \right)$. $1$st row: ${\rho} = 4$, $2$nd row: ${\rho} = 5$, $3$rd row: ${\rho} = 6$. Various colors refer to various parameter sets listed on Table \ref{table: Feasible parameter sets for inflation for negative omega for alpha = 4}, \ref{table: Feasible parameter sets for inflation for negative omega for alpha = 5} and \ref{table: Feasible parameter sets for inflation for negative omega for alpha = 6} correspondingly. Dotted lines represent the evolutions of $\epsilon$ while the solid lines represent the counterpart of ${\eta}_{{\sigma}{\sigma}}$. }
\label{fig: Epsilon EtaSigmaSigma at omega1 smaller than 0}
\end{figure}

\begin{figure}[h!]
\centering
\includegraphics[width=72.5mm, height=55mm]{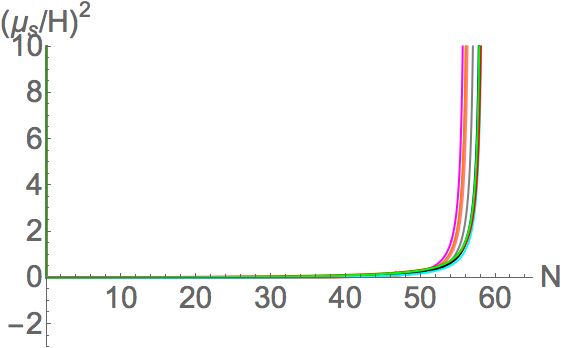} 
\includegraphics[width=72.5mm, height=55mm]{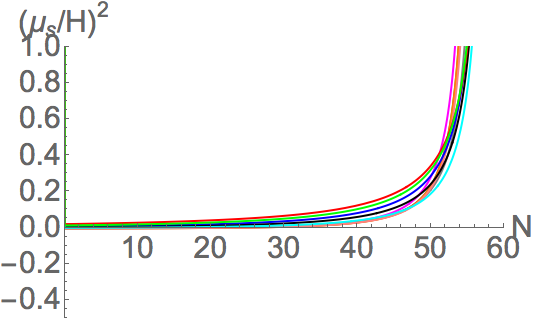} 
\includegraphics[width=72.5mm, height=55mm]{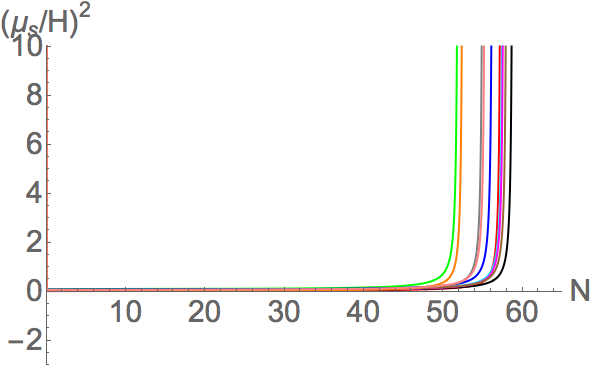} 
\includegraphics[width=72.5mm, height=55mm]{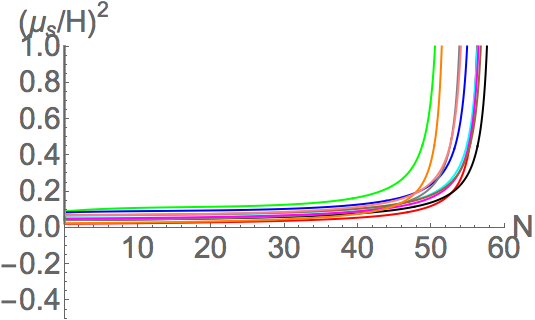} 
\includegraphics[width=72.5mm, height=55mm]{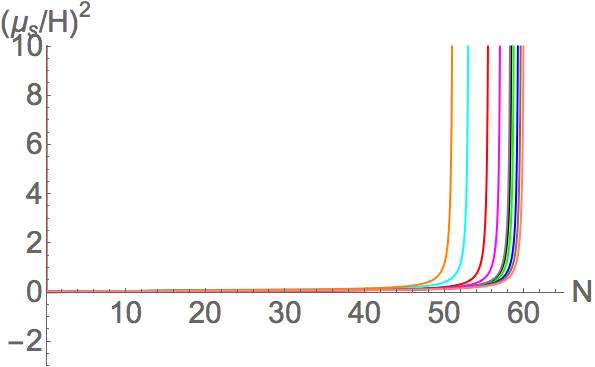} 
\includegraphics[width=72.5mm, height=55mm]{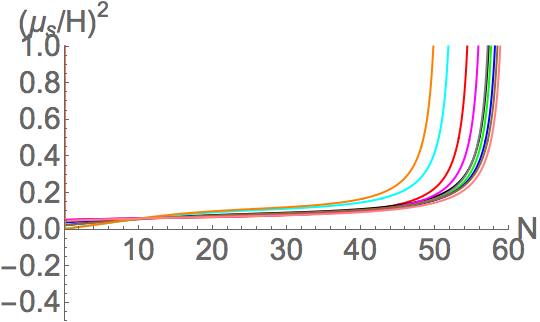} 
\caption{Square of effective mass per Hubble parameter $\left( {\mu}_{s}/H \right)^{2}$ at ${\omega}_{1} < 0 \; \left( {\iota}_{1} < 0 \right)$. $1$st row: ${\rho} = 4$, $2$nd row: ${\rho} = 5$, $3$rd row: ${\rho} = 6$. The graphs on the right hand side (R.H.S.) are the close shots of that on the left hand side (L.H.S.) near zero. Various colors refer to various parameter sets listed on Table \ref{table: Feasible parameter sets for inflation for negative omega for alpha = 4}, \ref{table: Feasible parameter sets for inflation for negative omega for alpha = 5} and \ref{table: Feasible parameter sets for inflation for negative omega for alpha = 6} correspondingly. }
\label{fig: Square of effective mass per Hubble parameter at omega1 smaller than 0}
\end{figure}

\begin{figure}[h!]
\centering
\includegraphics[width=72.5mm, height=55mm]{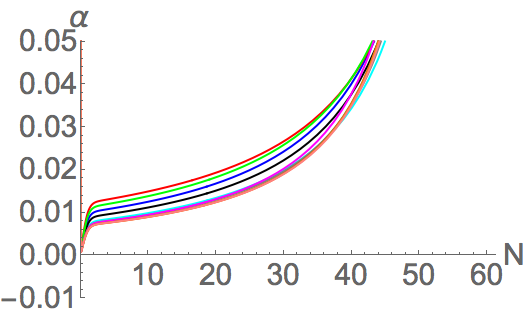} 
\includegraphics[width=72.5mm, height=55mm]{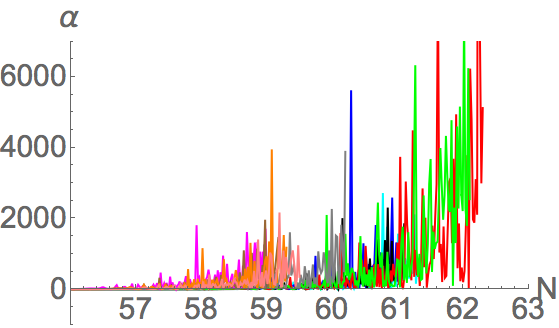} 
\includegraphics[width=72.5mm, height=55mm]{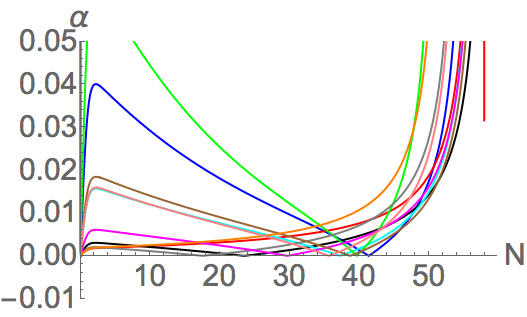} 
\includegraphics[width=72.5mm, height=55mm]{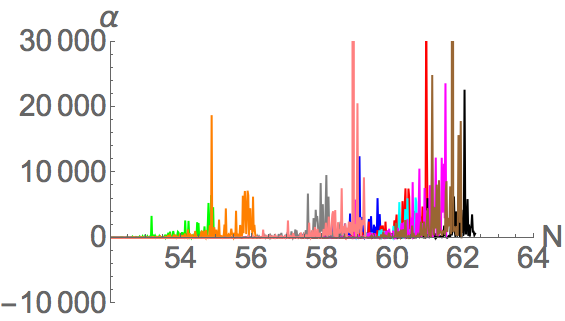} 
\includegraphics[width=72.5mm, height=55mm]{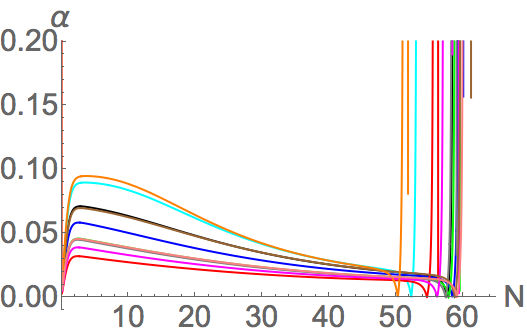} 
\includegraphics[width=72.5mm, height=55mm]{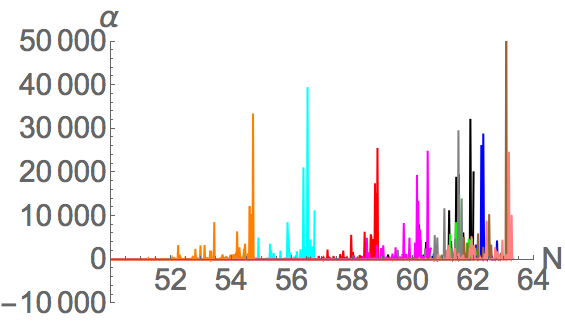} 
\caption{2 times of turn rate per Hubble parameter ${\alpha} = 2 {\omega}/H$ at ${\omega}_{1} < 0 \; \left( {\iota}_{1} < 0 \right)$. $1$st row: ${\rho} = 4$, $2$nd row: ${\rho} = 5$, $3$rd row: ${\rho} = 6$. The graphs on the right hand side (R.H.S.) are the close shots of that on the left hand side (L.H.S.). Various colors refer to various parameter sets listed on Table \ref{table: Feasible parameter sets for inflation for negative omega for alpha = 4}, \ref{table: Feasible parameter sets for inflation for negative omega for alpha = 5} and \ref{table: Feasible parameter sets for inflation for negative omega for alpha = 6} correspondingly. Left graphs show the overall evolutions, while right graphs are the closed shots at the end of inflation. }
\label{fig: 2 times of turn rate per Hubble parameter at omega1 smaller than 0}
\end{figure}

\begin{figure}[h!]
\centering
\includegraphics[width=72.5mm, height=55mm]{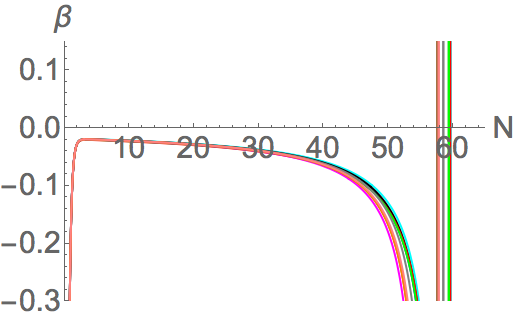} 
\includegraphics[width=72.5mm, height=55mm]{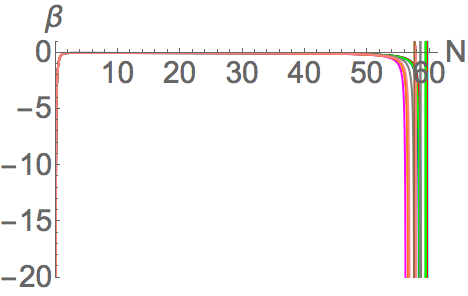} 
\includegraphics[width=72.5mm, height=55mm]{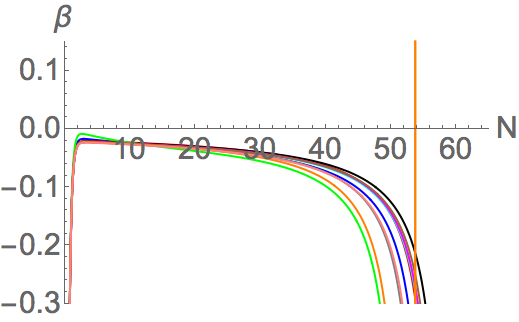} 
\includegraphics[width=72.5mm, height=55mm]{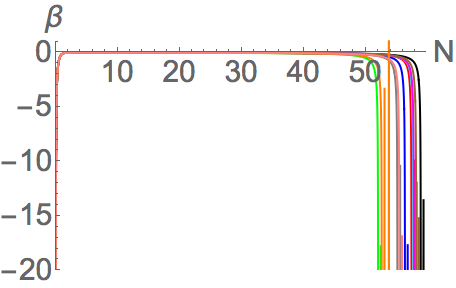} 
\includegraphics[width=72.5mm, height=55mm]{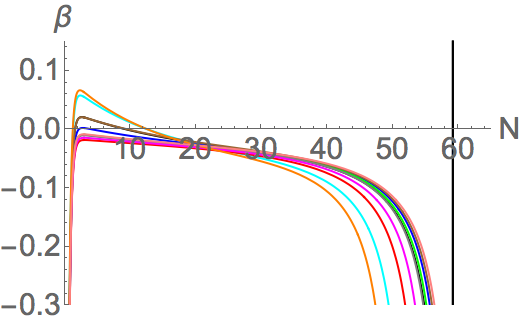} 
\includegraphics[width=72.5mm, height=55mm]{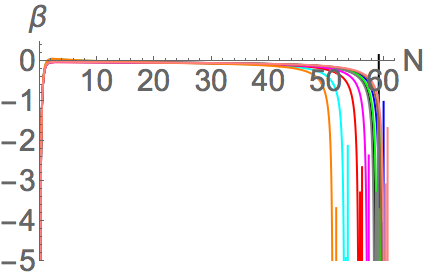} 
\caption{$\beta$ at ${\omega}_{1} < 0 \; \left( {\iota}_{1} < 0 \right)$. $1$st row: ${\rho} = 4$, $2$nd row: ${\rho} = 5$, $3$rd row: ${\rho} = 6$. The graphs on the right hand side (R.H.S.) are the close shots of that on the left hand side (L.H.S.). Various colors refer to various parameter sets listed on Table \ref{table: Feasible parameter sets for inflation for negative omega for alpha = 4}, \ref{table: Feasible parameter sets for inflation for negative omega for alpha = 5} and \ref{table: Feasible parameter sets for inflation for negative omega for alpha = 6} correspondingly. }
\label{fig: Beta at omega1 smaller than 0}
\end{figure}

\begin{sidewaystable}[h!]
\begin{center}
\begin{tabular}{ |c|c|c|c|c|c|c|c|c|c|c|c|c|c| }
\hline
$\text{Color}$ & ${\iota}_{1}$ & ${\iota}_{2}/ {\iota}_{1}$ & ${M}_{3/2+}/ {10}^{-7} {M}_{\text{pl}}$ & ${T}_{R \text{end}}$ & ${T}_{I \text{end}}$ & ${T}_{R \text{hc}}$ & ${T}_{I \text{hc}}$ & ${N}_{\text{end}}$ & ${N}_{\text{stop}}$ & ${\beta}_{\text{iso}}$ & $\cos{\left( \Delta \right)}$ \\
\hline
$\text{Red}$ & $-17$ & $7$ & $3.97837$ & $14.1667$ & $59.5$ & $104$ & $90$ & $57.1196$ & $61.1125$ & $3.38902 \times {10}^{-38}$ & $4.03578 \times {10}^{-5}$ \\
\hline
$\text{Green}$ & $-17$ & $7$ & $3.97837$ & $14.1667$ & $59.5$ & $85$ & $108$ & $51.7453$ & $54.9271$ & $3.61843 \times {10}^{-37}$ & $1.98477 \times {10}^{-5}$ \\
\hline
$\text{Blue}$ & $-17$ & $7$ & $3.97837$ & $14.1667$ & $59.5$ & $90$ & $107$ & $56.0768$ & $59.6890$ & $8.03244 \times {10}^{-37}$ & $3.27246 \times {10}^{-5}$ \\
\hline
$\text{Black}$ & $-17$ & $7$ & $3.97837$ & $14.1667$ & $59.5$ & $100$ & $98$ & $58.6238$ & $62.3304$ & $7.06245 \times {10}^{-41}$ & $1.10696 \times {10}^{-4}$ \\
\hline
$\text{Cyan}$ & $-17$ & $7$ & $3.97837$ & $14.1667$ & $59.5$ & $95$ & $103$ & $57.4204$ & $60.9148$ & $1.34988 \times {10}^{-37}$ & $1.23669 \times {10}^{-4}$ \\
\hline
$\text{Gray}$ & $-19$ & $6$ & $3.36704$ & $95$ & $57$ & $110$ & $98$ & $54.8516$ & $58.5579$ & $3.8036 \times {10}^{-39}$ & $7.51476 \times {10}^{-5}$ \\
\hline
$\text{Magenta}$ & $-21$ & $5$ & $2.89767$ & $87.5$ & $52.5$ & $121$ & $102$ & $57.5109$ & $61.5113$ & $7.10984 \times {10}^{-38}$ & $2.02087 \times {10}^{-4}$ \\
\hline
$\text{Brown}$ & $-23$ & $4$ & $2.52805$ & $76.6667$ & $46$ & $128$ & $106$ & $57.8894$ & $61.9445$ & $2.08549 \times {10}^{-37}$ & $6.69631 \times {10}^{-5}$ \\
\hline
$\text{Orange}$ & $-27$ & $5$ & $1.98762$ & $112.5$ & $67.5$ & $160$ & $115$ & $52.3259$ & $56.1155$ & $6.41039 \times {10}^{-37}$ & $2.44855 \times {10}^{-5}$ \\
\hline
$\text{Pink}$ & $-29$ & $6$ & $1.78559$ & $145$ & $87$ & $160$ & $160$ & $55.1624$ & $59.2207$ & $3.48978 \times {10}^{-36}$ & $7.42136 \times {10}^{-5}$ \\
\hline
\end{tabular}
\end{center}
\caption{Feasible parameter sets for inflation for ${\omega}_{1} < 0$ (or ${\iota}_{1} < 0$) and ${\rho} = 5$. Note that the $F$ term SUSY breaking scale is $0$ for all real values of parameters and fields. Recall the unit of each variable is the following. $\left[ {\iota}_{1} \right] = {M}_{\text{pl}}$, $\left[ {M}_{3/2+} \right] = {M}_{\text{pl}}$, $\left[ {T}_{R \text{end}} \right] = \left[ {T}_{I \text{end}} \right] = \left[ {T}_{R \text{hc}} \right] = \left[ {T}_{I \text{hc}} \right] = {M}_{\text{pl}}$. ${N}_{\text{end}}$ is the e-folding number when ${\epsilon} = 1$, which represents the end of inflation. ${N}_{\text{stop}}$ means the e-folding number that we stop the numerical calculation. }
\label{table: Feasible parameter sets for inflation for negative omega for alpha = 5}

\begin{center}
\begin{tabular}{ |c|c|c|c|c|c|c|c|c|c|c|c|c|c| }
\hline
$\text{Color}$ & ${\iota}_{1}$ & ${\iota}_{2}/ {\iota}_{1}$ & ${M}_{3/2+}/ {10}^{-8} {M}_{\text{pl}}$ & ${T}_{R \text{end}}$ & ${T}_{I \text{end}}$ & ${T}_{R \text{hc}}$ & ${T}_{I \text{hc}}$ & ${N}_{\text{end}}$ & ${N}_{\text{stop}}$ & ${\beta}_{\text{iso}}$ & $\cos{\left( \Delta \right)}$ \\
\hline
$\text{Gray}$ & $-18$ & $9$ & $6.85871$ & $121.5$ & $81$ & $60$ & $100$ & $58.2101$ & $61.6281$ & $8.06083 \times {10}^{-35}$ & $6.77743 \times {10}^{-5}$ \\
\hline
$\text{Magenta}$ & $-20$ & $8$ & $5.55556$ & $120$ & $80$ & $67$ & $100$ & $56.9523$ & $60.5887$ & $7.4834 \times {10}^{-35}$ & $8.14647 \times {10}^{-5}$ \\
\hline
$\text{Brown}$ & $-22$ & $7$ & $4.59137$ & $115.5$ & $77$ & $70$ & $105$ & $59.5551$ & $63.1895$ & $2.44999 \times {10}^{-35}$ & $1.01176 \times {10}^{-5}$ \\
\hline
$\text{Orange}$ & $-24$ & $6$ & $3.85802$ & $108$ & $72$ & $70$ & $105$ & $50.9481$ & $54.7494$ & $2.71462 \times {10}^{-35}$ & $1.01176 \times {10}^{-5}$ \\
\hline
$\text{Pink}$ & $-26$ & $6$ & $3.28731$ & $117$ & $78$ & $87$ & $106$ & $59.907$ & $63.3246$ & $5.44139 \times {10}^{-37}$ & $4.42432 \times {10}^{-5}$ \\
\hline
$\text{Red}$ & $-30$ & $5$ & $2.46914$ & $22.5$ & $75$ & $101$ & $103$ & $55.4472$ & $58.8651$ & $4.7257 \times {10}^{-37}$ & $8.54936 \times {10}^{-5}$ \\
\hline
$\text{Green}$ & $-30$ & $5$ & $2.46914$ & $22.5$ & $75$ & $100$ & $107$ & $58.682$ & $62.0998$ & $3.2964 \times {10}^{-37}$ & $3.88947 \times {10}^{-5}$ \\
\hline
$\text{Blue}$ & $-30$ & $5$ & $2.46914$ & $22.5$ & $75$ & $98$ & $110$ & $59.2057$ & $62.8400$ & $1.28312 \times {10}^{-34}$ & $2.44867 \times {10}^{-5}$ \\
\hline
$\text{Black}$ & $-30$ & $5$ & $2.46914$ & $22.5$ & $75$ & $95$ & $113$ & $58.3676$ & $62.0021$ & $1.44804 \times {10}^{-36}$ & $1.58228 \times {10}^{-5}$ \\
\hline
$\text{Cyan}$ & $-30$ & $5$ & $2.46914$ & $22.5$ & $75$ & $89$ & $116$ & $52.9635$ & $56.7473$ & $1.58416 \times {10}^{-34}$ & $9.00233 \times {10}^{-6}$ \\
\hline
\end{tabular}
\end{center}
\caption{Feasible parameter sets for inflation for ${\omega}_{1} < 0$ (or ${\iota}_{1} < 0$) and ${\rho} = 6$. Note that the $F$ term SUSY breaking scale is $0$ for all real values of parameters and fields. Recall the unit of each variable is the following. $\left[ {\iota}_{1} \right] = {M}_{\text{pl}}$, $\left[ {M}_{3/2+} \right] = {M}_{\text{pl}}$, $\left[ {T}_{R \text{end}} \right] = \left[ {T}_{I \text{end}} \right] = \left[ {T}_{R \text{hc}} \right] = \left[ {T}_{I \text{hc}} \right] = {M}_{\text{pl}}$. ${N}_{\text{end}}$ is the e-folding number when ${\epsilon} = 1$, which represents the end of inflation. ${N}_{\text{stop}}$ means the e-folding number that we stop the numerical calculation. }
\label{table: Feasible parameter sets for inflation for negative omega for alpha = 6}
\end{sidewaystable}

\noindent Similar to the cases ${\omega}_{1} > 0 $, in Figure \ref{fig: Alpha 456 and omega1 smaller than 0}, we show the possible positive values of ${T}_{R\text{hc}}$, ${T}_{I \text{hc}}$ and ${\iota}_{2}/ {\iota}_{1}$ based on the Planck observation constraints listed in Table \ref{table:Planck data 2018 slow roll potential parameters and spectral indices} at various ${\rho} = 4, 5, 6$. The "C" shape tube regions gradually change the color from red to green as ${\iota}_{1}$ gradually decreases with a constant increment (For details, please refer to the description of Figure \ref{fig: Alpha 456 and omega1 smaller than 0}). We also extract some of the possible parameters satisfying the e-folding constraints $50 \leq {N}_{\text{end}} - {N}_{\text{hc}} \leq 60$ as listed in Table \ref{table: Feasible parameter sets for inflation for negative omega for alpha = 4}, \ref{table: Feasible parameter sets for inflation for negative omega for alpha = 5} and \ref{table: Feasible parameter sets for inflation for negative omega for alpha = 6}, and their corresponding evolution paths are plotted in Figure \ref{fig: evolution paths at omega1 smaller than 0}. One can see that for ${\rho} = 4$, the paths move towards their corresponding minimum point (dS vacuum point) in a nearly a straight line, turn to the minimum point with a larger turn rate and then oscillate around the minimum point. However, the extent of turning is smaller than that of the evolution paths at ${\omega}_{1} > 0$ and ${\rho} = 4$. As ${\rho}$ increases gradually up to $6$, the paths change its shape into the form like turning greatly in the initial stage, moving towards their corresponding minimum point and finally oscillating around the minimum point. \\

\vspace{3mm}

\noindent Next, in Figure \ref{fig: Epsilon EtaSigmaSigma at omega1 smaller than 0}, we can see the evolutions of $\epsilon$ and ${\eta}_{{\sigma}{\sigma}}$ of all the listed parameter sets in Table \ref{table: Feasible parameter sets for inflation for negative omega for alpha = 4}, \ref{table: Feasible parameter sets for inflation for negative omega for alpha = 5} and \ref{table: Feasible parameter sets for inflation for negative omega for alpha = 6}. Basically, starting from very small positive values, all the $\epsilon$ (dotted) lines evolve and keep at a close-to-zero value, surge to 3 after $50$ e-folding and then oscillate between $0$ and $3$. Also, ${\eta}_{{\sigma}{\sigma}}$ lines start from a value very close to zero, evolve around zero and then sharply increase at the end of inflation. In Figure \ref{fig: Square of effective mass per Hubble parameter at omega1 greater than 0}, after dropping sharply from a large positive value to a small positive value, the effective mass square of the entropy perturbation $\left( {\mu}_{s}/H \right)^{2}$ remains light and physical. This shows that the trajectory rolls from the plateau, where the field space curvature orthogonal to the trajectory is small in the initial stage, to the minimum point, where the entropy mass square largely grows.

\subsection{Suppose ${\rho} = 3$. }
\begin{figure}[h!]
\centering
\includegraphics[width=72.5mm, height=65mm]{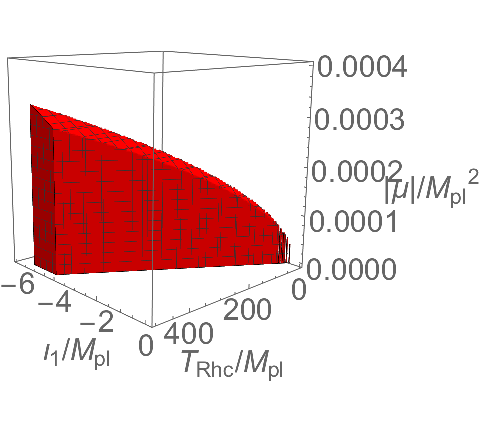} 
\includegraphics[width=72.5mm, height=65mm]{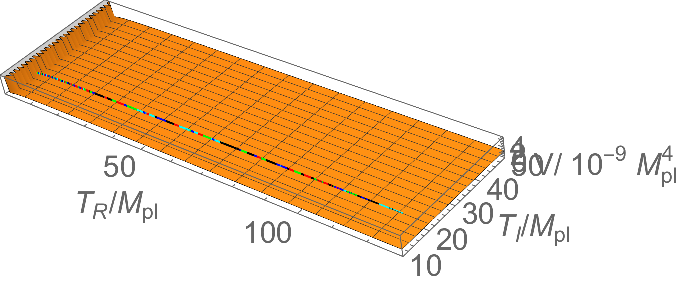} 
\caption{Left: A region of possible values of ${T}_{R \text{hc}}/ {M}_{\text{pl}}$, ${\iota}_{1}/ {M}_{\text{pl}}$ and $\left| {\mu} \right|/ {M}_{\text{pl}}^{2}$ for inflation based on the constraints listed in Table \ref{table:Planck data 2018 slow roll potential parameters and spectral indices}. Right: The evolution of background fields on the potential surface. The colors of the lines correspond to the parameter sets listed in Table \ref{table:Planck data 2018 slow roll potential parameters and spectral indices}. }
\label{fig: 3D for alpha 3}
\end{figure}

\begin{figure}[h!]
\centering
\includegraphics[width=130mm, height=35mm]{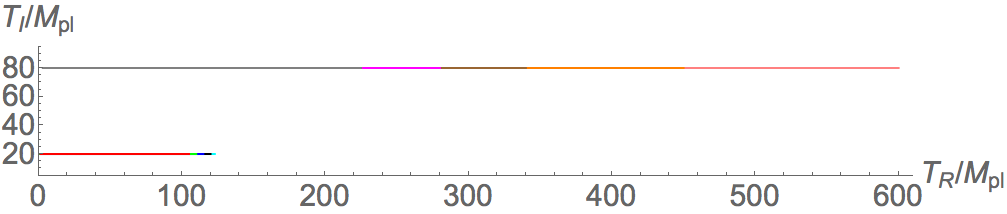} 
\caption{Evolution paths of inflation dynamics at ${\rho} = 3$. Various colors refer to various parameter sets listed on Table \ref{table: Feasible parameter sets for inflation for alpha 3}. }
\label{fig: evolution paths for alpha 3}
\end{figure}

\begin{figure}[h!]
\centering
\includegraphics[width=72.5mm, height=60mm]{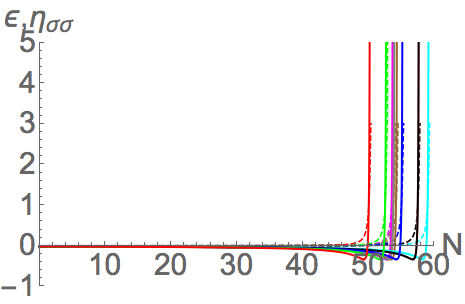} 
\caption{Evolutions of slow-roll parameters ${\epsilon}$ and ${\eta}_{{\sigma}{\sigma}}$ for ${\rho} = 3$. Various colors refer to various parameter sets listed on Table \ref{table: Feasible parameter sets for inflation for alpha 3}. Dotted lines represent the evolutions of $\epsilon$ while the solid lines represent the counterpart of ${\eta}_{{\sigma}{\sigma}}$. }
\label{fig: evolution of Epsilon and Eta Sigma Sigma for alpha 3}
\end{figure}

\begin{figure}[h!]
\centering
\includegraphics[width=72.5mm, height=60mm]{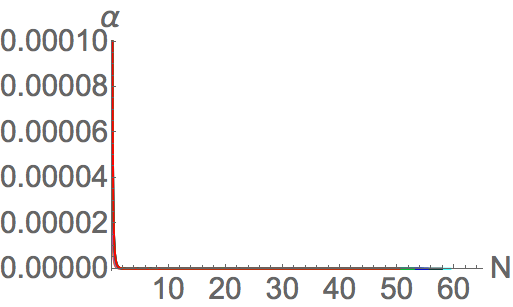} 
\includegraphics[width=72.5mm, height=60mm]{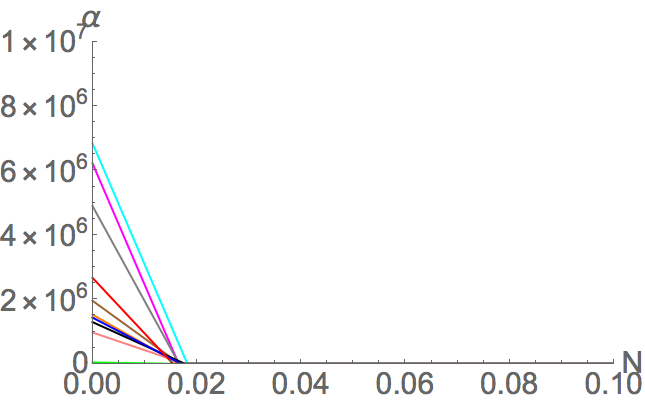} 
\caption{Left: Evolutions of the turn rate per Hubble parameter ${\alpha} \left( N \right) = 2 {\omega} \left( N \right)/ H \left( N \right)$. Right: A close shot of evolutions of $\alpha \left( N \right)$ for ${\rho} = 3$. Various colors refer to various parameter sets listed on Table \ref{table: Feasible parameter sets for inflation for alpha 3}. }
\label{fig: evolutions of alpha for alpha 3}
\end{figure}

\begin{figure}[h!]
\centering
\includegraphics[width=72.5mm, height=60mm]{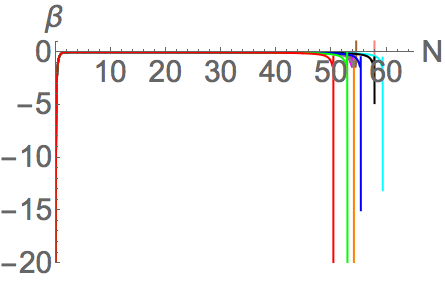} 
\includegraphics[width=72.5mm, height=60mm]{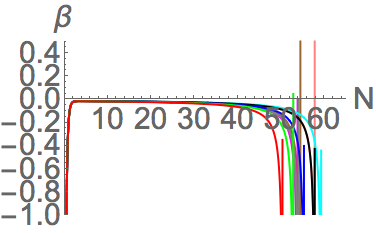} 
\caption{Left: Evolutions of ${\beta} \left( N \right)$. Right: A closed shot of evolutions of $\beta \left( N \right)$ for ${\rho} = 3$. Various colors refer to various parameter sets listed on Table \ref{table: Feasible parameter sets for inflation for alpha 3}. }
\label{fig: evolutions of beta for alpha 3}
\end{figure}

\noindent The total potential is given by Eq.(\ref{Total potential at alpha 3}). The possible initial ${T}_{R \text{hc}}$, ${\iota}_{1} = {\omega}_{1}/ \left| {\mu} \right|^{2}$ and $\left| {\mu} \right|$ are shown in Figure \ref{fig: 3D for alpha 3}. Also, the parameters of possible inflation are given by Table \ref{table: Feasible parameter sets for inflation for alpha 3}. The evolutions of the background fields ${T}_{R}$ and ${T}_{I}$ are evaluated by their E.O.M.s Eq. (\ref{EOMs for alpha = 3}). Basically, the paths are straight lines from the initial starting points. Next, for slow-roll parameters as shown in Figure \ref{fig: evolution of Epsilon and Eta Sigma Sigma for alpha 3}, starting from a value very close to zero, they keep its low value and become significant after $50$ e-folding, which is consistent with the usual inflation mechanism. Apart from these, from Figure \ref{fig: evolutions of alpha for alpha 3}, we can see that turn rates drop rapidly and keep nearly zero as the background fields roll off the ridge. This shows that the fields turn for a little while and roll straight off the ridge. Furthermore, in Figure \ref{fig: evolutions of mus per hubble for alpha 3}, we can know that the evolutions of the effective mass square of the entropy perturbation $\left( {\mu}_{s}/ H \right)^{2}$ remains nearly zero.

\section{Discussion}
\label{Discussion}
\subsection{The scale of the potential}
\noindent Referring to Figure \ref{fig: Alpha 456 and omega1 greater than 0} and \ref{fig: Alpha 456 and omega1 smaller than 0}, one can see that the evolution paths exist at different energy scales. In fact, there are two main parameters affecting the potential scale. One is $\left| {\mu} \right|$, and another is ${\rho}$. Referring to Eq.(\ref{F term of total potential}), (\ref{D term of total potential when omega1 greater than zero}) and (\ref{D term of total potential when omega1 smaller than zero}), one can see that the potential is directly proportional to $\left| {\mu} \right|^{2}$, while the powers of $\left( \frac{{M}_{\text{pl}}}{2 {T}_{R}} \right)^{\rho}$ in Eq.(\ref{F term of total potential}), $\left(\frac{\left( {\rho}^{2} - 7 \rho + 4 \right) {M}_{\text{pl}}}{- \left( \rho - 3 \right) \rho  {\iota}_{1} }\right)^{\rho -1}$ in Eq.(\ref{D term of total potential when omega1 greater than zero}) and $\left(\frac{\left( {\rho}^{2} - 7 \rho + 4 \right) {M}_{\text{pl}}}{- \left( \rho - 3 \right) \rho  {\iota}_{1} }\right)^{\rho -1}$ in Eq.(\ref{D term of total potential when omega1 smaller than zero}) show ${\rho}$ is related to the power of some field values and parameters. Since $\left| {\mu} \right|$ is held in a fixed scale $O \left( {10}^{-4} \right) {M}_{\text{pl}}$, the difference lies in various ${\rho}$. Since the scale of field values is $O \left( {10}^{2} \right) {M}_{\text{pl}}$, the potential scale decreases by ${10}^{-2}$ times as ${\rho}$ increases by $1$. \\

\subsection{Turn rate for ${\rho} \neq 3$}
\noindent In Figure \ref{fig: evolution paths at omega1 greater than 0} and \ref{fig: evolution paths at omega1 smaller than 0}, one can see that there are path patterns with various extent of turning. This can be tracked by Figure \ref{fig: 2 times of turn rate per Hubble parameter at omega1 greater than 0} (and \ref{fig: 2 times of turn rate per Hubble parameter at omega1 smaller than 0}) respectively. For example, for ${\omega}_{1} > 0$ and ${\rho} = 4$, when the background fields roll off from the initial point, there is no rigorous turning until the end of inflation, leading to the evolution of $2 \omega/H$ close to zero. As they gradually approach to their corresponding minimum points, Hubble scale $H$ reduces gradually and the turning becomes rigorous, resulting in the gradual rise of $2 \omega/H$. Finally, when they oscillate around their minimum points, the turning becomes sharply rigorous, leading to the sudden surge of $2 \omega/H$. Similar arguments can also be made for other cases by tracking the evolution paths and the variations of $2 \omega/H$. 

\subsection{The turning for ${\rho} = 3$ and the afterward evolution}
\begin{figure}[h!]
\centering
\includegraphics[width=72.5mm, height=50mm]{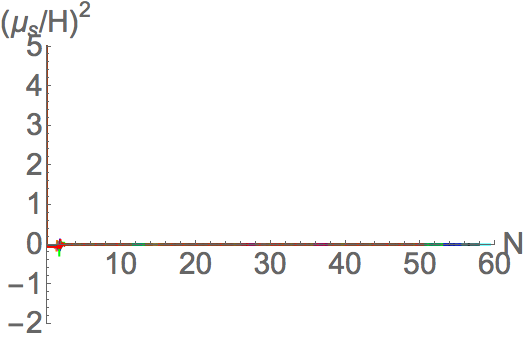} 
\includegraphics[width=72.5mm, height=50mm]{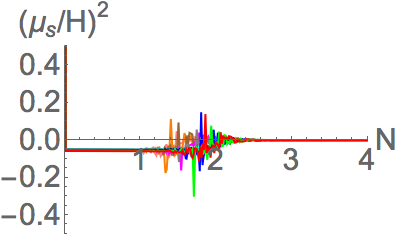} 
\caption{Left: Evolutions of $\left( {\mu}_{s}/ H \right)^{2}$ for ${{\rho}} = 3$. Right: A close shot from $0$ to $4$ e-foldings. Various colors refer to various parameter sets listed on Table \ref{table: Feasible parameter sets for inflation for alpha 3}. }
\label{fig: evolutions of mus per hubble for alpha 3}
\end{figure}

\begin{figure}[h!]
\centering
\includegraphics[width=72.5mm, height=60mm]{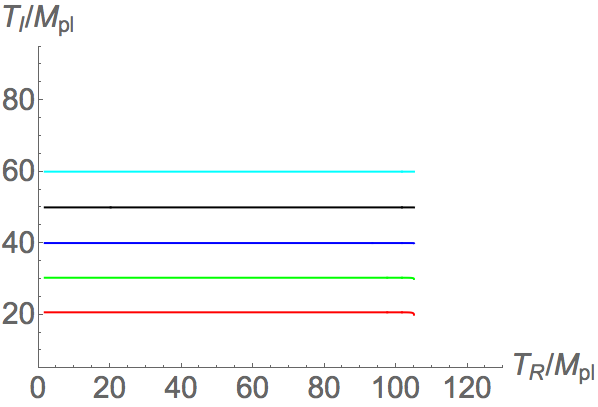} 
\includegraphics[width=72.5mm, height=60mm]{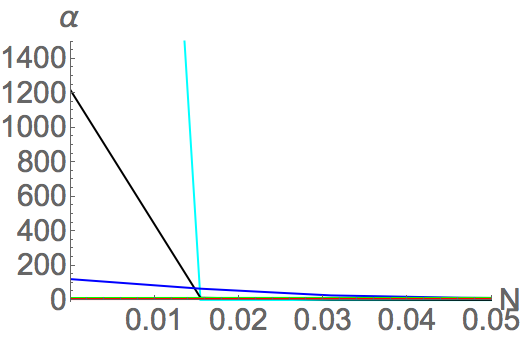} 
\caption{Field evolution at different $\alpha \left( N \right)$ (Left) and their corresponding initial turn rate per Hubble parameter (Right). Here, the field value ${T}_{I}$ does not affect the strength of $\alpha \left( N \right)$. Thus, for obvious observation, we evaluate at different ${T}_{I}$. Various colors refer to various parameter sets listed on Table \ref{table: Parameter sets showing the significant turning for alpha 3}. }
\label{fig: evolutions of field evolution and alpha for alpha 3}
\end{figure}

\noindent From Figure \ref{fig: evolutions of mus per hubble for alpha 3}, since the initial speeds of background fields have the small scale ${10}^{-5} {M}_{\text{pl}}$, it is not obvious to see the turning. Hence, we demonstrate the inflation path with the greatest possible initial speed of ${T}_{I}$. The parameter sets are shown in Table \ref{table: Parameter sets showing the significant turning for alpha 3} and their corresponding evolution paths are shown in Figure \ref{fig: evolutions of field evolution and alpha for alpha 3}. We can significantly see that the background fields move along ${T}_{I}$ a little while and turn towards the decreasing ${T}_{R}$ direction. However, ${\alpha} \left( N \right)$ is reduced as the initial speed of background fields increases. That is because the turn rate per Hubble parameter is mainly inversely proportional to the resultant field rate. Hence, even though it is not significant to see the turning in the field evolution graph, the initial ${\alpha} \left( N \right)$ is comparatively large, and vice versa. After that, the background fields keep its ${T}_{I}$ coordinate when they continue rolling. This is because Hubble parameter and $\dot{T}_{R}/{T}_{R}$ in Eq.(\ref{EOMs for alpha = 3}) act as friction terms to stabilize the curve at a constant ${T}_{I}$ coordinate.

\begin{sidewaystable}[h!]
\begin{center}
\begin{tabular}{ |c|c|c|c|c|c|c|c|c|c|c|c|c|c|c| }
\hline
$\text{Color}$ & $\left| {\mu} \right|/ {M}_{\text{pl}}^{2}$ & ${\iota}_{1}/ {M}_{\text{pl}}$ & ${M}_{3/2+}/ {M}_{\text{pl}}$ & ${T}_{R \text{end}}/ {M}_{\text{pl}} $ & ${T}_{R \text{hc}}/ {M}_{\text{pl}}$ & ${T}_{I \text{hc}}/ {M}_{\text{pl}}$ & ${N}_{\text{end}}$ & ${N}_{\text{stop}}$ & ${\beta}_{\text{iso}}$ & $\cos{\left( \Delta \right)}$ \\
\hline
$\text{Gray}$ & ${10}^{-4}$ & $-2$ & $1.36083 \times {10}^{-5}$ & $3$ & $225$ & $80$ & $53.7825$ & $54.1041$ & $7.54744 \times {10}^{-36}$ & $3.28888 \times {10}^{-6}$ \\
\hline
$\text{Magenta}$ & ${10}^{-4}$ & $-2.5$ & $1.21716 \times {10}^{-5}$ & $3.75$ & $280$ & $80$ & $53.5343$ & $53.8559$ & $1.36149 \times {10}^{-34}$ & $2.63095 \times {10}^{-5}$ \\
\hline
$\text{Brown}$ & ${10}^{-4}$ & $-3$ & $1.11111 \times {10}^{-5}$ & $4.5$ & $340$ & $80$ & $54.1961$ & $54.5177$ & $1.68237 \times {10}^{-37}$ & $2.19281 \times {10}^{-6}$ \\
\hline
$\text{Orange}$ & ${10}^{-4}$ & $-4$ & $9.6225 \times {10}^{-6}$ & $6$ & $450$ & $80$ & $53.7825$ & $54.1041$ & $2.19583 \times {10}^{-35}$ & $1.64444 \times {10}^{-6}$ \\
\hline
$\text{Pink}$ & ${10}^{-4}$ & $-5$ & $8.60663 \times {10}^{-6}$ & $7.5$ & $600$ & $80$ & $57.5058$ & $57.8274$ & $7.62281 \times {10}^{-36}$ & $1.31667 \times {10}^{-6}$ \\
\hline 
$\text{Red}$ & ${10}^{-4}$ & $-1$ & $1.9245 \times {10}^{-5}$ & $1.5$ & $105$ & $20$ & $50.061$ & $50.3826$ & $7.23759 \times {10}^{-34}$ & $6.57141 \times {10}^{-6}$ \\
\hline
$\text{Green}$ & ${10}^{-4}$ & $-1$ & $1.9245 \times {10}^{-5}$ & $1.5$ & $110$ & $20$ & $52.5417$ & $52.9298$ & $5.27718 \times {10}^{-32}$ & $6.57574 \times {10}^{-6}$ \\
\hline
$\text{Blue}$ & ${10}^{-4}$ & $-1$ & $1.9245 \times {10}^{-5}$ & $1.5$ & $115$ & $20$ & $55.0234$ & $55.3450$ & $8.40563 \times {10}^{-35}$ & $6.57969 \times {10}^{-6}$ \\
\hline
$\text{Black}$ & ${10}^{-4}$ & $-1$ & $1.9245 \times {10}^{-5}$ & $1.5$ & $120$ & $20$ & $57.5058$ & $57.8274$ & $1.87223 \times {10}^{-35}$ & $6.58331 \times {10}^{-6}$ \\
\hline
$\text{Cyan}$ & ${10}^{-4}$ & $-1$ & $1.9245 \times {10}^{-5}$ & $1.5$ & $123$ & $20$ & $58.9956$ & $59.3172$ & $2.43366 \times {10}^{-33}$ & $6.58534 \times {10}^{-6}$ \\
\hline
\end{tabular}
\end{center}
\caption{Feasible parameter sets for inflation for ${\rho} = 3$. In these sets, ${T'}_{R \text{hc}} = {T'}_{I \text{hc}} = {10}^{-5} {M}_{\text{pl}}$. ${N}_{\text{end}}$ is the e-folding number when ${\epsilon} = 1$, which represents the end of inflation. ${N}_{\text{stop}}$ means the e-folding number that we stop the numerical calculation. }
\label{table: Feasible parameter sets for inflation for alpha 3}

\begin{center}
\begin{tabular}{ |c|c|c|c|c|c|c|c|c|c|c|c|c|c|c| }
\hline
$\text{Color}$ & $\left| {\mu} \right|/ {M}_{\text{pl}}^{2}$ & ${\iota}_{1}/ {M}_{\text{pl}}$ & ${T}_{R \text{end}}/ {M}_{\text{pl}} $ & ${T}_{R \text{hc}}/ {M}_{\text{pl}}$ & ${T}_{I \text{hc}}/ {M}_{\text{pl}}$ & ${T'}_{R \text{hc}}/ {M}_{\text{pl}}$ & ${T'}_{I \text{hc}}/ {M}_{\text{pl}}$ & ${N}_{\text{end}}$ & ${N}_{\text{stop}}$ & ${\beta}_{\text{iso}}$ & $\cos{\left( \Delta \right)}$ \\
\hline
$\text{Cyan}$ & ${10}^{-4}$ & $-1$ & $1.5$ & $105$ & $60$ & ${10}^{-5}$ & ${10}^{-3}$ & $50.061$ & $50.3826$ & $9.20002 \times {10}^{-36}$ & $3.28604 \times {10}^{-4}$ \\
\hline
$\text{Black}$ & ${10}^{-4}$ & $-1$ & $1.5$ & $105$ & $50$ & ${10}^{-5}$ & ${10}^{-2}$ & $50.061$ & $50.3826$ & $1.82773 \times {10}^{-30}$ & $3.2857 \times {10}^{-3}$ \\
\hline
$\text{Blue}$ & ${10}^{-4}$ & $-1$ & $1.5$ & $105$ & $40$ & ${10}^{-5}$ & ${10}^{-1}$ & $50.061$ & $50.3826$ & $4.48765 \times {10}^{-39}$ & $3.28483 \times {10}^{-2}$ \\
\hline
$\text{Green}$ & ${10}^{-4}$ & $-1$ & $1.5$ & $105$ & $30$ & ${10}^{-5}$ & $1$ & $50.0607$ & $50.3823$ & $1.87863 \times {10}^{-9}$ & $3.203 \times {10}^{-1}$ \\
\hline
$\text{Red}$ & ${10}^{-4}$ & $-1$ & $1.5$ & $105$ & $20$ & ${10}^{-5}$ & $2$ & $50.0599$ & $50.3815$ & $6.73078 \times {10}^{-7}$ & $5.7951 \times {10}^{-1}$ \\
\hline
\end{tabular}
\end{center}
\caption{Parameter sets for showing the significant turning in ${\rho} = 3$ case. The difference of each parameter sets lie in ${T'}_{I \text{hc}}/ {M}_{\text{pl}}$, the initial rate of ${T}_{I}$ and the initial coordinate of ${T}_{I}$. ${N}_{\text{end}}$ is the e-folding number when ${\epsilon} = 1$, which represents the end of inflation. ${N}_{\text{stop}}$ means the e-folding number that we stop the numerical calculation. }
\label{table: Parameter sets showing the significant turning for alpha 3}
\end{sidewaystable}

\subsection{The smallness of ${\beta}_{\text{iso}}$ and $\cos{\left( {\Delta} \right)}$}
\noindent To understand the smallness of ${\beta}_{\text{iso}}$ and $\cos{\left( {\Delta} \right)}$, we should consider the reason of smallness of ${T}_{\mathcal{SS}}$ and ${T}_{\mathcal{RS}}$, since they are related by Eq.(\ref{betaiso from Nhc to Nend}) and (\ref{cosine Delta from Nhc to Nend}). By Eq.(\ref{TRS integration in terms of e-foldings}) and (\ref{TSS integration in terms of e-foldings}) (or by Eq.(\ref{TRS integration in terms of cosmic time}) and (\ref{TSS integration in terms of cosmic time})), we can see that ${T}_{\mathcal{SS}}$ and ${T}_{\mathcal{RS}}$ depend on the strengths of ${\alpha}$ (turn rate per Hubble parameter) and ${\beta}$. By Figure \ref{fig: 2 times of turn rate per Hubble parameter at omega1 greater than 0}, \ref{fig: Beta at omega1 greater than 0}, \ref{fig: 2 times of turn rate per Hubble parameter at omega1 smaller than 0} and \ref{fig: Beta at omega1 smaller than 0}, we can see ${\alpha} \left( N \right)$ is nearly zero, while ${\beta} \left( N \right)$ keeps negative from ${N}_{\text{hc}}$ to ${N}_{\text{end}}$. These show that ${T}_{\mathcal{SS}}$ becomes exponentially suppressed and there is very small contribution by each $dN$ in the integration of ${T}_{\mathcal{RS}}$, leading to smallness of ${T}_{\mathcal{SS}}$ and ${T}_{\mathcal{RS}}$, and so are ${\beta}_{\text{iso}}$ and $\cos{\left( {\Delta} \right)}$.

\section{Conclusion}
\label{Conclusion}
\noindent We studied the inflation dynamics of generalized dilaton-axion models with a new Fayet-Iliopoulos (FI) term not requiring gauging R-symmetry, where both dilaton ${T}_{R}$ and its axionic partner ${T}_{I}$ are responsible for contributing the evolution of inflation. We found that to obtain dS vacuum at the end of inflation, integer ${\rho}$ in the generalized K\"ahler potential is constrained at different cases of ${\omega}_{1}$. Particularly, when ${\rho} = 3$, the axionic partner does not contribute to the inflation as it is absent in the total potential. We also evaluated the feasible initial conditions to produce successful inflation based on Planck observation and e-folding constraint and showed different evolution paths with various turning patterns of the background field. This shows that apart from slow roll parameters, spectral index and inflation energy scale, the change of turn rates can also be one of the fingerprints to identify the specific inflation pattern in double field model verification. Finally, we gave the predictions of iso-curvature mode contributions such as ${\beta}_{\text{iso}}$ and $\cos{\Delta}$. 

\section{Acknowledgement}
\noindent The author thanks Prof. Hiroyuki ABE very much for suggestion and useful discussions. The author is supported by AY 2019 Scholarship for Young Doctoral Students, Waseda University.

\appendix
\section{Christoffel symbols and curvatures of field spaces}
\noindent Given that the metric of the field space in the Einstein frame is given by
\begin{equation}
\mathcal{G}_{IJ} = \frac{{\rho} {M}_{\text{pl}}^{2}}{2 {T}_{R}^{2}} {\delta}_{IJ} \quad \forall \; I, J \in \left\{1,2 \right\}, \\
\end{equation}
\noindent the non-zero Christoffel symbols defined by
\begin{equation}
{\Gamma}^{I}_{JK} = \frac{1}{2} \mathcal{G}^{IL} \left( \mathcal{G}_{LJ,K} + \mathcal{G}_{LK,J} - \mathcal{G}_{JK,L} \right), \\
\end{equation}
\noindent are
\begin{equation}
- {\Gamma}^{1}_{11} = {\Gamma}^{1}_{22} = - {\Gamma}^{2}_{12} = - {\Gamma}^{2}_{21} = \frac{1}{{T}_{R}}. \\
\end{equation}
\noindent The non-zero curvature tensors defined by
\begin{equation}
{R}^{I}_{JKL} = {\partial}_{K} {\Gamma}^{I}_{JL} + {\Gamma}^{I}_{CK} {\Gamma}^{C}_{JL} - \left( {\partial}_{L} {\Gamma}^{I}_{JK} + {\Gamma}^{I}_{CL} {\Gamma}^{C}_{JK} \right), \\
\end{equation}
\noindent are
\begin{equation}
- {R}^{1}_{212} = {R}^{1}_{221} = {R}^{2}_{112} = - {R}^{2}_{121} = \frac{1}{{T}_{R}^{2}}, \\
\end{equation}
\noindent which gives the Ricci tensor defined by ${R}_{JL} = \mathcal{G}^{K}_{I} {R}^{I}_{JKL} = {R}^{I}_{JIL}$ as
\begin{equation}
{R}_{11} = {R}_{22} = - \frac{1}{{T}_{R}^{2}}, \\
\end{equation}
\noindent and the curvature scalar, defined by $R \equiv \mathcal{G}^{IJ} {R}_{IJ}$, as
\begin{equation}
R = - \frac{4}{{\rho} {M}_{\text{pl}}^{2}}. \\
\end{equation}

\section{A formalism of Double Field Inflation}
\label{A formalism of Double Field Inflation}
\noindent In this section, we follow the derivation in \cite{1210.7487} and \cite{1310.8285}. Note that in the Jordan frame, the Lagrangian is
\begin{equation}
{S}_{\text{Jordan}} = \int d^{4} x \sqrt{-\tilde{g}} \left[ f\left({\phi}^{I}\right) \tilde{R} - \frac{1}{2} \tilde{\mathcal{G}}_{IJ} \tilde{g}^{{\mu}{\nu}} \tilde{\triangledown}_{\mu} {\phi}^{I} \tilde{\triangledown}_{\nu} {\phi}^{J} - \tilde{V}\left( {\phi}^{I} \right) \right]. \\
\end{equation}
\noindent where $f\left( {\phi}^{I} \right)$ is the non-minimal coupling function and $\tilde{V}\left( {\phi}^{I} \right)$ is the potential for the scalar fields in the Jordan frame. To change the equation in Jordan frame into the counterpart in Einstein frame, we define a spacetime metric in the Einstein frame ${g}_{{\mu}{\nu}}\left( x \right)$ as 
\begin{equation}
{g}_{{\mu}{\nu}}\left( x \right) = {\Omega}^{2} \left( x \right) \tilde{g}_{{\mu}{\nu}}\left( x \right), \\
\end{equation}
\noindent where the conformal factor ${\Omega}^{2} \left( x \right)$ is given by
\begin{equation}
{\Omega}^{2} \left( x \right) = \frac{2}{{M}^{2}_{\text{pl}}} f \left( {\phi}^{I} \left( x \right) \right). \\
\end{equation}
\noindent Then, the action in Jordan frame becomes that in Einstein frame, which is given by
\begin{equation}
{S}_{\text{Einstein}} = \int d^{4} x \sqrt{-{g}} \left[ \frac{{M}^{2}_{\text{pl}}}{2} {R} - \frac{1}{2} {\mathcal{G}}_{IJ} {g}^{{\mu}{\nu}} {\triangledown}_{\mu} {\phi}^{I} {\triangledown}_{\nu} {\phi}^{J} - {V}\left( {\phi}^{I} \right) \right]. \\
\end{equation}
\noindent and the potential in the Einstein frame becomes
\begin{equation}
\label{Action in Einstein frame}
V \left( {\phi}^{I} \right) = \frac{\tilde{V} \left( {\phi}^{I} \right)}{{\Omega}^{4} \left( x \right)} = \frac{{M}^{4}_{\text{pl}}}{4 f^2 \left( {\phi}^{I} \right)} \tilde{V} \left( {\phi}^{I} \right). \\
\end{equation}
\noindent The coefficients $\mathcal{G}_{{I}{J}}$ of the non-canonical kinetic terms in the Einstein frame depend on the non-minimal coupling function $f\left( {\phi}^{I} \right)$ and its derivatives. They are given by
\begin{equation}
\label{Field space metric transformation}
\mathcal{G}_{{I}{J}} \left( {\phi}^{K} \right) = \frac{{M}^{2}_{\text{pl}}}{2 f \left( {\phi}^{L} \right)} \left[ \tilde{\mathcal{G}}_{{I}{J}} \left( {\phi}^{K} \right) + \frac{3}{f \left( {\phi}^{L} \right)} {f}_{,I} {f}_{,J} \right], \\
\end{equation}
\noindent where ${f}_{,I}=\frac{{\partial}{f}}{{\partial}{\phi}^{I}}$. Varying the action in Einstein frame with respect to ${g}_{{\mu}{\nu}} \left( x \right)$, we have the Einstein equations
\begin{equation}
{R}_{{\mu}{\nu}} - \frac{1}{2}{g}_{{\mu}{\nu}} R = \frac{1}{{M}^{2}_{\text{pl}}} {T}_{{\mu}{\nu}}, \\
\end{equation}
\noindent where 
\begin{equation}
{T}_{{\mu}{\nu}} = \mathcal{G}_{{I}{J}} {\partial}_{\mu} {\phi}^{I} {\partial}_{\nu} {\phi}^{J} - {g}_{{\mu}{\nu}} \left[ \frac{1}{2} \mathcal{G}_{{K}{L}} {\partial}_{\gamma} {\phi}^{K} {\partial}^{\gamma} {\phi}^{L} + V\left( {\phi}^{K} \right) \right]. \\
\end{equation}
\noindent Varying Eq. (\ref{Action in Einstein frame}) with respect to ${\phi}^{I}$, we obtain the equation of motion for ${\phi}^{I}$
\begin{equation}
\square {\phi}^{I} + {g}^{{\mu}{\nu}} {\Gamma}^{I}_{{J}{K}} {\partial}_{\mu} {\phi}^{J} {\partial}_{\nu} {\phi}^{K} - \mathcal{G}^{{I}{K}} {V}_{,K} = 0, \\ 
\end{equation}
\noindent where $\square {\phi}^{I} = {g}^{{\mu}{\nu}} {\phi}^{I}_{;{\mu}{\nu}}$ and ${\Gamma}^{I}_{{J}{K}}$ is the Christoffel symbol for the field space manifold in terms of $\mathcal{G}_{{I}{J}}$ and its derivative. Expanding each scalar field to the first order around its classical background value, 
\begin{equation}
{\phi}^{I} \left( {x}^{\mu} \right) = {\varphi}^{I} \left( t \right) + {\delta} {\phi}^{I} \left( {x}^{\mu} \right), \\
\end{equation}
\noindent and perturbing a spatially flat Friedmann-Robertson-Walker (FRW) metric,
\begin{equation}
ds^2={g}_{{\mu}{\nu}} d{x}^{\mu} d{x}^{\nu} = - \left(1+2A \right) dt^2 + 2 a \left( {\partial}_{i} B \right) d{x}^{i} dt + a^2 \left[ \left( 1-2{\psi} \right) {\delta}_{ij} + 2 {\partial}_{i} {\partial}_{j} E \right] d{x}^{i} d{x}^{j}, \\
\end{equation}
\noindent where $a\left( t \right)$ is the scale factor. To the zeroth order, the $00$ and $ij$ components of the Einstein equations become
\begin{equation}
H^2 = \frac{1}{3 {M}^{2}_{\text{pl}}} \left[ \frac{1}{2} \mathcal{G}_{{I}{J}} \dot{\varphi}^{I} \dot{\varphi}^{J} + V \left( {\varphi}^{I} \right) \right], \\
\end{equation}
\begin{equation}
\dot{H} = - \frac{1}{2 {M}^{2}_{\text{pl}}} \mathcal{G}_{{I}{J}} \dot{\varphi}^{I} \dot{\varphi}^{J}, \\
\end{equation}
\noindent where $H = \frac{\dot{a} \left( t \right)}{a \left( t \right)}$ is the Hubble parameter, and the field field space metric is calculated at the zeroth order, $\mathcal{G}_{{I}{J}} = \mathcal{G}_{{I}{J}} \left( {\varphi}^{K} \right)$. Introducing the number of e-folding $N=\ln{a}$ with $d N = H dt$, the above Einstein equation becomes
\begin{equation}
3 {M}^{2}_{\text{pl}} - \frac{1}{2} \mathcal{G}_{{I}{J}} {{\varphi}^{I}}' {{\varphi}^{J}}' = \frac{V \left( {\varphi}^{I} \right)}{H^2}, \\
\end{equation}
\begin{equation}
\frac{H'}{H} = - \frac{1}{2 {M}^{2}_{\text{pl}}} \mathcal{G}_{{I}{J}} {{\varphi}^{I}}' {{\varphi}^{J}}', \\
\end{equation}
\noindent where the prime $'$ means the derivative with respect to $N$. For any vector in the field space $A^{I}$, we define a covariant derivative with respect to the field-space metric as usual by
\begin{equation}
\mathcal{D}_{J} {A}^{I} = {\partial}_{J} {A}^{I} + {\Gamma}^{I}_{{J}{K}} {A}^{K}, \\
\end{equation}
\noindent and the time derivative with respect to the cosmic time $t$ is given by
\begin{equation}
\mathcal{D}_{t} {A}^{I} \equiv \dot{\varphi}^{J} \mathcal{D}_{J} {A}^{I} = \dot{A}^{I} + {\Gamma}^{I}_{{J}{K}} \dot{\varphi}^{J} {A}^{K} = H \left( {{A}^{I}}' + {\Gamma}^{I}_{{J}{K}} {{\varphi}^{J}}' {A}^{K} \right). \\
\end{equation}
\noindent Now, we define the length of the velocity vector for the background fields as
\begin{equation}
|\dot{\varphi}^{I}| \equiv \dot{\sigma} = \sqrt{\mathcal{G}_{{P}{Q}} \dot{\varphi}^{P} \dot{\varphi}^{Q}} \quad \Rightarrow \quad |{{\varphi}^{I}}'| \equiv {\sigma}' = \sqrt{\mathcal{G}_{{P}{Q}} {{\varphi}^{P}}' {{\varphi}^{Q}}'}. 
\end{equation}
\noindent Introducing the unit vector of the velocity vector of the background fields
\begin{equation}
\hat{\sigma}^{I} \equiv \frac{\dot{\varphi}^{I}}{\dot{\sigma}} = \frac{{{\varphi}^{I}}'}{{\sigma}'} = \frac{{{\varphi}^{I}}'}{\sqrt{\mathcal{G}_{{P}{Q}} {{\varphi}^{P}}' {{\varphi}^{Q}}'}} 
\end{equation}
\noindent the $00$ and $ij$ components of the Einstein equations become
\begin{equation}
H^2 = \frac{1}{3 {M}^{2}_{\text{pl}}} \left[ \frac{1}{2} \dot{\sigma}^2 + V \right] \quad \Leftrightarrow \quad 3 {M}^{2}_{\text{pl}} - \frac{1}{2} {{\sigma}'}^{2} = \frac{V \left( {\varphi}^{I} \right)}{H^2} \quad \Leftrightarrow \quad \frac{V}{{M}_{\text{pl}}^{2} H^2} = \left( 3 - {\epsilon} \right), \\
\end{equation}
\begin{equation}
\dot{H} = - \frac{1}{2 {M}^{2}_{\text{pl}}} \dot{\sigma}^2 \quad \Leftrightarrow \quad \frac{H'}{H} = - \frac{1}{2 {M}^{2}_{\text{pl}}} {{\sigma}'}^{2} \quad \Leftrightarrow \quad \frac{\dot{\sigma}^{2}}{{M}_{\text{pl}}^{2} H^2} = 2 {\epsilon}, \\
\end{equation}
\noindent and the equation of motion of ${\phi}^{I}$ in the zeroth order is 
\begin{equation}
\ddot{\sigma} + 3 H \dot{\sigma} + {V}_{,\sigma} = 0 \quad \Leftrightarrow \quad \frac{\ddot{\sigma}}{H \dot{\sigma}} = - 3 - \frac{3 - {\epsilon}}{2 {\epsilon}}\frac{d}{d N} \left( \ln{V} \right), \\
\end{equation}
\noindent where
\begin{equation}
{V}_{,\sigma} \equiv \hat{\sigma}^{I} {V}_{,I}. \\
\end{equation}
\noindent and ${\epsilon}$ is the first order Hubble slow-roll parameter defined in Eq.(\ref{1st order Hubble slow-roll parameter}). Now, we define a quantity $\hat{s}^{{I}{J}}$ to obtain the field component orthogonal to $\hat{\sigma}^{I}$
\begin{equation}
\hat{s}^{{I}{J}} \equiv \mathcal{G}^{{I}{J}} - \hat{\sigma}^{I} \hat{\sigma}^{J}, \\
\end{equation}
\noindent which obeys the following relations with $\hat{\sigma}^{I}$
\begin{equation}
\begin{split}
\hat{\sigma}_{I} \hat{\sigma}^{I} = 1, \\
\hat{s}^{{I}{J}} \hat{s}_{{I}{J}} = \mathcal{N} - 1, \\
\hat{s}^{I}_{\; A} \hat{s}^{A}_{\; J} = \hat{s}^{I}_{\; J}, \\
\hat{\sigma}_{I} \hat{s}^{{I}{J}} = 0 \quad \forall J. \\
\end{split}
\end{equation}
\noindent The slow-roll parameters are given by
\begin{equation}
\label{1st order Hubble slow-roll parameter}
\epsilon \equiv - \frac{\dot{H}}{H^2} = \frac{3 \dot{\sigma}^2}{\dot{\sigma}^2 + 2 V} \quad \Leftrightarrow \quad \frac{ \dot{\sigma}^{2} }{V} = \frac{2 {\epsilon}}{3 - {\epsilon}}, \\
\end{equation}
\noindent and
\begin{equation}
{\eta}_{{\sigma}{\sigma}} \equiv {M}^{2}_{\text{pl}} \frac{\mathcal{M}_{{\sigma}{\sigma}}}{V} \quad \text{and} \quad {\eta}_{{s}{s}} \equiv {M}^{2}_{\text{pl}} \frac{\mathcal{M}_{{s}{s}}}{V}, \\
\end{equation}
\noindent where $\mathcal{M}^{I}_{J}$ is the effective mass squared matrix given by
\begin{equation}
\label{Effective mass-squared matrix}
\begin{split} 
{\mathcal{M}}^{I}_{J} \equiv&\; \mathcal{G}^{IK} \left( \mathcal{D}_{J} \mathcal{D}_{K} V \right) - \mathcal{R}^{I}_{LMJ} \dot{\varphi}^{L} \dot{\varphi}^{M}, \\
{\mathcal{M}}_{{\sigma}{J}} \equiv&\; \hat{\sigma}_{I} \mathcal{M}^{I}_{J} = \hat{\sigma}^{K} \left( \mathcal{D}_{K} \mathcal{D}_{J} V \right), \\
{\mathcal{M}}_{{\sigma}{\sigma}} \equiv&\; \hat{\sigma}_{I} \hat{\sigma}^{J} \mathcal{M}^{I}_{J} = \hat{\sigma}^{K} \hat{\sigma}^{J} \left( \mathcal{D}_{K} \mathcal{D}_{J} V \right), \\
{\mathcal{M}}_{{s}{J}} \equiv&\; \hat{s}_{I} \mathcal{M}^{I}_{J} = \hat{s}_{I} \left( \mathcal{G}^{IK} \left( \mathcal{D}_{J} \mathcal{D}_{K} V \right) - \mathcal{R}^{I}_{LMJ} \dot{\varphi}^{L} \dot{\varphi}^{M} \right), \\
{\mathcal{M}}_{{s}{s}} \equiv& \; \hat{s}_{I} \hat{s}^{J} \mathcal{M}^{I}_{J} = \hat{s}_{I} \hat{s}^{J} \left( \mathcal{G}^{IK} \left( \mathcal{D}_{J} \mathcal{D}_{K} V \right) - \mathcal{R}^{I}_{LMJ} \dot{\varphi}^{L} \dot{\varphi}^{M} \right), \\
\end{split}
\end{equation}
\noindent and $\hat{s}^{I}$ is defined in the following argument. Now we define the turn-rate vector ${\omega}^{I}$ as the covariant rate of change of the unit vector $\hat{\sigma}^{I}$
\begin{equation}
{\omega}^{I} \equiv \mathcal{D}_{t} \hat{\sigma}^{I} = - \frac{1}{\dot{\sigma}} {V}_{,K} \hat{s}^{{I}{K}} = \frac{-1}{H {\sigma}'} {V}_{,K} \hat{s}^{{I}{K}}. \\
\end{equation}
\noindent Since ${\omega}^{I} \propto \hat{s}^{{I}{K}}$, we have
\begin{equation}
{\omega}^{I} \hat{\sigma}_{I} = 0. \\
\end{equation}
\noindent We can also find 
\begin{equation}
\mathcal{D}_{t} \hat{s}^{{I}{J}} = - \hat{\sigma}^{I} {\omega}^{J} - \hat{\sigma}^{J} {\omega}^{I}. \\
\end{equation}
\noindent Also, we introduce a new unit vector $\hat{s}^{I}$ pointing in the direction of the turn-rate, ${\omega}^{I}$, and a new projection operator ${\gamma}^{{I}{J}}$
\begin{equation}
\hat{s}^{I} \equiv \frac{{\omega}^{I}}{\omega}, \\
\end{equation}
\begin{equation}
{\gamma}^{{I}{J}} \equiv {\mathcal{G}}^{{I}{J}} - \hat{\sigma}^{I} \hat{\sigma}^{J} - \hat{s}^{I} \hat{s}^{J}. \\
\end{equation}
\noindent where ${\omega} = |{\omega}^{I}|$ is the magnitude of the turn-rate vector. The new unit vector $\hat{s}^{I}$ and the new projection operator ${\gamma}^{{I}{J}}$ also satisfy
\begin{equation}
\begin{split}
\hat{s}^{{I}{J}} =& \hat{s}^{I} \hat{s}^{J} + {\gamma}^{{I}{J}}, \\
{\gamma}^{{I}{J}} {\gamma}_{{I}{J}} =& \mathcal{N} -2, \\
\hat{s}^{{I}{J}} \hat{s}_{J} =& \hat{s}^{I}, \\
\hat{\sigma}_{I} \hat{s}^{I} =& \hat{\sigma}_{I} {\gamma}^{{I}{J}} = \hat{s}_{I} {\gamma}^{{I}{J}} = 0 \quad \forall J.
\end{split}
\end{equation}
\noindent We then find 
\begin{equation}
{\mathcal{D}}_{t} \hat{s}^{I} = - {\omega} \hat{\sigma}^{I} - {\Pi}^{I}, \\
{\mathcal{D}}_{t} {\gamma}^{{I}{J}} = \hat{s}^{I} {\gamma}^{J} + \hat{s}^{J} {\gamma}^{I}, \\
\end{equation}
\noindent where
\begin{equation}
{\Pi}^{I} \equiv \frac{1}{\omega} {\mathcal{M}}_{{\sigma}{K}} {\gamma}^{{I}{K}}, \\
\end{equation}
\noindent and hence
\begin{equation}
\hat{\sigma}_{I} {\Pi}^{I} = \hat{s}_{I} {\Pi}^{I} = 0, \\
\end{equation}
\noindent Now, we define the curvature and entropic perturbations as follows
\begin{equation}
\label{Definition of curvature perturbation}
\mathcal{R} = {\psi} + \frac{H}{\dot{\sigma}} \hat{\sigma}_{J} {\delta} {\phi}^{J} = \frac{H}{\dot{\sigma}} {Q}_{\sigma}, \\
\end{equation}
\begin{equation}
\label{Definition of entropic perturbation}
\mathcal{S} = \frac{H}{\dot{\sigma}} {Q}_{s}, \\
\end{equation}

\noindent whose E.O.M.s are given by \cite{1310.8285}
\begin{equation}
\begin{split}
\ddot{Q}_{\sigma} + 3 H \dot{Q}_{\sigma} + \left[ \left( \frac{k}{a} \right)^{2} + \mathcal{M}_{{\sigma}{\sigma}} - {\omega}^{2} - \frac{1}{{M}_{\text{pl}}^{2} a^3} \frac{d}{dt} \left( \frac{a^3 \dot{\sigma}^{2}}{H} \right) \right] {Q}_{\sigma} =&\; 2 \frac{d}{dt} \left( {\omega} {Q}_{s} \right) - 2 \left( \frac{{V}_{, \sigma}}{\dot{\sigma}} + \frac{\dot{H}}{H} \right) {\omega} {Q}_{s}, \\
\ddot{Q}_{s} + 3 H \dot{Q}_{s} + \left[ \left( \frac{k}{a} \right)^{2} + \mathcal{M}_{ss} + 3 {\omega}^{2} \right] {Q}_{s} =&\; 4 {M}_{\text{pl}}^{2} \frac{\omega}{\dot{\sigma}} \frac{k^2}{a^2} {\Psi}, \\
\end{split}
\end{equation}
\noindent where $\Psi$ is the gauge-invariant Bardeen potential \cite{{astro-ph/0507632}, {0809.4944}}, $\mathcal{M}_{{\sigma}{\sigma}}$ and $\mathcal{M}_{ss}$ are given by Eq.(\ref{Effective mass-squared matrix}) and

\begin{equation}
\label{Effective mass squared of entropy perturbations}
{\mu}_{s}^{2} = \mathcal{M}_{ss} + 3 {\omega}^{2}, \\
\end{equation}
\noindent is an effective square mass of entropy perturbations. After the first horizon crossing, the co-moving wave number $k$ obeys $\frac{k}{aH}<1$. Hence, the curvature and entropic perturbations satisfy the following equations
\begin{equation}
\label{An evolution equation of curvature perturbation}
\dot{\mathcal{R}} = {\alpha} H \mathcal{S} + O \left( \frac{k^2}{a^2 H^2} \right), \\
\end{equation}
\begin{equation}
\label{An evolution equation of entropic perturbation}
\dot{\mathcal{S}} = {\beta} H \mathcal{S} + O \left( \frac{k^2}{a^2 H^2} \right), \\
\end{equation}
\noindent which allows us to write the transfer functions
\begin{equation}
\label{TRS integration in terms of cosmic time}
{T}_{\mathcal{RS}} \left( {t}_{\text{hc}}, t \right) = \int^{t}_{{t}_{\text{hc}}} d {t}' {\alpha} \left( t' \right) H \left( t' \right) {T}_{\mathcal{SS}} \left( {t}_{\text{hc}}, t' \right), \\
\end{equation}
\begin{equation}
\label{TSS integration in terms of cosmic time}
{T}_{\mathcal{SS}} \left( {t}_{\text{hc}}, t \right) = \exp \left[ \int^{t}_{{t}_{\text{hc}}} {dt'} {\beta} \left( t' \right) H \left( t' \right) \right]. \\
\end{equation}
\noindent where ${t}_{\text{hc}}$ is the time of the first horizon crossing. Being changed from the cosmic time $t$ into the number of e-folding $N=\ln{a}$, where $dN=Hdt$\footnote{In some literatures like \cite{1310.8285}, ${N}_{*} = {N}_{\text{tot}} - N\left( t \right)$ is used and the corresponding differential equation becomes $d {N}_{*} = - H dt$. But, in this paper, we keep using $dN=Hdt$.}, ${T}_{\mathcal{RS}} \left( {t}_{\text{hc}}, t \right)$ and ${T}_{\mathcal{SS}} \left( {t}_{\text{hc}}, t \right)$ become 
\begin{equation}
\label{TRS integration in terms of e-foldings}
{T}_{\mathcal{RS}} \left( {N}_{\text{hc}}, N \right) = \int^{N}_{{N}_{\text{hc}}} {dN'} {\alpha} \left( N' \right) {T}_{\mathcal{SS}} \left( {N}_{\text{hc}}, N' \right), \\
\end{equation}
\noindent and
\begin{equation}
\label{TSS integration in terms of e-foldings}
{T}_{\mathcal{SS}} \left( {N}_{\text{hc}}, N \right) = \exp \left[ \int^{N}_{{N}_{\text{hc}}} {dN'} {\beta} \left( N' \right) \right]. \\
\end{equation}
\noindent Now that the E.O.M.s of curvature and entropic perturbations are \cite{1210.7487}
\begin{equation}
\label{Differential equation of curvature perturbation}
\dot{\mathcal{R}} = 2 {\omega} \mathcal{S} + O \left( \frac{k^2}{a^2 H^2} \right), \\
\end{equation}
\noindent and
\begin{equation}
\label{Differential equation of entropic perturbation}
\dot{Q}_{s} \simeq - \frac{{\mu}^{2}_{s}}{3 H} {Q}_{s}, \\
\end{equation}
\noindent where the effective squared mass of the entropy perturbation can be written as that relative to the Hubble scale as
\begin{equation}
{\mu}^{2}_{s} = {\mathcal{M}}_{ss} + 3 {\omega}^{2} \quad \Leftrightarrow \quad \frac{{\mu}^{2}_{s}}{H^2} = \left( 3 - {\epsilon} \right) {\eta}_{ss} + \frac{3}{4} {\alpha}^{2}, \\
\end{equation}
\noindent and $\simeq$ means slow-roll approximation and ${\alpha}$ is given in Eq.(\ref{alpha}). Comparing with Eq.(\ref{Definition of curvature perturbation}), (\ref{Definition of entropic perturbation}), (\ref{An evolution equation of curvature perturbation}) and (\ref{An evolution equation of entropic perturbation}) with Eq.(\ref{Differential equation of curvature perturbation}) and (\ref{Differential equation of entropic perturbation}) \cite{1210.7487}, we obtain
\begin{equation}
\label{alpha}
{\alpha} \left( t \right) = \frac{2 {\omega} \left( t \right)}{H \left( t \right)} \quad \Leftrightarrow \quad {\alpha} \left( N \right) = \frac{2 {\omega} \left( N \right)}{H \left( N \right)}, \\
\end{equation}
\noindent and 
\begin{equation}
{\beta} = - \frac{{\mu}^{2}_{s}}{3 H^2} - {\epsilon} - \frac{\ddot{\sigma}}{H \dot{\sigma}} = - {\eta}_{ss} \left( 1 - \frac{1}{3}{\epsilon} \right) + \left( 3 - {\epsilon} \right) + \frac{3 - {\epsilon}}{2 {\epsilon}} \frac{d}{d N} \left({\ln{V}} \right) - \frac{1}{4} {\alpha}^{2}, \\
\end{equation}
\noindent Note that the power spectrum for the gauge invariant curvature perturbation is given by
\begin{equation}
\langle \mathcal{R} \left( \bold{k}_{1} \right) \mathcal{R} \left( \bold{k}_{2} \right) \rangle = \left( 2 \pi \right)^{3} {\delta}^{\left( 3 \right)} \left( \bold{k}_{1} + \bold{k}_{2} \right) {P}_{\mathcal{R}} \left( {k}_{1} \right), \\
\end{equation}
\noindent where ${P}_{\mathcal{R}} \left( k \right) = |\mathcal{R}|^{2}$. The dimensionless power spectrum is 
\begin{equation}
{\mathcal{P}}_{\mathcal{R}} = \frac{k^3}{2 {\pi}^2} |\mathcal{R}|^{2}, \\
\end{equation}
\noindent and the spectral index is defined as
\begin{equation}
{n}_{s} \equiv 1 + \frac{d \ln {\mathcal{P}}_{\mathcal{R}}}{d \ln {k}_{\text{hc}}}, \\
\end{equation}
\noindent where ${k}_{\text{hc}} = a \left( {t}_{\text{hc}} \right) H \left( {t}_{\text{hc}} \right)$ represents the pivot scale at the first horizon crossing ${t}_{\text{hc}}$, which is related to the cosmic time $t$ by
\begin{equation}
\frac{d \ln{k}}{dt} = \frac{d \left(aH \right)}{d t} = \frac{\dot{a}}{a} + \frac{\dot{H}}{H} = H \left( 1 + \frac{\dot{H}}{H^2} \right) = \left( 1 - {\epsilon} \right) H. \\
\end{equation}
Using the transfer function, we can relate the power spectra of adiabatic and entropic perturbations at time ${t}_{\text{hc}}$ to its value at some later time $t > {t}_{\text{hc}}$ with the corresponding pivot scale $k$ as
\begin{equation}
\begin{split}
{\mathcal{P}}_{\mathcal{R}} \left( k \right) =& \; {\mathcal{P}}_{\mathcal{R}} \left( {k}_{\text{hc}} \right) \left[ 1 + {T}^{2}_{\mathcal{RS}} \left( {t}_{\text{hc}}, t \right) \right], \\
{\mathcal{P}}_{\mathcal{S}} \left( k \right) =& \; {\mathcal{P}}_{\mathcal{R}} \left( {k}_{\text{hc}} \right) {T}^{2}_{\mathcal{SS}} \left( {t}_{\text{hc}}, t \right), \\
\end{split}
\end{equation}
\noindent The transfer functions are given by
\begin{equation}
\begin{split}
\frac{1}{H \left( {t}_{\text{hc}} \right)} \frac{{\partial} {T}_{\mathcal{RS}} \left( {t}_{\text{hc}}, t \right) }{{\partial} {t}_{\text{hc}}} &=\; - {\alpha} \left( {t}_{\text{hc}} \right) - {\beta} \left( {t}_{\text{hc}} \right) {T}_{\mathcal{SS}} \left( {t}_{\text{hc}}, {t} \right), \\
\frac{1}{H \left( {t}_{\text{hc}} \right)} \frac{{\partial} {T}_{\mathcal{SS}} \left( {t}_{\text{hc}}, t \right) }{{\partial} {t}_{\text{hc}}} &=\; - {\beta} \left( {t}_{\text{hc}} \right) {T}_{\mathcal{SS}} \left( {t}_{\text{hc}}, t \right). \\
\end{split}
\end{equation}
\noindent In term of the number of e-folding $N$, the above differential equation becomes
\begin{equation}
\begin{split}
\frac{{\partial} {T}_{\mathcal{RS}} \left( {N}_{\text{hc}}, N \right) }{ {\partial} {N}_{\text{hc}} } &=\; - {\alpha} \left( {N}_{\text{hc}} \right) - {\beta} \left( {N}_{\text{hc}} \right) {T}_{\mathcal{SS}} \left( {N}_{\text{hc}}, N \right), \\
\frac{{\partial} {T}_{\mathcal{SS}} \left( {N}_{\text{hc}}, N \right) }{{\partial} {N}_{\text{hc}}} &=\; - {\beta} \left( {N}_{\text{hc}} \right) {T}_{\mathcal{SS}} \left( {N}_{\text{hc}}, N \right). \\
\end{split}
\end{equation}
\noindent The spectral index for the power spectrum of the adiabatic fluctuations becomes
\begin{equation}
{n}_{s} \simeq {n}_{s} \left( {t}_{\text{hc}} \right) + \frac{1}{H} \left( \frac{{\partial} {T}_{\mathcal{RS}}}{{\partial} {t}_{\text{hc}}} \right) \sin{2 \Delta}, \\
\end{equation}
\noindent where
\begin{equation}
{n}_{s} \left( {t}_{\text{hc}} \right) = 1 - 6 {\epsilon} \left( {t}_{\text{hc}} \right) + 2 {\eta}_{{\sigma}{\sigma}} \left( {t}_{\text{hc}} \right), \\
\end{equation}
\noindent and the trigonometric functions for ${T}_{\mathcal{RS}}$ are defined as
\begin{equation}
\label{Trigo Delta}
\sin{\Delta} \equiv \; \frac{1}{\sqrt{1 + T^{2}_{\mathcal{RS}}}}, \quad \cos{\Delta} \equiv \frac{{T}_{\mathcal{RS}}}{\sqrt{1 + {T}^{2}_{\mathcal{RS}}}}, \quad \tan{\Delta} \equiv \; \frac{1}{T_{\mathcal{RS}}}. \\
\end{equation}
\noindent The iso-curvature fraction is given by
\begin{equation}
{\beta}_{\text{iso}} \equiv \frac{{\mathcal{P}}_{\mathcal{S}}}{{\mathcal{P}}_{\mathcal{R}} + {\mathcal{P}}_{\mathcal{S}}} = \frac{T^{2}_{\mathcal{SS}}}{1 + T^{2}_{\mathcal{SS}} + T^{2}_{\mathcal{RS}}}, \\
\end{equation}
\noindent which can be used for compared with the recent observables in Planck collaboration. Also, the tensor-to-scalar ratio is given by
\begin{equation}
r \simeq \frac{16 {\epsilon}}{1 + T^{2}_{\mathcal{RS}}}. \\
\end{equation}

\end{document}